\documentclass[aps,reprint,nofootinbib,nobibnotes,notitlepage,superscriptaddress,onecolumn,prd,
 amsmath,amssymb
]{revtex4-2}

\usepackage[caption=false]{subfig}
\usepackage{braket}
\usepackage{float}
\usepackage{lipsum}
\usepackage{graphicx}
\usepackage{dcolumn}
\usepackage{bm}
\usepackage{natbib}
\usepackage{hyperref}
\hypersetup{
	colorlinks = true,
    linkcolor = Red,
    urlcolor  = Red,
    citecolor = Red
}

\usepackage{microtype}

\usepackage{verbatim}
\usepackage[amssymb]{SIunits}
\usepackage{tabularx}
\usepackage[dvipsnames]{xcolor}
\usepackage{wasysym}

\usepackage{tikz}

\def\be{\begin{equation}}
\def\ee{\end{equation}}

\usepackage{color}
\definecolor{darkgreen}{RGB}{0,120,0}

\definecolor{darkgreen}{RGB}{0,120,0}
\newcommand{\resub}[1]{{#1}}

\newcommand{\av}[1]{\left\langle{#1}\right\rangle}

\newcommand{\C}{\mathsf{C}}
\newcommand{\Si}{\mathsf{S}^{-1}}
\newcommand{\F}{\mathcal{F}}

\newcommand{\Ci}{\mathsf{C}^{-1}}

\newcommand{\R}{\mathcal{R}}

\newcommand{\hn}{\hat{\vec n}}

\newcommand{\tjo}[3]{\begin{pmatrix} {#1} & {#2} & {#3}\\ 0 & 0 & 0\end{pmatrix}}
\newcommand{\tj}[6]{\begin{pmatrix} {#1} & {#2} & {#3}\\ {#4} & {#5} & {#6}\end{pmatrix}}

\def\beq{\begin{eqnarray}}
\def\eeq{\end{eqnarray}}

\let\vec\mathbf

\usepackage{empheq}

\def\mnras{\rm{MNRAS}}

\def\aap{\rm{A\&A}}

\def\apj{\rm{ApJ}}                 
\def\jcap{\rm{JCAP}}

\begin{document}

\title{Optimal Estimation of the Binned Mask-Free Power Spectrum, Bispectrum, and Trispectrum on the Full Sky: Tensor Edition}

\author{Oliver~H.\,E.~Philcox}
\email{ohep2@cantab.ac.uk}
\affiliation{Center for Theoretical Physics, Department of Physics,
Columbia University, New York, NY 10027, USA}
\affiliation{Simons Society of Fellows, Simons Foundation, New York, NY 10010, USA}

\begin{abstract} 
    \noindent We derive optimal estimators for the binned two-, three-, and four-point correlators of statistically isotropic tensor fields defined on the sphere, in the presence of arbitrary beams, inpainting, and masking. This is a conceptually straightforward extension of the associated scalar field estimators \citep{Philcox:2023uwe}, but upgraded to include spin-$2$ fields such as Cosmic Microwave Background polarization and galaxy shear, and parity-violating physics in all correlators. All estimators can be realized using spin-weighted spherical harmonic transforms and Monte Carlo summation and are are implemented in the public code \href{https://github.com/oliverphilcox/PolyBin}{\textsc{PolyBin}}, with computation scaling, at most, with the total number of bins. We perform a suite of validation tests verifying that the estimators are unbiased and, in limiting regimes, minimum variance. These facilitate general binned analyses of higher-point functions, and allow constraints to be placed on various pheomena, such as non-separable inflationary physics (novelly including polarized trispectra), non-linear evolution in the late Universe, and cosmic parity-violation.
\end{abstract}

\maketitle



\section{Introduction}\label{sec: intro}
Cosmology abounds with scalar fields: the density of galaxies, the temperature of the Cosmic Microwave Background (CMB), and the ionization fraction of the Universe, are a few such examples. Although these fields exist in three-dimensions, geometry forces us to observe them only on the surface of a two-sphere, whose radius corresponds to their distance, or, equivalently, age. Whilst the analysis of such observables has yielded significant information about the Universe's structure, composition and evolution, there is more information to be wrought if one additionally considers tensor fields. Such quantities (which are again observed on the two-sphere) contain also directional information, with canonical examples being the polarization of the CMB and the elliptical shapes of galaxies \citep{Spergel:1997vq,Seljak:1996gy,Bartelmann:1999yn,Zaldarriaga:2002qt,Kamionkowski:1996ks}. Many tensor fields exhibit symmetry under $90\degree$ rotations; these are known as spin-$2$ fields \citep{Kamionkowski:1996ks}, and their study has yielded important constraints on the dynamics of inflation (through $E$ and $B$ modes in the CMB) and the growth of structure (through gravitational lensing) \citep[e.g.,][]{Planck:2018jri,Planck:2019kim,2020A&A...641A...6P,Lewis:2006fu,Hildebrandt:2016iqg,DES:2021wwk,Dalal:2023olq}. Moreover, there exist cosmological fields with other spins, including the spin-$1$ orientation field of galaxies \citep[e.g.,][]{Motloch:2021mfz}, which shows $180\degree$ rotational symmetries, and a plurality of examples beyond cosmology, arising in disciplines such as oceanography and atmospheric physics. 

Information can be extracted from stochastic scalar and tensor fields such as the above through careful measurement and modeling of their statistical properties. This is principally performed using $N$-point functions, which encapsulate the correlations between values of the field at $N$ points on the sphere \citep{peebles80,Bartelmann:1999yn}. On large scales, the standard cosmological model asserts that the asymptotic distribution of these fields is Gaussian, and thus fully described by its two-point function or power spectrum \citep{2020A&A...641A...6P}. This provides a direct window to study the physics of the early Universe, with higher-point functions, such as the bispectrum and trispectrum allowing for tests of our assumptions, as well as parametrization of gravitationally-induced non-linearity. For this reason, it has been commonplace for CMB experiments to study three- and four-point functions of temperature fluctuations, allowing for a direct probe of the physics of inflation, and, on small-scales, gravitational lensing \citep[e.g.,][]{Planck:2019kim,Planck:2018lbu,Planck:2018jri,2006JCAP...05..004C,Fergusson:2010gn,Fergusson:2014gea,2009PhRvD..80d3510F,2012JCAP...12..032F,Smidt:2010ra}. Many of these include polarization data, though this has not yet been considered for large-scale trispectra. Similarly, the distribution of matter in the late Universe has begun to be probed using non-Gaussian statistics, both in photometric and spectroscopic surveys \citep[e.g.,][]{Bernardeau:2011tc,Dodelson:2005rf,Rizzato:2018whp,Schneider:2002ze,Zaldarriaga:2002qt,2001ApJ...546..652S,2015MNRAS.451..539G,2018MNRAS.478.4500P,2018MNRAS.478.4500P,Cabass:2022oap,Cabass:2022wjy,Cabass:2022ymb,Philcox:2021kcw,Ivanov:2023qzb,Philcox:2022hkh,Hou:2022wfj,Philcox:2021hbm,2020JCAP...05..005D},  though the latter usually works in deprojected three-dimensional space.

To perform accurate analyses of cosmological fields on the two-sphere, we require robust estimators for the various statistics. In general, this is non-trivial, since the fields in question are rarely observed uniformly or even across the entire sky. Furthermore, experimental noise and telescope beams impact different scales by different amounts, and there are often gaps in the survey observations due to the observer effects such as Milky Way extinction and bright stars. This imprints a complex `window function' (or `mask') onto the observations, leading to leakage between different scales and polarization components. Whilst robust methods exist for accounting for such effects in the power spectrum \citep{Hivon:2001jp}, there has been comparably little discussion for higher-order statistics, particularly when tensors are involved (though see \citep{2000PhRvD..62j3004G,2011MNRAS.412.1993M,Bucher:2015ura,Yadav:2007ny,Duivenvoorden:2019ses,Fergusson:2014gea,Gruetjen:2015sta,Munshi:2009wy,Fergusson:2010gn,2009PhRvD..80d3510F,Shiraishi:2014roa,2015arXiv150200635S,2011MNRAS.417....2S,Kamionkowski:2010me}). Much of this is pragmatic: many of the current analyses of CMB bispectra and trispectra look only for the amplitude of a single inflationary component \resub{(parameterized by some ``non-linear'' amplitude, such as $f_{\rm NL}$ or $g_{\rm NL}$)}, whose biases can be well established using Monte Carlo simulations \citep{Planck:2019kim,Duivenvoorden:2019ses}. If we wish to study more complex inflationary models (which are not separable in harmonic space \citep[e.g.,][]{Arkani-Hamed:2015bza}) or to include gravitational contributions to the model, it is advantageous to instead measure the full statistic in some set of bins (rather than just \resub{a characteristic non-linear amplitude, such as $f_{\rm NL}$}) \citep[cf.,][]{Bucher:2015ura,Sohn:2023fte}. In this case, it is important to account for the various observational effects listed above, such that the output spectra can be reliably compared to models.

In \citep{Philcox:2023uwe} we introduced a formalism for efficient and accurate computation of $N$-point functions sourced by scalar fields on the two-sphere. Starting from the theoretical likelihood for the observed field, this derived optimal estimators for scalar polyspectra, which, by definition, had minimum variance in the Gaussian limit (following previous work including \citep{1997PhRvD..55.5895T,1998ApJ...499..555T,Hamilton:2005kz,Hamilton:2005ma,Hamilton:1999uw,2011MNRAS.417....2S,2015arXiv150200635S,2021PhRvD.103j3504P,Philcox:2021ukg}). A slight generalization of these allowed for polyspectra to be efficiently computed using arbitrary weighting schemes (allowing for any linear masking, inpainting, deprojection, \textit{et cetera}), utilizing iterated spherical harmonic transforms \citep{Philcox:2023uwe}, which become highly efficient on large scales. An important feature of such estimates is that they are unbiased, \textit{i.e.}\ they deconvolve all mask- and beam-induced effects such that the output spectra can be directly compared to theoretical models. The physical utility of these has already been demonstrated: in \citep{PhilcoxCMB}, we performed the first measurement of the parity-odd component of the \textit{Planck} CMB trispectrum, placing strong constraints on parity-breaking models of the early Universe, \resub{and refuting (scalar) inflationary interpretation of the recent claims of parity-violation in galaxy survey data \citep{Philcox:2022hkh,Hou:2022wfj,Cabass:2022oap}.} 

This work considers the analogous problem applied to tensor fields. In particular, we construct estimators for the power spectrum, bispectrum, and trispectrum of correlated spin-$0$ and spin-$2$ fields on the two-sphere, in the presence of masking, inpainting, and arbitrary linear weighting schemes. These are significantly more nuanced than their scalar equivalents, due to the introduction of additional variables ($E$, and $B$ fields), and the corresponding proliferation of output spectra. However, these can still be computed efficiently on large-scales, with computation time scaling at most linearly with the number of bins. We additionally allow computation of spectra sourced by both parity-conserving and parity-breaking physics; the first time, to our knowledge, that this has been performed for higher-point tensor observables. \resub{The latter could be occur due to a range of new physics in inflation (such as couplings to vector fields, Chern-Simons interactions, loops, and beyond \citep{Shiraishi:2016mok,CyrilCS,Gluscevic:2010vv,Bartolo:2014hwa,Lue:1998mq,Cabass:2022rhr}), or via some form of chiral gravity modifying the observables at late times \citep{Cabass:2022oap}.} In all cases, our estimators are unbiased (such that they can be straightforwardly compared to theoretical models), and can be implemented using spin-weighted spherical harmonic transforms. We additionally provide a general purpose implementation in the \href{https://github.com/oliverphilcox/PolyBin}{\textsc{PolyBin}} code \citep{PolyBin}.\footnote{Available at \href{https://github.com/oliverphilcox/PolyBin}{GitHub.com/OliverPhilcox/PolyBin}.} This should allow for a range of novel theoretical tests, both for CMB and weak lensing analyses, such as a first measurement of the large-scale polarized trispectrum.

\vskip 4pt
The remainder of this paper is as follows. In \S\ref{sec: defs} we set out our conventions for spinning fields on the two-sphere and the various correlators, before discussing the general form of $N$-point function estimators in \S\ref{sec: estimators}. \S\ref{sec: Cl},\,\ref{sec: bl},\,\&\ref{sec: tl} discuss practical computation of the power spectrum, bispectrum, and trispectrum respectively, before the estimators are validated with a suite of tests in \ref{sec: validation}. Finally, we conclude in \S\ref{sec: summary}. We caution that the remainder of this paper is (quite necessarily) mathematically dense. Much of the material builds upon the spin-$0$ work presented in \citep{Philcox:2023uwe}, but we recapitulate the relevant material here for completeness.

\section{Tensors and Correlators}\label{sec: defs}
\subsection{Definition}
A spin-$s$ field defined on the two-sphere, ${}_sa(\hn)$, can be expanded in spin-weighted spherical harmonics as
\beq
    {}_sa(\hn) = \sum_{\ell m}{}_sa_{\ell m}\,{}_sY_{\ell m}(\hn), \qquad {}_sa_{\ell m} = \int d\hn\,\left[{}_sY_{\ell m}(\hn)\right]^*\,{}_sa(\hn).
\eeq
Assuming the underlying fields to be real with $\left[{}_sa(\hn)\right]^* = {}_{-s}a(\hn)$, the spin-components satisfy
\beq
    \left[{}_sa_{\ell m}\right]^* = (-1)^{s+m}{}_{-s}a_{\ell (-m)}, \qquad \mathbb{P}\left[{}_sa_{\ell m}\right] = (-1)^{\ell}{}_{-s}a_{\ell m},
\eeq
where $\mathbb{P}$ is the parity operator. For spin-$\pm2$ fields, it is conventional to expand in $E$ and $B$ modes:
\beq
    {}_{\pm 2}a_{\ell m} = -\left[a^E_{\ell m}\pm ia^B_{\ell m}\right],
\eeq
where $a^E$ and $a^B$ give parity-even and parity-odd contributions respectively (both of which are real in map-space) and the negative sign appears by convention. These satisfy
\beq\label{eq: conjugation-relations-TEB}
    \left[a^{E,B}_{\ell m}\right]^* = (-1)^{m}a^{E,B}_{\ell (-m)}, \qquad \mathbb{P}\left[a^{E,B}_{\ell m}\right] = p_{E,B}(-1)^{\ell}a^{E,B}_{\ell m},
\eeq
where $p_{E}=1$ and $p_B=-1$ specifies the parity. A similar decomposition is possible for spin-$1$ fields, in terms of gradient (parity-odd) and curl (parity-even) contributions.

For the cosmic microwave background, there are three fields of interest: the spin-$0$ temperature, ${}_0a = T$, and the spin-$\pm2$ polarization fields ${}_{\pm2}a = Q\pm iU$. These are related to $T$, $E$, and $B$ fields via a linear transformation
\beq\label{eq: QU-EB-matrix}
    \begin{pmatrix}{}_{\,\,\,\,0}a_{\ell m}\\ {}_{-2}a_{\ell m}\\ {}_{+2}a_{\ell m}\end{pmatrix} = \begin{pmatrix}1 & 0 &0 \\ 0 & -1 & -i\\ 0 & -1 & i\end{pmatrix}\begin{pmatrix}a^T_{\ell m}\\a^E_{\ell m}\\ a^B_{\ell m}\end{pmatrix}\quad\Rightarrow\quad{}_sa_{\ell m} \equiv \sum_X{}_s\mathcal{R}_Xa^X_{\ell m},
\eeq
where $X\in\{T,E,B\}$, and the transformation matrix (which is not a rotation) will be of use later. An analogous form can be written for galaxy positions and shapes, whence ${}_0a = \delta_g$ (the galaxy density), and ${}_{\pm2}a = \gamma_1\pm i\gamma_2$ (the galaxy shear).

\subsection{Power Spectra}
\subsubsection{Definition}
The correlators of spin-$2$ fields are fundamentally described by the power spectra of $T,E,B$:
\beq
    \av{a^X_{\ell_1m_1}a^{Y*}_{\ell_2m_2}} = C_{\ell_1}^{XY}\,\delta^{\rm K}_{\ell_1\ell_2}\delta^{\rm K}_{m_1m_2}\qquad X,Y\in\{T,E,B\},
\eeq
assuming rotation and translation invariance and noting that auto-spectra include noise. We additionally ignore any spatially-varying window function, but will return to this later. In terms of the spin fields in harmonic space, we have the ideal correlators
\beq\label{eq: s1s2-power}
    \av{{}_{s_1}a_{\ell_1m_1}\left[{}_{s_2}a_{\ell_2m_2}\right]^*} = {}_{s_1s_2}C_{\ell_1}\,\delta^{\rm K}_{\ell_1\ell_2}\delta^{\rm K}_{m_1m_2},
\eeq
where the (Hermitian) ${}_{ss'}C_\ell$ matrix of correlators can be written in terms of $T,E,B$ as
\beq
    {}_{ss'}C_\ell = \sum_{XY}\left({}_s\mathcal{R}_X\right)C_\ell^{XY}\left({}_{s'}\mathcal{R}_Y\right)^\dag = \begin{pmatrix} C_\ell^{TT} & -\left(C_\ell^{TE}-i\,C_\ell^{TB}\right) & -\left(C_\ell^{TE}+i\,C_\ell^{TB}\right)\\ -\left(C_\ell^{TE}+i\,C_\ell^{TB}\right) & C_\ell^{EE}+C_\ell^{BB}& C_\ell^{EE}-C_\ell^{BB}+2i\,C_\ell^{EB}\\ -\left(C_\ell^{TE}-i\,C_\ell^{TB}\right) & C_\ell^{EE}-C_\ell^{BB}-2i\,C_\ell^{EB}& C_\ell^{EE}+C_\ell^{BB}\end{pmatrix}
\eeq
where we note that $C_\ell^{XY}=C_\ell^{YX}$ are real. The full set of non-trivial correlators involve the following fields
\beq\label{eq: Cl-pairs}
    U_2\equiv \{TT, TE, EE, BB\} + \{TB, EB\},
\eeq
separating out parity-even and parity-odd spectra, with the latter contributing to the spin-$(s,s')$ covariance multiplied by $i$. In general, the two-point covariance is non-trivial, but significantly simplifies if we assume parity-conservation and zero intrinsic $B$ mode:
\beq
    \left({}_{ss'}C_\ell\right)^{\rm ideal} = \begin{pmatrix} C_\ell^{TT} & -C_\ell^{TE} & -C_\ell^{TE}\\ -C_\ell^{TE} & C_\ell^{EE}& C_\ell^{EE}\\ -C_\ell^{TE} & C_\ell^{EE}& C_\ell^{EE}\end{pmatrix}+\begin{pmatrix} N^T_\ell & 0 &0 \\ 0 & 2N^P_\ell & 0\\ 0 & 0 & 2N^P_\ell\end{pmatrix}.
\eeq
We have additionally separated out contributions from noise, assuming the impact on the temperature and $E/B$ fields to be independent and described by spectra $N_\ell^T$ and $N_\ell^P$ respectively.

\subsubsection{Application}
To form optimal estimators, we will require both the pixel-space covariance matrix and its inverse. From \eqref{eq: s1s2-power}, these are given by
\beq\label{eq: pixel-covariance}
    {}_{s_1s_2}\mathsf{C}(\hn_1,\hn_2) &\equiv& \av{{}_{s_1}a(\hn_1)[{}_{s_2}a(\hn_2)]^*} = \sum_{\ell m}{}_{s_1}Y_{\ell m}(\hn_1)[{}_{s_1s_2}C_\ell]{}_{s_2}Y^*_{\ell m}(\hn_2)\\\nonumber
    {}_{s_1s_2}\mathsf{C}^{-1}(\hn_1,\hn_2) &=& \sum_{\ell m}{}_{s_1}Y_{\ell m}(\hn_1)[{}_{s_1s_2}(C^{-1})_\ell]{}_{s_2}Y^*_{\ell m}(\hn_2)
\eeq
in terms of spherical harmonics. The latter can be written in terms of the matrix of $T/E/B$ correlators $\mathbb{C}_\ell \equiv \{C_\ell^{XY}\}$
\beq
    {}_{s_1s_2}\mathsf{C}^{-1}(\hn_1,\hn_2) &=& \sum_{\ell m}{}_{s_1}Y_{\ell m}(\hn_1)\,{}_{s_1s_2}[\R^{-\dag}\mathbb{C}_\ell^{-1}\R^{-1}]\,{}_{s_2}Y^*_{\ell m}(\hn_2).
\eeq
Whilst this may seem unduly complicated, it can be straightforwardly applied to any map in question, noting that $\R$ and $\mathbb{C}$ are just $3\times 3$ matrices. Acting on some map ${}_s x(\hn)$, we find
\beq
    {}_{s_1}[\mathsf{C}^{-1}x](\hn_1) &=& \sum_{\ell m}{}_{s_1}Y_{\ell m}(\hn_1)\,{}_{s_1s_2}[\R^{-\dag}\mathbb{C}_\ell^{-1}\R^{-1}]\,{}_{s_2}x_{\ell m},
\eeq
replacing the $\hn_2$ integral by a harmonic transform of $x$. The first $\R^{-1}$ operator transforms ${}_sx_{\ell m}$ into the $T/E/B$ representation $x_{\ell m}^X$, which is then Wiener filtered by $\mathbb{C}^{-1}_\ell$ and transformed back to the spin representation by $\R^{-1}$, before being projected into real-space by an inverse harmonic transform. Similarly, the inner product can be written
\beq
    x^\dag\mathsf{C}^{-1}y &\equiv& \sum_{s_1s_2}\int d\hn_1d\hn_2 \,{}_{s_1}x^*(\hn_1)\,{}_{s_1s_2}\mathsf{C}^{-1}(\hn_1,\hn_2)\,{}_{s_2}y(\hn_2)\\\nonumber
    &=& \sum_{\ell m}\sum_{XY}[x^X_{\ell m}]^*(\mathbb{C}_\ell)^{-1}_{XY}y^Y_{\ell m},
\eeq
which is straightforward to compute.

\subsection{Bispectra}
The bispectra of three isotropic and homogeneous harmonic space fields can be written
\beq\label{eq: reduced-bis}
    \av{a^X_{\ell_1m_1}a^Y_{\ell_2m_2}a^Z_{\ell_3m_3}} \equiv w^{\ell_1\ell_2\ell_3}_{m_1m_2m_3}b^{XYZ}_{\ell_1\ell_2\ell_3}, \qquad \{X,Y,Z\}\in\{T,E,B\},
\eeq
where $b$ is the reduced bispectrum and $|\ell_1-\ell_2|\leq\ell_3\leq\ell_1+\ell_2$. In this work, we adopt the following weights (which differ from those of \citep{Philcox:2023uwe})
\beq\label{eq: bis-weighting}
    w^{\ell_1\ell_2\ell_3}_{m_1m_2m_3} &=& \sqrt{\frac{(2\ell_1+1)(2\ell_2+1)(2\ell_3+1)}{4\pi}}\tj{\ell_1}{\ell_2}{\ell_3}{m_1}{m_2}{m_3}\left[\frac{1}{3}\tj{\ell_1}{\ell_2}{\ell_3}{-1}{-1}{2}+\text{2 cyc.}\right]\\\nonumber
    &=& \frac{1}{3}\int d\hn\,\left[{}_{+1}Y_{\ell_1m_1}(\hn){}_{+1}Y_{\ell_2m_2}(\hn){}_{-2}Y_{\ell_3m_3}(\hn)+\text{2 cyc.}\right],
\eeq
expressing in terms of the Gaunt integral in the second line. Usually, one fixes $w^{\ell_1\ell_2\ell_3}_{m_1m_2m_3}$ to the Gaunt symbol: however, this nulls any contributions from parity-odd trispectra (as discussed in \citep{Shiraishi:2014roa}), since it is zero for odd $\ell_1+\ell_2+\ell_3$. Here, the only restriction is that $\ell_i\geq 2$, which is usually true (after monopole and dipole subtraction). Furthermore, the bispectrum is symmetric under interchange of any $\{\ell_i,X\}$ pairs; similarly the weights are symmetric under exchange of any pair of $\{\ell_i,m_i\}$. Given these symmetries, the bispectrum is fully specified by the following components: 
\beq\label{eq: Bl-triplets}
    U_3\equiv \{TTT,TTE,TEE,EEE,TBB,EBB\}+\{TTB,TEB,EEB,BBB\},
\eeq
where the two groups have $p_{XYZ}=p_Xp_Yp_Z=1$ and $-1$ respectively. An additional result is that not all $\ell$-bins are independent. To fully specify the spectra we require the triplets of \eqref{eq: Bl-triplets} with 
\beq\label{eq: bin-sym-bis}
    \ell_1\leq \ell_2\quad (X=Y), \qquad \ell_2\leq \ell_3\quad (Y=Z),
\eeq
with no restrictions on $\ell_i$ (except from the triangle conditions) else.\footnote{Our convention differs from that of \textit{Planck} \citep{Planck:2019kim}, which asserted the $X=Y=Z$ conditions for all correlators, but included also permutations e.g., $TET$ in their analysis. Both approaches are equivalent, though we find it useful to minimize the number of output spectra, particularly for higher-order correlators.}

Finally, we note that, due to the symmetries of the weighting, the bispectrum has the following properties under conjugation and parity:
\beq
    \left(b_{\ell_1\ell_2\ell_3}^{XYZ}\right)^* = (-1)^{\ell_1+\ell_2+\ell_3}b^{XYZ}_{\ell_1\ell_2\ell_3},\qquad \mathbb{P}\left[b_{\ell_1\ell_2\ell_3}^{XYZ}\right] = p_{XYZ}(-1)^{\ell_1+\ell_2+\ell_3}b^{XYZ}_{\ell_1\ell_2\ell_3}
\eeq
using \eqref{eq: conjugation-relations-TEB}. As such, configurations of fields with $p_{XYZ}=1$ (the first group in \ref{eq: Bl-triplets}) have real bispectra for even $\ell_1+\ell_2+\ell_3$ and imaginary else, whilst the situation reverses for $p_{XYZ}=-1$. Note that rotational invariance forbids scalar physics from generating a parity-odd three-point function, thus $TTT$ (\textit{et cetera}) do not contain contributions with odd $\ell_1+\ell_2+\ell_3$, and cannot be used to probe primordial scalar parity-violation. Parity-even spectra with a single $B$ mode can be generated from scalars however and appear at second order in cosmic perturbation theory \citep{Coulton:2021yjq}. As such, scalar physics will generate only four spectra at leading order (ignoring, e.g., lensing): ($b^{TTT},b^{TTE},b^{TEE},b^{EEE}$).

In our case, we require the bispectrum of the spin-$s$ components directly, since these can be straightforwardly related to the real-space windowed quantities. Noting that ${}_sa_{\ell m} = {}_s\R_X a^X_{\ell m}$, in the notation of \eqref{eq: QU-EB-matrix}, we can write
\beq\label{eq: spin-to-TEB-bis}
    \av{{}_{s_1}a_{\ell_1 m_1}\,{}_{s_2}a_{\ell_2 m_2}\,{}_{s_3}a_{\ell_3 m_3}} = \sum_{XYZ}w^{\ell_1\ell_2\ell_3}_{m_1m_2m_3}\left({}_{s_1}\R_X\right)\left({}_{s_2}\R_Y\right)\left({}_{s_3}\R_Z\right)b^{XYZ}_{\ell_1\ell_2\ell_3}.
\eeq
Unlike the trispectrum, we do not require permutations in this definition, since the weights are symmetric under $(\ell,m)$ exchange.

\subsection{Trispectra}
The relation for trispectra follows similarly. In this case, we write the general definition
\beq\label{eq: tl-definition}
    \av{a^X_{\ell_1m_1}a^Y_{\ell_2m_2}a^Z_{\ell_3m_3}a^W_{\ell_4m_4}} \equiv \sum_{LM}(-1)^Mw^{L(-M)}_{\ell_1\ell_2m_1m_2}w^{LM}_{\ell_3\ell_4m_3m_4}t^{XY,ZW}_{\ell_1\ell_2,\ell_3\ell_4}(L) + \text{23 perms.}, \quad \{X,Y,Z,W\}\in\{T,E,B\},
\eeq
where the permutations are over the joint set $\{X,\ell,m\}$, and the relevant weights are similar to before;
\beq\label{eq: tl-weights}
    w^{LM}_{\ell_1\ell_2m_1m_2} &=& \sqrt{\frac{(2\ell_1+1)(2\ell_2+1)(2L+1)}{4\pi}}\tj{\ell_1}{\ell_2}{L}{m_1}{m_2}{M}\tj{\ell_1}{\ell_2}{L}{-1}{-1}{2}\\\nonumber
    &=& \int d\hn\,{}_{+1}Y_{\ell_1 m_1}(\hn){}_{+1}Y_{\ell_2m_2}(\hn){}_{-2}Y_{LM}(\hn),
\eeq
requiring $\ell_i\geq 1$ and $L\geq 2$. The trispectrum has the following symmetries:
\beq\label{eq: tl-syms}
    t^{XY,ZW}_{\ell_1\ell_2,\ell_3\ell_4}(L) = t^{YX,ZW}_{\ell_2\ell_1,\ell_3\ell_4}(L) = t^{ZW,XY}_{\ell_3\ell_4,\ell_1\ell_2}(L),
\eeq
with other pairs non-trivially related due to the degeneracy in the definition of tetrahedron diagonals. This additionally requires the triangle conditions $|\ell_1-\ell_2|\leq L\leq \ell_1+\ell_2$, $|\ell_3-\ell_4|\leq L\leq \ell_3+\ell_4$.

The relevant trispectra are defined by the following 21 quadruplets of fields
\beq\label{eq: Tl-quads}
    U_4&\equiv&\,\,\{TTTT,TTTE,TTEE,TTBB,TETE,TEEE,TEBB,TBTB,TBEB,EEEE,EEBB,EBEB,BBBB\}\nonumber\\
    &&\,+\{TTTB,TTEB,TETB,TEEB,TBEE,TBBB,EEEB,EBBB\},
\eeq
with $p_{XYZW}=1$ and $-1$ respectively. As before, the trispectrum symmetries of \eqref{eq: tl-syms} imply that we can apply the following restrictions on $\ell_i$ to yield a complete set of bins
\beq\label{eq: tl-bin-conditions}
    &&\ell_1\leq \ell_2\quad (X=Y), \qquad \ell_3\leq \ell_4\quad (Z=W)\\\nonumber
    &&\ell_1\leq \ell_3\quad (X=Z \cup\,Y=W),\qquad \ell_2\leq \ell_4 \,\,\,\textit{if}\,\,\,\ell_1=\ell_3 \quad (X=Z \cup\,Y=W).
\eeq

Here, the trispectrum satisfies
\beq
    \left(t_{\ell_1\ell_2,\ell_3\ell_4}^{XY,ZW}(L)\right)^* = (-1)^{\ell_1+\ell_2+\ell_3+\ell_4}t_{\ell_1\ell_2,\ell_3\ell_4}^{XY,ZW}(L), \qquad \mathbb{P}\left[t_{\ell_1\ell_2,\ell_3\ell_4}^{XY,ZW}(L)\right] = p_{XYZW}(-1)^{\ell_1+\ell_2+\ell_3+\ell_4}t_{\ell_1\ell_2,\ell_3\ell_4}^{XY,ZW}(L)
\eeq
such that spectra with $p_{XYZW}=1$ are parity-even for even $\ell_1+\ell_2+\ell_3+\ell_4$ and parity-odd for odd $\ell_1+\ell_2+\ell_3+\ell_4$ whilst this is reversed for $p_{XYZW}=-1$. In linear CMB theory, scalar physics only sources spectra containing $T$ and $E$ modes at leading order, which reduces the number of non-trivial spectra to six (though $B$-modes appear at second order and above, and through lensing). Finally, in terms of the spin-weighted fields we can write
\beq\label{eq: tl-unbinned-def}
    \av{{}_{s_1}a_{\ell_1m_1}{}_{s_2}a_{\ell_2m_2}{}_{s_3}a_{\ell_3m_3}{}_{s_4}a_{\ell_4m_4}} &\equiv& \sum_{LM}(-1)^Mw^{L(-M)}_{\ell_1\ell_2m_1m_2}w^{LM}_{\ell_3\ell_4m_3m_4}\left({}_{s_1}\R_X\right)\left({}_{s_2}\R_Y\right)\left({}_{s_3}\R_Z\right)\left({}_{s_4}\R_W\right)t^{XY,ZW}_{\ell_1\ell_2,\ell_3\ell_4}(L)\\\nonumber
    &&\,\quad\,+\,\text{23 perms.}
\eeq

\subsection{Binning}\label{subsec: binning}
To form estimators, we must relate the continuous spectra defined above to their binned forms (which we measure directly). This is somewhat non-trivial in the presence of polarization, since only some of the various correlators are independent, with, for example, the $BT$ power spectrum being fully defined by $TB$.

For the power spectrum, we begin by expressing $C_\ell^{XY}$ in terms of the binned quantities of interest: $C^u(b)$, where $b$ specifies an $\ell$-bin and $u\in U_2$ specifies a pair of fields from the set given in \eqref{eq: Cl-pairs}. For arbitrary $XY$ pairs, we can write
\beq\label{eq: C-ell-binned}
    C^{XY}_\ell = \sum_{b,u\in U_2}\frac{C^u(b)}{\Delta^u_2}\left[\delta_{\rm K}^{u,XY}+\delta_{\rm K}^{u,YX}\right]\Theta_\ell(b).\,
\eeq
where $\Theta_\ell(b)$ is unity if $\ell$ is in bin $b$ and zero else. Here, the symmetry factor $\Delta^u_2$ is two if both components of $u$ are equal and unity else (\textit{i.e.}\ $\Delta_u^2 = 1+\delta_{\rm K}^{u_1u_2}$ for $u\equiv u_1u_2$). In each case, the sum over pairs and bins picks out the relevant spectrum, for example, $C_\ell^{BT}$ is mapped to $-C_\ell^{TB}$. Strictly, the above result is valid only in the narrow bin limit, but this makes little difference for the higher-point functions. The above result can be straightforwardly inserted into \eqref{eq: s1s2-power}, and allows extraction of the derivatives of the ${}_sa_{\ell m}$ covariance with respect to the desired bandpowers. 

A similar procedure is possible for the bispectrum. In this case, we sum over all triplets of fields $s\in S_3$ \eqref{eq: Bl-triplets}:
\beq\label{eq: b-ell-binned}
    b^{XYZ}_{\ell_1\ell_2\ell_3} = \frac{1}{2}\sum_{\vec b,u\in U_3,\chi\in \pm1}\frac{b^u_\chi(\vec b)}{\Delta^u_3(\vec b)}\left[\delta_{\rm K}^{u,XYZ}\Theta_{\ell_1}(b_1)\Theta_{\ell_2}(b_2)\Theta_{\ell_3}(b_3)+\text{5 perms.}\right]\left[1+\chi \,p_u(-1)^{\ell_1+\ell_2+\ell_3}\right],
\eeq
where the allowed values of the bins $\vec b\equiv \{b_1,b_2,b_3\}$ are set by the symmetries of \eqref{eq: bin-sym-bis} (\textit{i.e.}\ $b_1\leq b_2$ if $u_1=u_2$ and $b_2\leq b_3$ if $u_2=u_3$, for $u\equiv u_1u_2u_3$).\footnote{In the narrow-bin limit, bins with $b_1=b_2=b_3$ have even $\ell_1+\ell_2+\ell_3$, thus cannot yield signals for $\chi\,p_u=-1$. In \citep{Philcox:2023uwe}, such bins were dropped; here, we keep them for consistency between parity-even and parity-odd definitions, but note that they will have large variances. A similar conclusion holds for certain trispectrum components.} Here the degeneracy factor $\Delta^u_3(\vec b)$ is defined by
\beq
    \Delta_3^u(\vec b) = \begin{cases} 6 & b_1=b_2=b_3 \text{ and } u_1 = u_2 = u_3\\2 & b_1=b_2 \text{ and } u_1=u_2\\ 2 & b_2=b_3 \text{ and } u_2 = u_3\\ 1 & \text{else,}\end{cases}
\eeq
where the conditions should be read sequentially. Finally, we have inserted a parity-factor $\left[1+\chi\,p_u(-1)^{\ell_1+\ell_2+\ell_3}\right]/2$; this allows us to separate parity-even and parity-odd components which source even and odd $\ell_1+\ell_2+\ell_3$ (depending on the spectrum of interest). Here, we denote conventional parity-conserving contributions by $b_+^u(\vec b)$ and those sourced by parity-odd effects (e.g., chiral gravitational waves) by $b_-^u(\vec b)$. Note that this is not required for the power spectrum, since spectra with $p_u = -1$ ($TB$ and $EB$) can only be sourced by parity-breaking physics (and likewise for $p_u=1$ and parity-conserving physics).

Finally, the binned trispectrum for fields $X,Y,Z,W$ can be written as a sum over all trispectra listed in \eqref{eq: Tl-quads}, for bins $\vec b\equiv\{b_1,b_2,b_3,b_4\}$ in $\ell_i$ and bin $B$ in $L$:
\beq\label{eq: binned-trispectrum}
    \av{a^{X}_{\ell_1m_1}a^Y_{\ell_2m_2}a^Z_{\ell_3m_3}a^W_{\ell_4m_4}} &=& \frac{1}{2}\sum_{LM}(-1)^Mw^{L(-M)}_{\ell_1\ell_2m_1m_2}w^{LM}_{\ell_3\ell_4m_3m_4}\sum_{\vec b,u\in U_4,\chi\in\pm1}\frac{t^u_\chi(\vec b,B)}{\Delta^u_4(\vec b)}\Theta_L(B)\left[1+\chi \,p_u(-1)^{\ell_1+\ell_2+\ell_3+\ell_4}\right]\nonumber\\
    &&\quad\,\times\,\left[\delta_{\rm K}^{u,XYZW}\Theta_{\ell_1}(b_1)\Theta_{\ell_2}(b_2)\Theta_{\ell_3}(b_3)\Theta_{\ell_4}(b_4)+\text{7 perms.}\right]+\left(2\leftrightarrow3\right)+\left(2\leftrightarrow4\right),
\eeq
where we include the $m$ variables in this definition (unlike \eqref{eq: b-ell-binned}), to yield the correct behavior under permutations. This can be straightforwardly related to that of the spin-$s$ fields ${}_sa_{\ell m}$ via \eqref{eq: tl-unbinned-def}. Here, the eight permutations in the bracket are over exchanges of the following sets of fields-and-bins, $\{b_i,u_i\}$:
\beq
    \{1234\},\,\{1243\},\,\{2134\},\,\{2143\},\,\{3412\},\,\{3421\},\,\{4312\},\,\{4321\},
\eeq
and the final two terms permute all relevant $\ell_i,m_i,u_i$. As before, the sum over bins in \eqref{eq: binned-trispectrum} is limited to non-trivial bins, as defined by \eqref{eq: tl-bin-conditions}. In this case, the degeneracy factor is defined as
\beq
    \Delta_4^u(\vec b) = \begin{cases} 8 & b_1=b_2=b_3=b_4 \text{ and } u_1=u_2=u_3=u_4\\
    4 & b_1=b_2 \text{ and } b_3=b_4 \text{ and } u_1=u_2 \text{ and } u_3 = u_4\\
    2 & b_1=b_2 \text{ and } u_1 = u_2\\
    2 & b_3=b_4 \text{ and } u_3 = u_4\\
    2 & b_1=b_3 \text{ and } b_2=b_4 \text{ and } u_1 = u_3 \text{ and } u_2 = u_4\\
    1 & \text{ else.}
    \end{cases}
\eeq
Finally, we insert a parity-factor, as for the bispectrum, to separate out contributions sourced by parity-even and parity-odd physics, here denoted by $t_+$ and $t_-$.

\section{Spectral Estimators}\label{sec: estimators}
\subsection{Ideal Case}

In this section, we derive optimal estimators for weakly non-Gaussian statistics, starting from a suitable expansion of the data likelihood and maximizing with respect to the signal of interest. Such estimators can be robustly defined and implemented, and, in the limit of Gaussian statistics, are minimum variance, \textit{i.e.}\ they saturate their Cram\'{e}r-Rao bound.

We begin with the Edgeworth expansion for the likelihood of data $d_i$ in some set of pixels $\hn$ and spins $s$ (jointly denoted by the index $i$). Assuming Einstein summation convention, we can write
\beq\label{eq: edgeworth}
    L[d] \propto \mathrm{exp}\left[-\frac{1}{2}d^{i*}\Ci_{ij}d^{j}-\frac{1}{2}\mathrm{Tr}\log\C\right]\left\{1 + \frac{1}{3!}\av{d_id_jd_k}\left(\mathcal{H}^{ijk}\right)^* + \frac{1}{4!}\av{d_id_jd_kd_l}\left(\mathcal{H}^{ijkl}\right)^*+\cdots\right\}.
\eeq
Here $\mathsf{C}_{ij} \equiv \av{d_id_j^*}$ is the map-space covariance matrix, whose form was given in \eqref{eq: pixel-covariance} as a function of two positions and two spins. We have additionally included the three- and four-point expectations $\av{d_id_jd_k}$ and $\av{d_id_jd_k}$ which can be similarly computed in terms of the bispectrum and trispectrum of \S\ref{sec: defs}. The objects denoted $\mathcal{H}_{ij\cdots}$ are \textit{Hermite tensors}, given by 
\beq
    \mathcal{H}_{ijk} &=& h_ih_jh_k - \big(h_i\av{h_jh_k}+\text{2 perms.}\big)\\\nonumber
    \mathcal{H}_{ijkl} &=& h_ih_jh_kh_l - \big(h_ih_j\av{h_kh_l}+\text{5 perms.}\big) + \big(\av{h_ih_j}\av{h_kh_l}+\text{2 perms.}\big),
\eeq
in terms of the Wiener-filtered data $h_i \equiv [\mathsf{C}^{-1}d]_i$. This expansion is similar to that used in previous works \citep{Philcox:2021ukg,Philcox:2023uwe}, but allows for complex data as required for fields with non-zero spin (e.g., $Q\pm iU$ polarization states).

To obtain estimators for the $N$-point correlators, we maximize the logarithm of \eqref{eq: edgeworth} with respect to some coefficients, $x^u_\chi(\vec b)$, contained only within the $(N>2)$-point function (e.g., the parity even/odd bandpowers in some set of bins and fields). This yields an estimator of the form
\beq\label{eq: gen-estimator-num1}
    \left[\mathcal{F}_N\widehat{x}\right]^u_\chi(\vec b) = \frac{1}{N!}\frac{\partial \av{d_{i_1}\cdots d_{i_N}}}{\partial x^u_\chi(\vec b)}\left(\mathcal{H}^{i_1\cdots i_N}\right)^*,
\eeq
involving a data-independent normalization matrix, defined as
\beq\label{eq: gen-estimator-fish1}
    \mathcal{F}^{uu'}_{N,\chi\chi'}(\vec b,\vec b') =  \frac{1}{N!}\left[\frac{\partial \av{d^{i_1}\cdots d^{i_N}}^*}{\partial x^u_\chi(\vec b)}\Ci_{i_1j_1}\cdots\Ci_{i_Nj_N}\frac{\partial \av{d^{j_1}\cdots d^{j_N}}}{\partial x^{u'}_{\chi'}(\vec b')}\right]^*.
\eeq
Assuming weak non-Gaussianity, this is an optimal estimator for $x^u_\chi(\vec b)$, \textit{i.e.}\ it is minimum variance. Furthermore, the choice of $\mathcal{F}_N$ ensures that the estimator is unbiased, such that $\mathbb{E}\left[\widehat{x}^u_\chi(\vec b)\right] = x^u_\chi(\vec b)$ (via the Hermite polynomial orthogonality).\footnote{We have implicitly assumed that the $N$-point correlators are fully described by the binned polyspectra. Any contributions from non-primordial sources, e.g., noise, are thus included within the binned correlators.} Furthermore, in the Gaussian limit, the covariance matrix of $\widehat{x}^u_\chi(\vec b)$ is given by the inverted normalization $\mathcal{F}_{N,\chi\chi'}^{-1,uu'}(\vec b,\vec b')$. Finally, denoting the Cholesky factorized Fisher matrix as $\mathcal{F}^{1/2}$, we note that \citep[e.g.,][]{Hamilton:2005kz,Hamilton:2005ma}
\beq\label{eq: rescaled-estimator-unit-gaussian}
    \left[\mathcal{F}_{N}^{1/2}\widehat{x}\right]^{u}_\chi(\vec b) \sim \mathcal{N}(\vec 0,\mathsf{I}),
\eeq
\textit{i.e.}\ the rescaled estimator obeys a standardized normal distribution in the limit of weak non-Gaussianity. This allows for efficient data compression, as in \citep{PhilcoxCMB}. 

\subsection{Realistic Case}
In practice, the estimators we will discuss in this work are a little more nuanced than the above. First, the data is often masked or windowed, such that the true map, $d$, is related to the observed map, $\tilde{d}$, by ${}_{s}\tilde{d}(\hn) = W(\hn){}_{s}d(\hn)$, where $W$ is some pixel-space mask which we will generally take to be spin-independent and real.\footnote{One could straightforwardly use a different mask for each spin and similarly deconvolve its effect from the output correlators.} As such, we should formulate the estimators of \eqref{eq: gen-estimator-num1}\,\&\,\eqref{eq: gen-estimator-fish1} in terms of masked fields $\tilde{d}$, and their respective correlators. Since the mask is multiplicative, this is straightforward to implement in practice.

Secondly, the above estimators require Wiener-filtered data, $h \equiv \Ci\tilde{d}$, where $\C = \av{\tilde{d}\tilde{d}^\dag}$. The full covariance matrix is high dimensional (containing $9N_{\rm pix}^2$ elements for spins $0,\pm2$) and thus hard to compute and harder-still to invert, particularly in the presence of translation-dependent noise and holes in the window (such as a Galactic sky mask). A more practical estimator is wrought by abandoning the exact Wiener filter and instead applying some custom weighting $\mathsf{S}^{-1}$, with $h$ redefined to $h \equiv \mathsf{S}^{-1}\tilde{d}$. The weighting is not required to be invertible (\textit{i.e.}\ $\mathsf{S}$ does not have to exist), and can be chosen to allow for simple implementation. In CMB analyses, a typical approach would be to choose the linear operator $\mathsf{S}^{-1}$ such that it (a) (non-invertibly) projects out undesired areas of the map (from e.g., the Milky Way and point sources), (b) filters the data in harmonic space by the ideal power spectrum (including noise). Assuming a diagonal-in-$\hn$ projection operator $\Pi$ and a diagonal-in-$\ell$ filtering by the idealized $T/E/B$-space covariance $\mathbb{C}_\ell$, this is applied to a map $x$ as
\beq
    {}_sx(\hn)\quad\to\quad\Pi(\hn){}_sx(\hn)\quad\to \quad{}_s[\Pi x]_{\ell m}\quad\to \quad\left[\mathbb{C}^{-1}_\ell\mathcal{R}^{-1} [\Pi x]\right]^X_{\ell m} \quad\to\quad {}_{s}\left[\R^{-\dag}\mathbb{C}^{-1}_\ell\mathcal{R}^{-1} [\Pi x]\right](\hn),
\eeq
which requires only two harmonic transforms to implement.

For a general filter $\mathsf{S}^{-1}$ and mask, the binned polyspectrum estimator becomes 
\beq\label{eq: gen-estimator}
    \left[\mathcal{F}_N\widehat{x}\right]^u_\chi(\vec b) &=& \frac{1}{N!}\frac{\partial \av{\tilde{d}_{i_1}\cdots \tilde{d}_{i_N}}}{\partial x^u_\chi(\vec b)}\left(\mathcal{H}^{i_1\cdots i_N}\right)^*\\\nonumber
    \mathcal{F}_{N,\chi\chi'}^{uu'}(\vec b,\vec b') &=&  \frac{1}{N!}\left[\frac{\partial \av{\tilde{d}^{i_1}\cdots \tilde{d}^{i_N}}^*}{\partial x^u_\chi(\vec b)}\mathsf{S}^{-1}_{i_1j_1}\cdots\mathsf{S}^{-1}_{i_Nj_N}\frac{\partial \av{\tilde{d}^{j_1}\cdots \tilde{d}^{j_N}}}{\partial x^{u'}_{\chi'}(\vec b')}\right]^*,
\eeq
where $h\equiv \mathsf{S}^{-1}\tilde{d}$ and $\mathcal{F}_N$ is Hermitian only if $\mathsf{S}^{-1}$ is. This remains unbiased, such that $\mathbb{E}[\widehat{x}^u_\chi(\vec b)]=x^u_\chi(\vec b)$ for any filter $\mathsf{S}^{-1}$ (due to the definition of $\mathcal{F}_N$). This implies that the estimator deconvolves the effect of any linear masking and inpainting contained within $\Si$. In the limit of $\mathsf{S}^{-1}\to \mathsf{C}^{-1}$, the estimator becomes optimal, such that $\mathcal{F}_N^{-1}$ is equal to the estimators covariance matrix; any difference second order in $\mathsf{S}-\mathsf{C}$ (if $\mathsf{S}^{-1}$ is invertible). The remainder of this work is dedicated to exploring how such estimators can be applied in practice.

\section{Power Spectrum}\label{sec: Cl}
The above derivation of the $N$-point spectrum estimators strictly applies only for $N>2$. From the Gaussian part of the likelihood \eqref{eq: edgeworth}, we can derive a similar estimator for the bandpower of some pair of fields $u$ in bin $b$:
\beq\label{eq: gen-Cl-estimator}
    \left[\mathcal{F}_2\widehat{C}\right]^u(b) &=& \frac{1}{2}h_i^*\frac{\partial\C^{ij}}{\partial C^u(b)}h_j\\\nonumber
    \mathcal{F}_{2}^{uu'}(b,b') &=&  \frac{1}{2}\mathrm{Tr}\left[\mathsf{S}^{-\dag}\frac{\partial\C}{\partial C^u(b)}\mathsf{S}^{-1}\frac{\partial\C}{\partial C^{u'}(b')}\right],
\eeq
again assuming a general weighting $\Si$ and denoting the windowed covariance as $\mathsf{C}_{ij} \equiv \av{\tilde{d}_i\tilde{d}_j^*}$. To derive this we have expanded the likelihood around some fiducial set of band-powers (since we cannot assume the fiducial value to be zero), the value of which exactly cancels in the estimator, once the $\mathrm{Tr}\log \mathsf{C}$ term is considered.\footnote{If we include an additive noise term with covariance $\mathsf{N}$, such that $\mathsf{C} = \sum_{b,u}C^u(b)\partial \mathsf{C}/\partial C^u(b) + \mathsf{N}$, the estimator numerator is modified to \beq
\left[\mathcal{F}_2\widehat{C}\right]^u(b) = \frac{1}{2}h_i^*\frac{\partial\C^{ij}}{\partial C^u(b)}h_j - \frac{1}{2}\mathrm{Tr}\left[\mathsf{S}^{-\dag}\frac{\partial C}{\partial C^u(b)}\Si\mathsf{N}\right].\eeq 
Here, we absorb the noise into the observed spectrum.} This bears strong similarities to the higher-point estimators of \eqref{eq: gen-estimator}, differing only due to the complex conjugate present in the definition of the power spectrum and the absence of a parity index.

\subsection{Numerator}
We now simplify this expression, starting first with the numerator. Starting with the explicit definition of $\mathsf{C}$ in pixel space (cf.,\,\ref{eq: pixel-covariance}), incorporating the window function, we can write:
\beq
    {}_{s_1s_2}\C(\hn_1,\hn_2) &=& W(\hn_1)W(\hn_2)\sum_{\ell m}{}_{s_1}Y_{\ell m}(\hn_1){}_{s_1s_2}[\R\,\mathbb{C} \,\R^\dag ]{}_{s_2}Y_{\ell m}^*(\hn_2)\\\nonumber
    &=& W(\hn_1)W(\hn_2)\sum_{\ell m XY}{}_{s_1}Y_{\ell m}(\hn_1){}_{s_1}\R_X\left(\sum_{b,u\in U_2}\frac{C^u(b)}{\Delta^u_2}\left[\delta_{\rm K}^{u,XY}+\delta_{\rm K}^{u,YX}\right]\Theta_\ell(b)B^{X}_\ell B^{Y}_\ell\right){}_{s_2}\R^\dag_Y{}_{s_2}Y_{\ell m}^*(\hn_2)
\eeq
using the definition of the binned power spectrum \eqref{eq: C-ell-binned} in the second line. We have additionally inserted an isotropic beam, $B_\ell$, which differ between temperature and polarization. Inserting this into the numerator of \eqref{eq: gen-Cl-estimator} and simplifying gives the explicitly real form
\beq\label{eq: numerator-simplified}
    \left[\mathcal{F}_2\widehat{C}\right]^u(b) &=& \frac{1}{2\Delta_2^u}\sum_{\ell m XY}\Theta_\ell(b)B^{X}_\ell B^{Y}_\ell[\R^\dag Wh]^{X*}_{\ell m}\left[\delta_{\rm K}^{u,XY}+\delta_{\rm K}^{u,YX}\right][\R^\dag Wh]^Y_{\ell m}
\eeq
where $[\R^\dag Wh]^X_{\ell m} \equiv \sum_{s}({}_s\R^\dag_X){}_s[Wh]_{\ell m}$, and we have replaced the integrals over $\hn_{1,2}$ by a harmonic transform. We may additionally compute the sum using only modes with $m\geq 0$ by adding a factor of $(1+\delta^{\rm K}_{m\geq 0})$, which evaluates to unity if $m=0$ and two else. In the absence of a mask and with an ideal weighting $\Si\to\mathsf{C}^{-1}$, $[\R^\dag h]^X_{\ell m} = [\mathbb{C}^{-1}a]^X_{\ell m}$; in general, the various $T/E/B$ components mix due to masking, however. Following these manipulations \eqref{eq: numerator-simplified} is relatively straightforward to compute: one performs harmonic transforms to yield the filtered maps $[\R^\dag Wh]_{\ell m}^X$ then combines pairwise to yield the estimator, summing over all $\ell$ modes in the bin and fields of interest.\footnote{In the practical implementation of these estimators, we replace $[\R^\dag Wh]_{\ell m}^X$ by $h_{\ell m}^X$ if the mask is uniform, thus avoiding any unnecessary harmonic transforms.} We note that, even without the mask, the $XY$ power spectrum does not reduce to the canonical form $a_{\ell m}^Xa _{\ell m}^{Y*}$ due to the non-trivial correlations between spectra (e.g., a non-zero $C_\ell^{TE}$ prior).

\subsection{Normalization}\label{sec: cl-norm}
Computing the Fisher matrix in \eqref{eq: gen-Cl-estimator} is non-trivial due to the trace operation, which na\"ively requires $\mathcal{O}(9N_{\rm pix}^2)$ operations to implement. To sidestep this, we adopt a similar procedure to \citep{Philcox:2023uwe,2011MNRAS.417....2S}, introducing a set of Gaussian random field (GRF) maps $a_i$ with known and invertible covariance $\mathsf{A} \equiv \av{aa^\dag}_a$:
\beq
    \mathcal{F}_{2}^{uu'}(b,b') &=&  \frac{1}{2}\av{a^\dag \mathsf{S}^{-\dag}\frac{\partial\C}{\partial C^u(b)}\mathsf{S}^{-1}\frac{\partial\C}{\partial C^{u'}(b')}\mathsf{A}^{-1}a}_a,
\eeq
where $\av{a^\dag \mathsf{A}^{-1}a}_a = \mathsf{1}$. Noting that the various covariances and weightings are linear operators, this can be computed in $\mathcal{O}(3N_{\rm pix})$ time, with the expectation (which replaces the trace) implemented as Monte Carlo average over realizations of $a_i$. Assuming $\mathsf{A}^{-1}$ is close to $\mathsf{S}^{-1}$, this converges quickly (roughly as $1+1/N_{\rm sim}$). The above expression can be rewritten succinctly as
\beq
    \mathcal{F}_{2}^{uu'}(b,b') &=&  \frac{1}{2}\av{\big(Q_2^{u}[\Si a](b)\big)^*\cdot[W\mathsf{S}^{-1}W]\cdot Q_2^{u'}[\mathsf{A}^{-1}a](b')}_a,
\eeq
defining
\beq\label{eq: Q2-deriv}
    {}_sQ_2^{u}[x](\hn;b) = {}_s\left[\frac{\partial \av{dd^\dag}}{\partial C^{u}(b)}Wx\right](\hn),
\eeq
where we have explicitly separated out the mask $W$. This is simply a product to two maps, $Q_2^*$ and $[W\Si W Q_2]$, summed over each pixel in real- or harmonic-space (via the Wiener-Khinchin theorem). To compute the normalization we thus perform the following set of operations:
\begin{enumerate}
    \item Generate a GRF $a$ from some fiducial covariance $\mathsf{A}$.
    \item Filter this map to obtain $\mathsf{A}^{-1}a$ and $\mathsf{S}^{-1}a$
    \item Compute $Q_2^u[x](b)$ maps for each bin of interest.
    \item Combine pairwise to form the Fisher matrix.
    \item Average over $N_{\rm sim}$ realizations of $a$.
\end{enumerate}

Inserting the explicit form for the two-point function \eqref{eq: C-ell-binned}, we find the following form for $Q_2$, expressed in harmonic space:
\beq
    {}_sQ_{2,\ell m}^{u}[x](b) &=& \frac{1}{\Delta_2^u}\sum_{XY}\Theta_\ell(b)B^{X}_\ell B^{Y}_\ell{}_s\R_X\left[\delta_{\rm K}^{u,XY}+\delta_{\rm K}^{u,YX}\right][\R^\dag Wx]^Y_{\ell m}\\\nonumber
    Q_{2,\ell m}^{u,X}[x](b) &=& \frac{1}{\Delta_2^u}\Theta_\ell(b)B^{u_1}_\ell B^{u_2}_\ell\left([\R^\dag Wx]^{u_2}_{\ell m}\delta_{\rm K}^{u_1X}+[\R^\dag Wx]^{u_1}_{\ell m}\delta_{\rm K}^{u_2X}\right),
\eeq
transforming to $T/E/B$ space in the second line. This is similar to the numerator term of \eqref{eq: numerator-simplified}, and can be simply implemented via spin-weighted spherical harmonic transforms.\footnote{If we had additionally included a noise term in the quadratic estimator, this could be computed in a similar manner, replacing the random fields $a$ by fields drawn from the noise covariance.}

\subsection{Ideal Limits}
We now consider the limiting form of the above estimators to gain some intuition for their form. In ideal scenarios, we can drop the mask and beam and assume $\Si = \mathsf{C}^{-1}$, which is translationally invariant. In this case, the $[\R^\dag Wh]_{\ell m}^X$ term becomes $h_{\ell m}^X \equiv [\mathbb{C}_\ell^{-1}a_{\ell m}]^X$ (since $\mathsf{C}^{-1}$ involves $\R^{-\dag}\mathbb{C}^{-1}\R^{-1}$), thus the numerator simplifies to
\beq
    \left[\mathcal{F}_2\widehat{C}\right]^u(b) &\to& \frac{1}{\Delta_2^u}\sum_{\ell m}\Theta_\ell(b)\mathrm{Re}\left[h^{u_1*}_{\ell m}h^{u_2}_{\ell m}\right],
\eeq
where the bracket takes the real part. This is just the standard power spectrum estimator, albeit with Wiener filtering through $\mathbb{C}^{-1}$.

In the same limit, the normalization can be written in terms of harmonic-space $T/E/B$ quantities as
\beq
    \mathcal{F}_{2}^{uu'}(b,b') &\to&  \frac{1}{2}\mathrm{Tr}\left[\mathbb{C}^{-1}\frac{\partial\mathbb{C}}{\partial C^u(b)}\mathbb{C}^{-1}\frac{\partial\mathbb{C}}{\partial C^{u'}(b')}\right]\\\nonumber
    &=&\frac{1}{2\Delta^u_2\Delta_2^{u'}}\sum_{\ell}(2\ell+1)\Theta_\ell(b)\Theta_\ell(b')\sum_{XYZW}\left[\mathbb{C}^{-1,XY}_\ell\left(\delta_{\rm K}^{u,YZ}+\delta_{\rm K}^{u,ZY}\right)\mathbb{C}^{-1,ZW}_\ell\left(\delta_{\rm K}^{u',WX}+\delta_{\rm K}^{u',XW}\right)\right]\\\nonumber
    &=&\frac{1}{\Delta^u_2\Delta_2^{u'}}\sum_{\ell}(2\ell+1)\Theta_\ell(b)\Theta_\ell(b')\left[\mathbb{C}^{-1,u_1u_1'}_\ell\mathbb{C}^{-1,u_2'u_2}_\ell+\mathbb{C}^{-1,u_2'u_1}_\ell \mathbb{C}^{-1,u_2u_1'}_\ell\right]
\eeq
inserting the explicit derivatives in the second line, and noting that the covariance is symmetric. Assuming that the bins are disjoint, this is diagonal in $b,b'$, but not in $u,u'$. This implies that each $\ell$-bin is uncorrelated, but there are non-trivial correlations between the different power spectrum estimators, since $\mathbb{C}^{-1}$ is not generally assumed to be diagonal. If we ignored such correlations, we would find a familiar form involving $\sum_\ell (2\ell+1)\Theta^2_\ell(b)/C_\ell^{XX}C_\ell^{ZZ}$. A subtle point is that the correlation matrix (defined by $\mathcal{F}_2^{-1}$ under ideal assumptions) is block diagonal, thus the spectra bifurcate into uncorrelated groups: parity-even $(p_u=1)$ and parity-odd ($p_u=-1$). This is a consequence of rotational symmetry and does not hold for the general case, since masks and filtering can induce leakage between parity-odd and parity-even components.

\section{Bispectrum}\label{sec: bl}
We now turn our attention to three-point functions. Computation proceeds analogously to the power spectrum, beginning with the general estimator for $N=3$ (cf.\,\ref{eq: gen-estimator}):
\beq\label{eq: gen-bis-estimator}
    \left[\mathcal{F}_3\widehat{b}\right]^u_\chi(\vec b) &=& \frac{1}{3!}\frac{\partial \av{\tilde{d}^{i_1}\tilde{d}^{i_2}\tilde{d}^{i_3}}}{\partial b^u_\chi(\vec b)}\bigg(h_{i_1}h_{i_2}h_{i_3}-\left[h_{i_1}\av{h_{i_2}h_{i_3}}+\text{2 perms.}\right]\bigg)^*\\\nonumber
    \mathcal{F}_{3,\chi\chi'}^{uu'}(\vec b,\vec b') &=&  \frac{1}{3!}\left[\frac{\partial \av{\tilde{d}^{i_1}\tilde{d}^{i_2}\tilde{d}^{i_3}}^*}{\partial b^u_\chi(\vec b)}\mathsf{S}^{-1}_{i_1j_1}\mathsf{S}^{-1}_{i_2j_2}\mathsf{S}^{-1}_{i_3j_3}\frac{\partial \av{\tilde{d}^{j_1}\tilde{d}^{j_2}\tilde{d}^{j_3}}}{\partial b^{u'}_{\chi'}(\vec b')}\right]^*.
\eeq
The principal difference between this and the power spectrum estimator is the presence of the linear term in the numerator: this can significantly reduce the large scale variance of the estimator \citep{Philcox:2023uwe} and may be computed via Monte Carlo methods. To this end, we introduce a set of maps $\{\alpha\}$ with covariance equal to that of the data, \textit{i.e.}\ $\av{\alpha\alpha^\dag} = \mathsf{C}$.\footnote{Note that the bispectrum estimator remains unbiased if $\av{\alpha\alpha^\dag}_\alpha\neq \mathsf{C}$, though its variance generically increases.} The linear term is then given by an average over $\alpha$;
\beq
     \left[\mathcal{F}_3\widehat{b}\right]^u_\chi(\vec b) &=& \frac{1}{3!}\frac{\partial \av{\tilde{d}^{i_1}\tilde{d}^{i_2}\tilde{d}^{i_3}}}{\partial b^u_\chi(\vec b)}\bigg(h_{i_1}h_{i_2}h_{i_3}-\left[h_{i_1}\av{[\Si \alpha]_{i_2}[\Si \alpha]_{i_3}}_\alpha+\text{2 perms.}\right]\bigg)^*,
\eeq
which can be estimated via Monte Carlo averaging.

\subsection{Numerator}
To massage the bispectrum estimator into a practically computable form, we first require the explicit definition of the map-space three-point correlation function in terms of the reduced bispectrum \eqref{eq: spin-to-TEB-bis}:
\beq\label{eq: 3-point-expansion}
    \av{{}_{s_1}\tilde{d}(\hn_1){}_{s_2}\tilde{d}(\hn_2){}_{s_3}\tilde{d}(\hn_3)} &=& \prod_{i=1}^3\left[W(\hn_i)\sum_{\ell_im_iX_i}B_{\ell_i}^{X_i}{}_{s_i}Y_{\ell_im_i}(\hn_i){}_{s_i}\R_{X_i}\right]w^{\ell_1\ell_2\ell_3}_{m_1m_2m_3}b^{X_1X_2X_3}_{\ell_1\ell_2\ell_3}\\\nonumber
    &=&\frac{1}{2}\prod_{i=1}^3\left[W(\hn_i)\sum_{\ell_im_iX_i}B_{\ell_i}^{X_i}{}_{s_i}Y_{\ell_im_i}(\hn_i){}_{s_i}\R_{X_i}\right]w^{\ell_1\ell_2\ell_3}_{m_1m_2m_3}\\\nonumber
    &&\,\times\,\sum_{\vec b,u\in U_3}\frac{b^u_\chi(\vec b)}{\Delta^u_3(\vec b)}\bigg[1+\chi\,p_u(-1)^{\ell_1+\ell_2+\ell_3}\bigg]\left[\delta_{\rm K}^{u,X_1X_2X_3}\Theta_{\ell_1}(b_1)\Theta_{\ell_2}(b_2)\Theta_{\ell_3}(b_3)+\text{5 perms.}\right],
\eeq
using the binned bispectrum definition \eqref{eq: b-ell-binned} in the second line, and inserting an isotropic beam $B_\ell^X$. Our next step is to insert this into the estimator numerator and simplify, which leads to the form:
\beq
    \left[\mathcal{F}_3\widehat{b}\right]^u_\chi(\vec b) &=& \frac{1}{2\Delta^u_3(\vec b)}\sum_{\ell_im_i}w^{\ell_1\ell_2\ell_3}_{m_1m_2m_3}\Theta_{\ell_1}(b_1)\Theta_{\ell_2}(b_2)\Theta_{\ell_3}(b_3)B_{\ell_1}^{u_1}B_{\ell_2}^{u_2}B_{\ell_3}^{u_3}\big[1+\chi\,p_u(-1)^{\ell_1+\ell_2+\ell_3}\big]\\\nonumber
    &&\,\times\,\bigg\{[\R^\dag Wh]^{u_1}_{\ell_1m_1}[\R^\dag Wh]^{u_2}_{\ell_2m_2}[\R^\dag Wh]^{u_3}_{\ell_3m_3}-\left([\R^\dag Wh]^{u_1}_{\ell_1m_1}\av{[\R^\dag W\alpha]^{u_2}_{\ell_2m_2}[\R^\dag W\alpha]^{u_3}_{\ell_3m_3}}_\alpha+\text{2 perms.}\right)\bigg\}^*
\eeq
where the spatial integrals (from the $i$ summations) have been replaced with harmonic transforms and we have invoked symmetry to absorb one of the permutation sums. This form of the estimator is not easy to implement, since it involves a coupled sum over three $(\ell,m)$ pairs, with a prohibitive $\mathcal{O}(\ell_{\rm max}^6)$ scaling. To proceed, we must rewrite the expression in a factorizable form, which is made possible by writing the weighting function $w^{\ell_1\ell_2\ell_3}_{m_1m_2m_3}$ as an integral over spin-weighted spherical harmonics, as in \eqref{eq: bis-weighting}. Following some simplification, we find:
\beq\label{eq: bis-num-simp}
    \left[\mathcal{F}_3\widehat{b}\right]^u_\chi(\vec b) &=& \widehat{\beta}^u_\chi[d,d,d](\vec b)-\left[\av{\widehat{\beta}^u_\chi[d,\alpha,\alpha](\vec b)}_\alpha + \text{2 perms.}\right]\\\nonumber
    \widehat{\beta}^u_\chi[\alpha,\beta,\gamma](\vec b) &\equiv& \frac{1}{6\Delta^u_3(\vec b)}\int d\hn\,\bigg\{{}_{+1}H^{u_1}[\alpha](b_1,\hn){}_{+1}H^{u_2}[\beta](b_2,\hn){}_{-2}H^{u_3}[\gamma](b_3,\hn)\\\nonumber
    &&\qquad\qquad\qquad\qquad\,+\,(\chi\,p_u){}_{+1}\overline{H}^{u_1}[\alpha](b_1,\hn){}_{+1}\overline{H}^{u_2}[\beta](b_2,\hn){}_{-2}\overline{H}^{u_3}[\gamma](b_3,\hn)\bigg\}+\text{2 perms.},
\eeq
defining the filtered maps
\beq\label{eq: H-map-def}
    {}_sH^X[x](\hn;b) = \sum_{\ell m}\Theta_\ell(b)B_\ell^X[\R^\dag W\Si x]^{X*}_{\ell m}{}_sY_{\ell m}(\hn), \quad {}_s\overline{H}^X[x](\hn;b) = \sum_{\ell m}(-1)^\ell\Theta_\ell(b)B_\ell^X[\R^\dag W\Si x]^{X*}_{\ell m} {}_sY_{\ell m}(\hn),
\eeq
which satisfy the conjugation relation
\beq\label{eq: H-s-conjugation-parity}
    \left[{}_sH^X[x](\hn;b)\right]^* = (-1)^{s}{}_{-s}H^X[x](\hn;b) = (-1)^s{}_s\overline{H}^X[x](-\hn;b).
\eeq
The conjugate property implies that $\{{}_{+1}H, -{}_{-1}H\}$ are a spin-$\pm1$ pair of fields (rather than ${}_{\pm1}H$). In \eqref{eq: bis-num-simp}, we symmetrize both over the choice of fields appearing in the expectation, and over the position of the external spin index (\textit{i.e.}\ over non-trivial exchanges of $\{u_1,b_1\}$ pairs). Via the conjugation relations, $\left(\widehat{\beta}^{u}_\chi\right)^* = \chi\,p_u\widehat{\beta}^{u}_\chi$, thus the estimator simplifies to:
\beq\label{eq: cubic-numerator-re-im}
    \widehat{\beta}^{u}_+[\alpha,\beta,\gamma](\vec b) &=&\frac{1}{3\Delta^u_3(\vec b)}\int d\hn\,\mathrm{Re}\left\{{}_{+1}H^{u_1}[\alpha](b_1,\hn){}_{+1}H^{u_2}[\beta](b_2,\hn){}_{-2}H^{u_3}[\gamma](b_3,\hn)+\text{2 perms.}\right\}\\\nonumber
    \widehat{\beta}^{u}_-[\alpha,\beta,\gamma](\vec b) &=&\frac{i}{3\Delta^u_3(\vec b)}\int d\hn\,\mathrm{Im}\left\{{}_{+1}H^{u_1}[\alpha](b_1,\hn){}_{+1}H^{u_2}[\beta](b_2,\hn){}_{-2}H^{u_3}[\gamma](b_3,\hn)+\text{2 perms.}\right\}.
\eeq
Despite all the numerous definitions, \eqref{eq: bis-num-simp} is straightforward to compute in $\mathcal{O}(3N_{\rm pix}N_{\rm it})$ operations, using the following set of operations:
\begin{enumerate}
    \item Draw a GRF $\alpha$ from the (assumed) true covariance $\mathsf{C}$.
    \item Filter both $\alpha$ and the data $d$ by the weighting scheme and the mask to form $W\Si d$, $W\Si \alpha$ maps, then use a spherical harmonic transform to compute $\R^\dag W\Si d$ and $\R^\dag W\Si \alpha$ in $T/E/B$ space.
    \item Compute the ${}_sH^X$ and ${}_s\overline{H}^X$ maps for each bin of interest with $X\in\{T,E,B\}$ and $s\in\{+1,-2\}$ via a further spin-weighted spherical harmonic transform.
    \item Compute $\hat{\beta}$ for each configuration of interest as a real-space summation.
    \item Iterate over $N_{\rm it}$ random fields $\alpha$ to form the Monte Carlo average and use this to assemble the estimator numerator.
\end{enumerate}

\subsection{Normalization}
The bispectrum Fisher matrix can be computed using similar techniques applied to the power spectrum (\S\ref{sec: cl-norm}) \citep[cf.,][]{2011MNRAS.417....2S}. First, we rewrite the $\mathcal{O}(N_{\rm pix}^3)$ map-space sum as a Monte Carlo average by introducing some fiducial covariance $\mathsf{A}$ and its inverse as follows:
\beq
    \Si_{i_2j_2}\Si_{i_3j_3} \to \Si_{i_2i_2'}\Si_{i_3i_3'}\times\frac{1}{2}\left(\mathsf{A}_{i_2'j_2'}\mathsf{A}_{i_3'j_3'}+\mathsf{A}_{i_2'j_3'}\mathsf{A}_{i_3'j_2'}\right)\times\mathsf{A}^{-1}_{j_2'j_2}\mathsf{A}^{-1}_{j_3'j_3},
\eeq
noting that the Fisher matrix is symmetric under $j_2\leftrightarrow j_3$. As in \S\ref{sec: Cl}, we write each covariance as the expectation of a random field $a^{(i)}$, $\mathsf{A} \equiv \av{a^{(i)}a^{(i)\dag}}$, allowing the sum to be rewritten as an expectation:
\beq
    \Si_{i_2j_2}\Si_{i_3j_3} \to \frac{1}{4}\Si_{i_2i_2'}\Si_{i_3i_3'}\av{a^{(1)}_{i_2'}a^{(1)}_{i_3'}a^{(1)*}_{j_2'}a^{(1)*}_{j_3'}+a^{(2)}_{i_2'}a^{(2)}_{i_3'}a^{(2)*}_{j_2'}a^{(2)*}_{j_3'}-a^{(1)}_{i_2'}a^{(1)}_{i_3'}a^{(2)*}_{j_2'}a^{(2)*}_{j_3'}-a^{(2)}_{i_2'}a^{(2)}_{i_3'}a^{(1)*}_{j_2'}a^{(1)*}_{j_3'}}_a\mathsf{A}^{-1}_{j_2'j_2}\mathsf{A}^{-1}_{j_3'j_3},
\eeq
where $a^{(1)}$ and $a^{(2)}$ are independent and identically distributed.\footnote{Whilst we could simply write the expectation as $\mathsf{A}_{i_2'j_2'}\mathsf{A}_{i_3'j_3'} = \av{a^{(1)}_{i_2'}a^{(1)}_{i_3'}a^{(2)*}_{j_2'}a^{(2)*}_{j_3'}}$, the form above makes more efficient use of the random fields.}  Applied to \eqref{eq: gen-bis-estimator}, we find
\beq
    \F^{uu'}_{3,\chi\chi'}(\vec b,\vec b') &=& \frac{1}{2}\left(F^{uu',1111}_{3,\chi\chi'}(\vec b,\vec b')+F^{uu',2222}_{3,\chi\chi'}(\vec b,\vec b')-F^{uu',1122}_{3,\chi\chi'}(\vec b,\vec b')-F^{uu',2211}_{3,\chi\chi'}(\vec b,\vec b')\right)\\\nonumber
    F^{uu',\mu\nu\rho\sigma}_{3,\chi\chi'}(\vec b,\vec b') &\equiv&    
    \frac{1}{12}\av{\frac{\partial \av{\tilde{d}^{i_1}\tilde{d}^{i_2}\tilde{d}^{i_3}}^*}{\partial b^u_\chi(\vec b)}\mathsf{S}^{-1}_{i_1j_1}[\Si a^{(\mu)}]_{i_2}[\Si a^{(\nu)}]_{i_3}\times[\mathsf{A}^{-1}a^{(\rho)}]^*_{j_2}[\mathsf{A}^{-1}a^{(\sigma)}]^*_{j_3}\frac{\partial \av{\tilde{d}^{j_1}\tilde{d}^{j_2}\tilde{d}^{j_3}}}{\partial b^{u'}_{\chi'}(\vec b')}}_a^*\\\nonumber
    &=& \frac{1}{12}\av{\left(Q_{3,\chi}^u[\Si a^{(\mu)},\Si a^{(\nu)}](\vec b)\right)^*\cdot[W\mathsf{S}^{-1}W]
    \cdot\left(Q_{3,\chi'}^{u'}[\mathsf{A}^{-1}a^{(\rho)},\mathsf{A}^{-1}a^{(\sigma)}]\right)}_a^*
\eeq
defining the pixel-space derivative maps
\beq\label{eq: Q-3-def}
    Q_{3,\chi,i}^u[x,y](\vec b) = \frac{\partial \av{d^{i}d^{j}d^{k}}}{\partial b^u_\chi(\vec b)}[Wx]^*_{j}[Wy]^*_{k}.
\eeq
As for the power spectrum, this is the inner product of two maps, for each combination of bins, fields, and parities.

Inserting the definition of the three-point function and simplifying, the $Q_3$ derivatives can be written in harmonic space
\beq
    {}_{s_1}Q_{3,\chi,\ell_1m_1}^u[x,y](\vec b) &=& \frac{1}{2\Delta^u_3(\vec b)}\sum_{\ell_2\ell_3m_2m_3X_1X_2X_3}{}_{s_1}\R_{X_1}w^{\ell_1\ell_2\ell_3}_{m_1m_2m_3}B_{\ell_1}^{X_1}B_{\ell_2}^{X_2}B_{\ell_3}^{X_3}\bigg[1+\chi\,p_u(-1)^{\ell_1+\ell_2+\ell_3}\bigg]\\\nonumber
    &&\,\times\,\left[\delta_{\rm K}^{u,X_1X_2X_3}\Theta_{\ell_1}(b_1)\Theta_{\ell_2}(b_2)\Theta_{\ell_3}(b_3)+\text{5 perms.}\right][\R^\dag Wx]^{X_2*}_{\ell_2m_2}[\R^\dag Wy]^{X_3*}_{\ell_3m_3}.
\eeq
As for the estimator numerator, this can be simplified by inserting the integral form of the weights:
\beq
    {}_{s_1}Q_{3,\chi,\ell_1m_1}^u[x,y](\vec b) &=& \bigg(\frac{\Theta_{\ell_1}(b_1)B_{\ell_1}^{u_1}}{6\Delta^u_3(\vec b)}{}_{s_1}\R_{u_1}\int d\hn\,\bigg\{{}_{+1}Y_{\ell_1m_1}(\hn){}_{+1}H^{u_2}[x](\hn;b_2){}_{-2}H^{u_3}[y](\hn;b_3)\\\nonumber
    &&\qquad\qquad\qquad\qquad\qquad\,+\,\chi\,p_u\,(-1)^{\ell_1}{}_{+1}Y_{\ell_1m_1}(\hn){}_{+1}\overline{H}^{u_2}[x](\hn;b_2){}_{-2}\overline{H}^{u_3}[y](\hn;b_3)\bigg\}\\\nonumber
    &&\,+\,\text{2 perms.}\bigg)\,+\,\text{5 perms.}
\eeq
where the first set of permutations are over the positions of the $\{+1,+1,-2\}$ spins and the second over the $\{b,u\}$ pairs. Using the parity properties of spin-weighted spherical harmonics, this can be further rewritten in $T/E/B$ space as
\beq\label{eq: Q3-simplified}
    Q_{3,\chi,\ell_1m_1}^{u,X}[x,y](\vec b) &=& \bigg(\frac{\Theta_{\ell_1}(b_1)B_{\ell_1}^{u_1}}{3\Delta^u_3(\vec b)}\delta_{\rm K}^{u_1X}\int d\hn\,\bigg\{-{}_{+1}Y^*_{\ell_1m_1}(\hn){}_{-1}H^{u_2}[x](\hn;b_2){}_{+2}H^{u_3}[y](\hn;b_3)\\\nonumber
    &&\qquad\qquad\qquad\qquad\qquad\,-\,(\chi\,p_u)\,{}_{-1}Y^*_{\ell_1m_1}(\hn){}_{+1}H^{u_2}[x](\hn;b_2){}_{-2}H^{u_3}[y](\hn;b_3)\bigg\}^*\\\nonumber
    &&\,+\,\text{2 perms.}\bigg)\,+\,\text{2 perms.},
\eeq
noting that two permutations are equivalent. We additionally note the conjugate relation:
\beq
    \left(Q_{3,\chi,\ell_1m_1}^{u,X}[x,y](\vec b)\right)^* = (\chi\,p_u)\,\times\,(-1)^{m_1}Q_{3,\chi,\ell_1(-m_1)}^{u,X}[x,y](\vec b),
\eeq
thus $Q_3$ is real (imaginary) in map-space if $\chi\,p_u=1$ ($\chi\,p_u=-1$). The above can be efficiently computed using spin-weighted spherical harmonic transforms, first to define the ${}_{\pm 1}H$ and ${}_{\pm 2}H$ fields, then to perform the $\hn$ integrals, noting that $Q_{\ell m}$ transforms as a triplet of real fields (\textit{i.e.}\ $T,Q,U$) if we add a factor of $i$ when $\chi\,p_u=-1$. As such, the derivative, and thus the full Fisher matrix, can be efficiently computed as a summation in real-space, or, for a local-in-$\ell$ weighting and $W=1$, harmonic space.

\subsection{Ideal Limits}
In the ideal limit of homogeneous noise, unit mask and beam, and ideal Wiener filtering, the bispectrum estimators significantly simplify. As before, we start by noting that $[\R^\dag Wh]^{X}_{\ell m} \to h_{\ell m}^X$, thus the bispectrum numerator can be written
\beq
    \left[\mathcal{F}_3\widehat{b}\right]^u_\chi(\vec b) &\to& \frac{1}{2\Delta^u_3(\vec b)}\sum_{\ell_im_i}w^{\ell_1\ell_2\ell_3}_{m_1m_2m_3}\Theta_{\ell_1}(b_1)\Theta_{\ell_2}(b_2)\Theta_{\ell_3}(b_3)\big[1+\chi\,p_u(-1)^{\ell_1+\ell_2+\ell_3}\big]\big(h^{u_1}_{\ell_1m_1}h^{u_2}_{\ell_2m_2}h^{u_3}_{\ell_3m_3}\big)^*\\\nonumber
    &=&\frac{\chi\,p_u}{\Delta^u_3(\vec b)}\sum_{\ell_1\ell_2\ell_3}\sum_{m_1m_2m_3}w^{\ell_1\ell_2\ell_3}_{m_1m_2m_3}\Theta_{\ell_1}(b_1)\Theta_{\ell_2}(b_2)\Theta_{\ell_3}(b_3)h^{u_1}_{\ell_1m_1}h^{u_2}_{\ell_2m_2}h^{u_3}_{\ell_3m_3},
\eeq
with the additional restriction $(-1)^{\ell_1+\ell_2+\ell_3} = \chi\,p_u$. Here, we have dropped the expectation terms, since this fixes $\ell_2=\ell_3$ which requires $\ell_1=0$ after summing over $m_i$. In the second line, we have replaced terms with their complex conjugates, finding a familiar bispectrum estimator: simply the product of three harmonic-space fields multiplied by the weights. For efficient implementation, one can express the weights in integral form as before:
\beq
    \left[\mathcal{F}_3\widehat{b}\right]^u_\chi(\vec b) &\to& \frac{1}{6\Delta^u_3(\vec b)}\int d\hn\,\bigg\{{}_{+1}H^{u_1}(\hn;b_1){}_{+1}H^{u_2}(\hn;b_2){}_{-2}H^{u_3}(\hn;b_3)\\\nonumber
    &&\qquad\qquad\qquad\qquad\qquad\,+\,(\chi\,p_u){}_{+1}\overline{H}^{u_1}(\hn;b_1){}_{+1}\overline{H}^{u_2}(\hn;b_2){}_{-2}\overline{H}^{u_3}(\hn;b_3)\bigg\}+\text{2 perms.},
\eeq
with the ideal definitions
\beq
    {}_{s}H^{X}(\hn;b) \to \sum_{\ell m}\Theta_\ell(b)h_{\ell m}^{X*}{}_{s}Y_{\ell m}(\hn), \qquad {}_{s}\overline{H}^{X}(\hn;b) \to \sum_{\ell m}(-1)^\ell\Theta_\ell(b)h_{\ell m}^{X*}{}_{s}Y_{\ell m}(\hn).
\eeq
This matches standard prescriptions \citep[e.g.,][]{Shiraishi:2014roa,Coulton:2019bnz}. Furthermore, from the conjugation properties of \eqref{eq: H-s-conjugation-parity}, this is explicitly real if $\chi\,p_u = 1$ and imaginary if $\chi\,p_u=-1$, and can be written succinctly in the form of \eqref{eq: cubic-numerator-re-im}.

For the normalization, we first rewrite in harmonic space in terms of $T/E/B$ fields:
\beq
    \mathcal{F}_{3,\chi\chi'}^{uu'}(\vec b,\vec b') &\to&  
    \frac{1}{3!}\sum_{\ell_im_iX_iX_i'}\left[\frac{\partial \av{d^{X_1}_{\ell_1m_1}d^{X_2}_{\ell_2m_2}d^{X_3}_{\ell_3m_3}}^*}{\partial b^u_\chi(\vec b)}\left(\mathbb{C}^{-1}_{\ell_1}\right)^{X_1X_1'}\left(\mathbb{C}^{-1}_{\ell_2}\right)^{X_2X_2'}\left(\mathbb{C}^{-1}_{\ell_3}\right)^{X_3X_3'}\frac{\partial \av{d^{X_1'}_{\ell_1m_1}d^{X_2'}_{\ell_2m_2}d^{X_3'}_{\ell_3m_3}}}{\partial b^{u'}_{\chi'}(\vec b')}\right]^*.
\eeq
Inserting the explicit polyspectrum derivatives, we find
\beq    
    \mathcal{F}_{3,\chi\chi'}^{uu'}(\vec b,\vec b') &\to& \frac{\chi\,p_u}{6\Delta_3^u(\vec b)\Delta_3^{u'}(\vec b')}\sum_{\ell_1\ell_2\ell_3}\left(\sum_{m_1m_2m_3}\left[w_{\ell_1\ell_2\ell_3}^{m_1m_2m_3}\right]^2\right)\sum_{X_iX_i'}\left[\mathbb{C}^{-1,X_1X_1'}_{\ell_1}\mathbb{C}^{-1,X_2X_2'}_{\ell_2}\mathbb{C}^{-1,X_3X_3'}_{\ell_3}\right]\\\nonumber
    &&\,\times\,\left[\delta_{\rm K}^{u,X_1X_2X_3}\Theta_{\ell_1}(b_1)\Theta_{\ell_2}(b_2)\Theta_{\ell_3}(b_3)+\text{5 perms.}\right]\left[\delta_{\rm K}^{u',X_1'X_2'X_3'}\Theta_{\ell_1}(b_1')\Theta_{\ell_2}(b_2')\Theta_{\ell_3}(b_3')+\text{5 perms.}\right],
\eeq
recalling that $\mathbb{C}^{-1}$ is real. This has the additional restriction $(-1)^{\ell_1+\ell_2+\ell_3}=\chi\,p_u = \chi'\,p_{u'}$, implying that there are no correlations between physics sourced by different parities. The $m_i$ summation yields
\beq
    \sum_{m_1m_2m_3}\left[w_{\ell_1\ell_2\ell_3}^{m_1m_2m_3}\right]^2 = \frac{1}{9}\frac{(2\ell_1+1)(2\ell_2+1)(2\ell_3+1)}{4\pi}\left[\tj{\ell_1}{\ell_2}{\ell_3}{-1}{-1}{2}+\text{2 cyc.}\right]^2,
\eeq
via Wigner $3j$ orthogonality. Finally, we can absorb one of the permutation summations, giving a factor of six, thus
\beq    
    \mathcal{F}_{3,\chi\chi'}^{uu'}(\vec b,\vec b') &\to& \frac{1}{9}\frac{(2\ell_1+1)(2\ell_2+1)(2\ell_3+1)}{4\pi}\left[\tj{\ell_1}{\ell_2}{\ell_3}{-1}{-1}{2}+\text{2 cyc.}\right]^2\frac{\chi\,p_u\,\delta_{\rm K}^{\chi\,p_u,\chi'\,p_{u'}}}{\Delta_3^u(\vec b)\Delta_3^{u'}(\vec b')}\sum_{\ell_1\ell_2\ell_3}\Theta_{\ell_1}(b_1')\Theta_{\ell_2}(b_2')\Theta_{\ell_3}(b_3')\nonumber\\
    &&\,\times\,\left[\mathbb{C}^{-1,u_1u_1'}_{\ell_1}\mathbb{C}^{-1,u_2u_2'}_{\ell_2}\mathbb{C}^{-1,u_3u_3'}_{\ell_3}\delta^{\rm K}_{b_1b_1'}\delta^{\rm K}_{b_2b_2'}\delta^{\rm K}_{b_3b_3'}+\text{5 perms.}\right],
\eeq
where the permutations are over $\{u_i,b_i\}$ pairs and again we restrict to $(-1)^{\ell_1+\ell_2+\ell_3}=\chi\,p_u$. Notably, this is not diagonal in fields, with, for example $TTT$ and $TEE$ non-trivially correlated (as for the power spectrum). Some covariances do vanish however, for example, there are no $TBB$-$TTT$ correlations if $C_\ell^{TB}=0$. Finally, we note that pairs of bispectra are correlated only if their bins $\vec b$ and $\vec b'$ are permutations of each other. Although \S\ref{subsec: binning} gave restrictions on which bins are required for each spectrum (which depend on the fields in question), there can be non-trivial correlations with $\vec b\neq \vec b'$, for example $b^{TTT}(b_1,b_2,b_3)$ and $b^{TTE}(b_1,b_3,b_2)$ are related. This lies in contrast with the spin-zero case, where the restriction of $b_1\leq b_2\leq b_3$ gave a factor $\delta^{\rm K}_{\vec b,\vec b'}$, and thus a diagonal ideal Fisher matrix.

\section{Trispectrum}\label{sec: tl}
Finally, we turn to the estimation of the polarized four-point function.\footnote{Whilst one \textit{could} extend the techniques of this work to higher-point correlators still, this would be somewhat academic since, if the likelihood is quasi-Gaussian, there is a strong descending hierarchy of information content in the $N$-point correlators. Quadspectra and beyond are similarly of little use for inflationary studies, since they are sourced only by three-particle interactions or higher-loops. In the non-linear regime, higher-order correlators can dominate; however, the $N$-point correlations functions are perturbative statistics at heart, and thus far from optimal summary statistics in this setting.} This is essentially analogous to the bispectrum estimator, but made somewhat more involved by the addition of new fields, internal momenta, and a more complex permutation structure. Our starting point is the general $N=4$ estimator, from \eqref{eq: gen-estimator}:
\beq\label{eq: gen-tris-estimator}
    \left[\mathcal{F}_4\widehat{t}\right]^u_\chi(\vec b,B) &=& \frac{1}{4!}\frac{\partial \av{\tilde{d}^{i_1}\cdots\tilde{d}^{i_4}}}{\partial t^u_\chi(\vec b,B)}\bigg(h_{i_1}h_{i_2}h_{i_3}h_{i_4}-\left[h_{i_1}h_{i_2}\av{h_{i_3}h_{i_4}}+\text{5 perms.}\right]+\left[\av{h_{i_1}h_{i_2}}\av{h_{i_3}h_{i_4}}+\text{2 perms.}\right]\bigg)^*\nonumber\\
    \mathcal{F}_{4,\chi\chi'}^{uu'}(\vec b,B,\vec b',B') &=&  \frac{1}{4!}\left[\frac{\partial \av{\tilde{d}^{i_1}\cdots\tilde{d}^{i_4}}^*}{\partial t^u_\chi(\vec b,B)}\mathsf{S}^{-1}_{i_1j_1}\cdots\mathsf{S}^{-1}_{i_4j_4}\frac{\partial \av{\tilde{d}^{j_1}\cdots\tilde{d}^{j_4}}}{\partial t^{u'}_{\chi'}(\vec b',B')}\right]^*,
\eeq
where the binning depends on four sides $\vec b \equiv \{b_1,b_2,b_3,b_4\}$ and a diagonal $B$, as specified in \S\ref{subsec: binning}. As in \S\ref{sec: bl}, the expectations in the numerator can be computed via Monte Carlo averaging over GRFs with fiducial covariance $\mathsf{C}$:
\beq
    \left[\mathcal{F}_4\widehat{t}\right]^u_\chi(\vec b,B) &=& \frac{1}{4!}\frac{\partial \av{\tilde{d}^{i_1}\cdots\tilde{d}^{i_4}}}{\partial t^u_\chi(\vec b,B)}\bigg(h_{i_1}h_{i_2}h_{i_3}h_{i_4}-\left[h_{i_1}h_{i_2}\av{[\Si\alpha]_{i_3}[\Si \alpha]_{i_4}}_\alpha+\text{5 perms.}\right]\\\nonumber
    &&\qquad\qquad\qquad\qquad\,+\,\left[\av{[\Si\alpha^{(1)}]_{i_1}[\Si\alpha^{(1)}]_{i_2}[\Si\alpha^{(2)}]_{i_3}[\Si\alpha^{(2)}]_{i_4}}_{\alpha}+\text{2 perms.}\right]\bigg)^*,
\eeq
where $\alpha^{(1)}$ and $\alpha^{(2)}$ are independent and identically distributed. As we show in \S\ref{subsec: tl-limits}, these terms vanish for parity-odd correlators in the ideal limit, but, for parity-even scenarios, perform an important r\^ole subtracting off the Gaussian part of the four-point correlator. As such, it is important that the random fields have a covariance close to the true value to avoid biasing the estimator.

\subsection{Numerator}\label{subsec: tl-num}
As before, we begin by considering the explicit form of the map-space four-point correlation function \eqref{eq: tl-unbinned-def}:
\beq
    \av{{}_{s_1}\tilde{d}(\hn_1)\cdots {}_{s_4}\tilde{d}(\hn_4)} &=& \prod_{i=1}^4\left[W(\hn_i)\sum_{\ell_im_iX_i}B_{\ell_i}^{X_i}{}_{s_i}Y_{\ell_im_i}(\hn_i){}_{s_i}\R_{X_i}\right]\sum_{LM}(-1)^Mw^{L(-M)}_{\ell_1\ell_2m_1m_2}w^{LM}_{\ell_3\ell_4m_3m_4}t^{X_1X_2,X_3X_4}_{\ell_1\ell_2,\ell_3\ell_4}(L)+\text{23 perms.}\nonumber\\
    &=&\frac{1}{2}\prod_{i=1}^4\left[W(\hn_i)\sum_{\ell_im_iX_i}B_{\ell_i}^{X_i}{}_{s_i}Y_{\ell_im_i}(\hn_i){}_{s_i}\R_{X_i}\right]\sum_{LM}(-1)^Mw^{L(-M)}_{\ell_1\ell_2m_1m_2}w^{LM}_{\ell_3\ell_4m_3m_4}\nonumber\\
    &&\,\times\,\sum_{\vec b,u\in U_4,\chi\in\pm1}\frac{t^u_\chi(\vec b,B)}{\Delta^u_4(\vec b)}\Theta_L(B)\left[1+\chi \,p_u(-1)^{\ell_1+\ell_2+\ell_3+\ell_4}\right]\nonumber\\
    &&\quad\,\times\,\left[\delta_{\rm K}^{u,XYZW}\Theta_{\ell_1}(b_1)\Theta_{\ell_2}(b_2)\Theta_{\ell_3}(b_3)\Theta_{\ell_4}(b_4)+\text{7 perms.}\right]+\left(2\leftrightarrow3\right)+\left(2\leftrightarrow4\right).
\eeq
Taking the derivative with respect to the trispectrum components and inserting into the estimator numerator yields
\beq
    \left[\mathcal{F}_4\widehat{t}\right]^u_\chi(\vec b,B) &=& \widehat{\tau}^{u}_{\chi}[d,d,d,d](\vec b,B)-\bigg[\av{\widehat{\tau}^{u}_{\chi}[d,d,\alpha,\alpha](\vec b,B)}_\alpha+\text{5 perms.}\bigg]\\\nonumber
    &&\qquad\,+\,\bigg[\av{\widehat{\tau}^{u}_{\chi}[\alpha^{(1)},\alpha^{(1)},\alpha^{(2)},\alpha^{(2)}](\vec b,B)}_\alpha+\text{2 perms.}\bigg]\\\nonumber
    \widehat{\tau}^{u}_{\chi}[\alpha,\beta,\gamma,\delta](\vec b,B) &=& 
    \frac{1}{2\Delta_4^u(\vec b)}\sum_{\ell_im_i}\sum_{LM}(-1)^{M}w^{L(-M)}_{\ell_1\ell_2m_1m_2}w^{LM}_{\ell_3\ell_4m_3m_4}\Theta_L(B)B_{\ell_1}^{u_1}B_{\ell_2}^{u_2}B_{\ell_3}^{u_3}B_{\ell_4}^{u_4}\left[1+\chi \,p_u(-1)^{\ell_1+\ell_2+\ell_3+\ell_4}\right]\\\nonumber
    &&\times\,\Theta_{\ell_1}(b_1)\Theta_{\ell_2}(b_2)\Theta_{\ell_3}(b_3)\Theta_{\ell_4}(b_4)[\R^\dag W\Si\alpha]^{u_1*}_{\ell_1m_1}[\R^\dag W\Si\beta]^{u_2*}_{\ell_2m_2}[\R^\dag W\Si\gamma]^{u_3*}_{\ell_3m_3}[\R^\dag W\Si\delta]^{u_4*}_{\ell_4m_4},
\eeq
replacing the map-space integrals by harmonic transforms and absorbing the permutations by symmetry. This factorizes into two terms, linked only by $L,M$:
\beq
    \widehat{\tau}^{u}_{\chi}[\alpha,\beta,\gamma,\delta](\vec b,B) &=& 
    \frac{1}{2\Delta_4^u(\vec b)}\sum_{LM}(-1)^{M}\Theta_L(B)\\\nonumber
    &&\times\,\left\{A^{u_1u_2}_{b_1b_2}[\alpha,\beta](L,-M)A^{u_3u_4}_{b_3b_4}[\gamma,\delta](L,M)+(\chi \,p_u)\overline{A}^{u_1u_2}_{b_1b_2}[\alpha,\beta](L,-M)\overline{A}^{u_3u_4}_{b_3b_4}[\gamma,\delta](L,M)\right\}
\eeq
defining the harmonic-space maps
\beq\label{eq: A-def}
    A^{u_1u_2}_{b_1b_2}[\alpha,\beta](L,M) &=& \sum_{\ell_1\ell_2m_1m_2}w^{LM}_{\ell_1\ell_2m_1m_2}\Theta_{\ell_1}(b_1)\Theta_{\ell_2}(b_2)B_{\ell_1}^{u_1}B_{\ell_2}^{u_2}[\R^\dag W\Si\alpha]^{u_1*}_{\ell_1m_1}[\R^\dag W\Si\beta]^{u_2*}_{\ell_2m_2}\\\nonumber
    \overline{A}^{u_1u_2}_{b_1b_2}[\alpha,\beta](L,M) &=& \sum_{\ell_1\ell_2m_1m_2}(-1)^{\ell_1+\ell_2+L}w^{LM}_{\ell_1\ell_2m_1m_2}\Theta_{\ell_1}(b_1)\Theta_{\ell_2}(b_2)B_{\ell_1}^{u_1}B_{\ell_2}^{u_2}[\R^\dag W\Si\alpha]^{u_1*}_{\ell_1m_1}[\R^\dag W\Si\beta]^{u_2*}_{\ell_2m_2}.
\eeq
This is a consequence of parametrizing the tetrahedron by sides and diagonals, which splits the shape up into two triangles with a shared diagonal. These have the symmetry property
\beq\label{eq: A-field-sym}
    \left[A^{u_1u_2}_{b_1b_2}[\alpha,\beta](L,M)\right]^* = (-1)^M\overline{A}^{u_1u_2}_{b_1b_2}[\alpha,\beta](L,-M),
\eeq
implying that the $\widehat{\tau}$ estimator can be rewritten as
\beq\label{eq: tau-re-im-form}
    \widehat{\tau}^{u}_{+}[\alpha,\beta,\gamma,\delta](\vec b,B) &=&    
    \frac{1}{2\Delta_4^u(\vec b)}\sum_{L\,M\geq 0}(1+\delta^{\rm K}_{M>0})\Theta_L(B)\,\mathrm{Re}\Bigg\{\left(\overline{A}^{u_1u_2}_{b_1b_2}[\alpha,\beta](L,M)\right)^*A^{u_3u_4}_{b_3b_4}[\gamma,\delta](L,M)\\\nonumber
    &&\qquad\qquad\qquad\qquad\qquad\qquad\qquad\qquad\,+\,\left(\overline{A}^{u_3u_4}_{b_3b_4}[\gamma,\delta](L,M)\right)^*A^{u_1u_2}_{b_1b_2}[\alpha,\beta](L,M)\Bigg\}\\\nonumber
    \widehat{\tau}^{u}_{-}[\alpha,\beta,\gamma,\delta](\vec b,B) &=& \frac{i}{\Delta_4^u(\vec b)}\sum_{L\,M\geq 0}(1+\delta^{\rm K}_{M>0})\Theta_L(B)\,\mathrm{Im}\Bigg\{\left(\overline{A}^{u_1u_2}_{b_1b_2}[\alpha,\beta](L,M)\right)^*A^{u_3u_4}_{b_3b_4}[\gamma,\delta](L,M)\\\nonumber
    &&\qquad\qquad\qquad\qquad\qquad\qquad\qquad\qquad\,+\,\left(\overline{A}^{u_3u_4}_{b_3b_4}[\gamma,\delta](L,M)\right)^*A^{u_1u_2}_{b_1b_2}[\alpha,\beta](L,M)\Bigg\},
\eeq
which is considerably simpler to implement, involving only modes with $M\geq 0$ (\textit{i.e.} those stored by \textsc{HEALPix}).

Next, we insert the integral forms of the trispectrum weights given in \eqref{eq: tl-weights}, yielding
\beq
    A^{u_1u_2}_{b_1b_2}[\alpha,\beta](L,M) &=& \int d\hn\,{}_{-2}Y_{LM}(\hn){}_{+1}H^{u_1}[\alpha](\hn;b_1){}_{+1}H^{u_2}[\beta](\hn;b_2)\\\nonumber
    &=& \left(\int d\hn\,{}_{-2}Y^*_{LM}(\hn){}_{-1}H^{u_1}[\alpha](\hn;b_1){}_{-1}H^{u_2}[\beta](\hn;b_2)\right)^*\\\nonumber
    \overline{A}^{u_1u_2}_{b_1b_2}[\alpha,\beta](L,M) &=& (-1)^{L}\int d\hn\,{}_{-2}Y_{LM}(\hn){}_{+1}\overline{H}^{u_1}[\alpha](\hn;b_1){}_{+1}\overline{H}^{u_2}[\beta](\hn;b_2)\\\nonumber
    &=& \left(\int d\hn\,{}_{+2}Y^*_{LM}(\hn){}_{+1}H^{u_1}[\alpha](\hn;b_1){}_{+1}H^{u_2}[\beta](\hn;b_2)\right)^*
\eeq
in terms of the $H$ maps of \eqref{eq: H-map-def} using the symmetry properties of \eqref{eq: H-s-conjugation-parity}. Despite the chain of definitions, this form is straightforward to implement in $\mathcal{O}(3N_{\rm it}N_{\rm pix})$ operations via the following prescription:
\begin{enumerate}
    \item Draw two GRFs $\alpha^{(1)}$ and $\alpha^{(2)}$ from the (close-to) true covariance $\mathsf{C}$.
    \item Filter both $\alpha$ and the data $d$ by the weighting scheme and the mask to form $W\Si d$, $W\Si \alpha$ maps, then use a spherical harmonic transform to compute $\R^\dag W\Si d$ and $\R^\dag W\Si \alpha$, as before.
    \item Compute the ${}_{\pm 1}H^X$ maps for each bin of interest with $X\in\{T,E,B\}$ via a further spin-weighted spherical harmonic transform.
    \item Compute $A(L,M)$ and $\overline{A}(L,M)$ harmonic-space maps for each pair of bins and fields.
    \item Combine these pairwise to form both $\tau[d,d,\alpha,\alpha]$ and $\tau[\alpha^{(1)},\alpha^{(1)},\alpha^{(2)},\alpha^{(2)}]$ and their permutations.
    \item Iterate over $N_{\rm it}/2$ pairs of random fields to form the Monte Carlo average and use this to assemble the estimator numerator.
\end{enumerate}

\subsection{Normalization}\label{subsec: tl-norm}
Finally, we consider the trispectrum Fisher matrix, which is computed similarly to the bispectrum piece. As before, we start by prudent introduction of the identity matrix in the following manner:
\beq
    \Si_{i_2j_2}\Si_{i_3j_3}\Si_{i_4j_4}+\text{5 perms.} = \Si_{i_2i_2'}\Si_{i_3i_3'}\Si_{i_4i_4'}\times \left[\mathsf{A}_{i_2'j_2'}\mathsf{A}_{i_3'j_3'}\mathsf{A}_{i_4'j_4'}+\text{5 perms.}\right]\times\mathsf{A}^{-1}_{j_2'j_2}\mathsf{A}^{-1}_{j_3'j_3}\mathsf{A}^{-1}_{j_4'j_4},
\eeq
where the permutations contribute equivalently to the Fisher matrix due to index interchange symmetry. Introducing two set of random fields $\alpha^{(1)},\alpha^{(2)}$, we can rewrite this as
\beq
    \Si_{i_2j_2}\Si_{i_3j_3}\Si_{i_4j_4}+\text{5 perms.} &=& \frac{1}{8}\Si_{i_2i_2'}\Si_{i_3i_3'}\Si_{i_4i_4'}\times \bigg\langle\left(a^{(1)}_{i_2'}a_{i_3'}^{(1)}a_{i_4'}^{(1)}a^{(1)*}_{j_2'}a^{(1)*}_{j_3'}a^{(1)*}_{j_4'}+a^{(2)}_{i_2'}a_{i_3'}^{(2)}a_{i_4'}^{(2)}a^{(2)*}_{j_2'}a^{(2)*}_{j_3'}a^{(2)*}_{j_4'}\right)\\\nonumber
    &&\qquad\,+\,9\left(a^{(1)}_{i_2'}a_{i_3'}^{(1)}a_{i_4'}^{(2)}a^{(1)*}_{j_2'}a^{(1)*}_{j_3'}a^{(2)*}_{j_4'}+a^{(1)}_{i_2'}a_{i_3'}^{(2)}a_{i_4'}^{(2)}a^{(1)*}_{j_2'}a^{(2)*}_{j_3'}a^{(2)*}_{j_4'}\right)\\\nonumber
    &&\qquad\,-\,6\left(a^{(1)}_{i_2'}a_{i_3'}^{(1)}a_{i_4'}^{(1)}a^{(1)*}_{j_2'}a^{(2)*}_{j_3'}a^{(2)*}_{j_4'}+a^{(2)}_{i_2'}a_{i_3'}^{(2)}a_{i_4'}^{(2)}a^{(1)*}_{j_2'}a^{(1)*}_{j_3'}a^{(2)*}_{j_4'}\right)\bigg\rangle_a\times\mathsf{A}^{-1}_{j_2'j_2}\mathsf{A}^{-1}_{j_3'j_3}\mathsf{A}^{-1}_{j_4'j_4},
\eeq
which makes most efficient use of the pairs of simulations, as discussed in \citep{2015arXiv150200635S}. This allows the Fisher matrix to be written in the form:
\beq
    \mathcal{F}_{4,\chi\chi'}^{uu'}(\vec b,B,\vec b',B') &=& \frac{1}{48}\bigg[\left(F^{uu',111,111}_{4,\chi\chi'}(\vec b,B,\vec b',B')+F^{uu',222,222}_{4,\chi\chi'}(\vec b,B,\vec b',B')\right)\\\nonumber
    &&\qquad\,+\,9\left(F^{uu',112,112}_{4,\chi\chi'}(\vec b,B,\vec b',B')+F^{uu',122,122}_{4,\chi\chi'}(\vec b,B,\vec b',B')\right)\\\nonumber
    &&\qquad\,-\,6\left(F^{uu',111,122}_{4,\chi\chi'}(\vec b,B,\vec b',B')+F^{uu',222,112}_{4,\chi\chi'}(\vec b,B,\vec b',B')\right)\bigg]\\\nonumber
    F^{uu',abc,def}_{4,\chi\chi'}(\vec b,B,\vec b',B') &=& \frac{1}{4!}\bigg\langle\frac{\partial \av{\tilde{d}^{i_1}\cdots \tilde{d}^{i_4}}^*}{\partial t^u_\chi(\vec b,B)}\Si_{i_1j_1}[\Si a^{(a)}]_{i_2}[\Si a^{(b)}]_{i_3}[\Si a^{(c)}]_{i_4}\\\nonumber
    &&\qquad\qquad\qquad\times\,[\mathsf{A}^{-1}a^{(d)}]^*_{j_2}[\mathsf{A}^{-1}a^{(e)}]^*_{j_3}[\mathsf{A}^{-1}a^{(f)}]^*_{j_4}\frac{\partial \av{\tilde{d}^{j_1}\cdots \tilde{d}^{j_4}}}{\partial t^{u'}_{\chi'}(\vec b',B')}\bigg\rangle^*_a\\\nonumber
    &=& \frac{1}{4!}\bigg\langle\left(Q^{u}_{4,\chi}[\Si a^{(a)},\Si a^{(b)},\Si a^{(c)}](\vec b,B)\right)^* \cdot[W\mathsf{S}^{-1}W]\\\nonumber
    &&\quad\qquad\,\cdot\left(Q^{u'}_{4,\chi'}[\mathsf{A}^{-1} a^{(d)},\mathsf{A}^{-1} a^{(e)},\mathsf{A}^{-1} a^{(f)}](\vec b',B')\right)\bigg\rangle^*_a,
\eeq
with the derivative maps
\beq
    Q^{u}_{4,\chi,i}[x,y,z](\vec b,B) = \frac{\partial\av{d^id^jd^kd^l}}{\partial t^u_\chi(\vec b,B)}[Wx]^*_{j}[Wy]^*_k[Wz]^*_l.
\eeq
Given the $Q_4$ maps, the Fisher matrix can thus be computed as an inner product (though we caution that there will be a large number of components, unless the binning is very coarse).

Inserting the binned trispectrum definition \eqref{eq: binned-trispectrum}, the derivative maps take the explicit form in harmonic space
\beq
    {}_{s_1}Q^{u}_{4,\chi,\ell_1m_1}[x,y,z](\vec b,B) &=& \frac{1}{2\Delta^u_4(\vec b)}\sum_{\ell_im_iX_i}{}_{s_1}\R_{X_1}\sum_{LM}(-1)^Mw^{L(-M)}_{\ell_1\ell_2m_1m_2}w^{LM}_{\ell_3\ell_4m_3m_4}\Theta_L(B)B_{\ell_1}^{X_1}B_{\ell_2}^{X_2}B_{\ell_3}^{X_3}B_{\ell_4}^{X_4}\\\nonumber
    &&\qquad\,\times\,\left[1+\chi \,p_u(-1)^{\ell_1+\ell_2+\ell_3+\ell_4}\right]\left[\delta_{\rm K}^{u,X_1X_2X_3X_4}\Theta_{\ell_1}(b_1)\Theta_{\ell_2}(b_2)\Theta_{\ell_3}(b_3)\Theta_{\ell_4}(b_4)+\text{7 perms.}\right]\\\nonumber
    &&\qquad\,\times\,[\R^\dag Wx]^{X_2*}_{\ell_2m_2}[\R^\dag Wy]^{X_3*}_{\ell_3m_3}[\R^\dag Wz]^{X_4*}_{\ell_4m_4}\,+\,\left(2\leftrightarrow3\right)+\left(2\leftrightarrow4\right).
\eeq
To simplify these further, we can replace the second trispectrum weight by its integral representation and rearrange, finding
\beq
    {}_{s_1}Q^{u}_{4,\chi,\ell_1m_1}[x,y,z](\vec b,B) &=& \bigg(\frac{\Theta_{\ell_1}(b_1)B_{\ell_1}^{u_1}}{2\Delta^u_4(\vec b)}{}_{s_1}\R_{u_1}\sum_{\ell_2m_2LM}(-1)^{M}w^{L(-M)}_{\ell_1\ell_2m_1m_2}\Theta_L(B)\Theta_{\ell_2}(b_2)B_{\ell_2}^{u_2}[\R^\dag Wx]^{u_2*}_{\ell_2m_2}\nonumber\\
    &&\,\times\,\left[A^{u_3u_4}_{b_3b_4}[y,z](L,M)+\chi\,p_u(-1)^{\ell_1+\ell_2+L}\overline{A}^{u_3u_4}_{b_3b_4}[y,z](L,M)\right]\nonumber\\
    &&\,+\,\text{7 perms.}\bigg)+\left(2\leftrightarrow3\right)+\left(2\leftrightarrow4\right),
\eeq
in terms of the $A$ functions of \eqref{eq: A-def}. Replacing also the first weighting function yields
\beq\label{eq: Q4-deriv-simp}
    {}_{s_1}Q^{u}_{4,\chi,\ell_1m_1}[x,y,z](\vec b,B) &=& \bigg(\frac{\Theta_{\ell_1}(b_1)B_{\ell_1}^{u_1}}{2\Delta^u_4(\vec b)}{}_{s_1}\R_{u_1}\int d\hn\bigg\{{}_{+1}Y_{\ell_1m_1}(\hn){}_{+1}H^{u_2}[x](\hn;b_2){}_{-2}A^{u_3u_4}_{b_3b_4}[y,z](\hn;B)\nonumber\\
    &&\qquad\qquad\qquad\qquad\qquad\,+\,(\chi\,p_u){}_{-1}Y_{\ell_1m_1}(\hn){}_{-1}H^{u_2}[x](\hn;b_2){}_{+2}A^{u_3u_4}_{b_3b_4}[y,z](\hn;B)\bigg\}\nonumber\\
    &&\,+\,\text{7 perms.}\bigg)+\left(x\leftrightarrow y\right)+\left(x\leftrightarrow z\right),
\eeq
invoking the $A$ symmetry given in \eqref{eq: A-field-sym}, noting that ${}_{s}\overline{H}^X[x](-\hn;b) = {}_{-s}H^X[x](\hn;b)$, and defining the spin-$\pm2$ maps
\beq
     {}_{+2}A^{u_3u_4}_{b_3b_4}[y,z](\hn;B) &=& \sum_{LM} \Theta_L(B){}_{+2}Y_{LM}(\hn)A^{u_3u_4*}_{b_3b_4}[y,z](L,M)\\\nonumber
    {}_{-2}A^{u_3u_4}_{b_3b_4}[y,z](\hn;B) &=& \sum_{LM} \Theta_L(B){}_{-2}Y_{LM}(\hn)\overline{A}^{u_3u_4*}_{b_3b_4}[y,z](L,M)
\eeq
(noting the complex conjugates). Here, the eight permutations in \eqref{eq: Q4-deriv-simp} are over pairs of bins and fields, preserving the $(b_1,b_2)$ and $(b_3,b_4)$ pairings. Finally, we can simplify a little further using the above symmetries to yield the $T/E/B$-space representation:
\beq\label{eq: Q4-simplified}
    Q^{u,X}_{4,\chi,\ell_1m_1}[x,y,z](\vec b,B) &=& -\bigg(\frac{\Theta_{\ell_1}(b_1)B_{\ell_1}^{u_1}}{2\Delta^u_4(\vec b)}\delta_{\rm K}^{u_1X}\int d\hn\,\bigg\{{}_{+1}Y^*_{\ell_1m_1}(\hn){}_{-1}H^{u_2}[x](\hn;b_2){}_{+2}A^{u_3u_4}_{b_3b_4}[y,z](\hn;B)\nonumber\\
    &&\qquad\qquad\qquad\qquad\qquad\,+\,(\chi\,p_u){}_{-1}Y^*_{\ell_1m_1}(\hn){}_{+1}H^{u_2}[x](\hn;b_2){}_{-2}A^{u_3u_4}_{b_3b_4}[y,z](\hn;B)\bigg\}^*\nonumber\\
    &&\,+\,\text{7 perms.}\bigg)+\left(x\leftrightarrow y\right)+\left(x\leftrightarrow z\right)
\eeq
analogous to the $Q_3$ form of \eqref{eq: Q3-simplified}. This can be computed as a further harmonic transform, noting that $\left(Q_{4,\chi,\ell_1m_1}^{u,X}\right)^* = (\chi\,p_u)(-1)^{m_1}Q_{4,\chi,\ell_1(-m_1)}^{u,X}$, thus $Q_4$ consists of three real fields in map-space, once we take the imaginary part of any parity-odd correlators.

Given the above manipulations, our route to computing the full Fisher matrix in $\mathcal{O}(3N_{\rm pix}N_{\rm it})$ operations becomes:
\begin{enumerate}
    \item Generate random fields $a^{(1)}$ and $a^{(2)}$ from the fiducial covariance $\mathsf{A}$ (which should be close to $\mathsf{C}$ for efficient convergence).
    \item Compute the $[\R^\dag W\Si a]$ and $[\R^\dag W \mathsf{A}^{-1}a]$ fields via harmonic transforms for both random fields.
    \item Compute ${}_{\pm 1}H$ for each field and bin of interest via a harmonic transform.
    \item Perform a second harmonic transform to compute $A$ and $\overline{A}$ for each pair of bins and fields, then transform to ${}_{+2}A$ and ${}_{-2}\overline{A}$, as a function of the diagonal $B$.
    \item Perform another transform to compute ${}_{s}Q$ in harmonic space, weight by the bin, and (if the mask is non-trivial), transform to real-space.
    \item Compute the Fisher matrix by performing an inner product for each pair of spectra in real- or harmonic-space.
    \item Assemble the Monte Carlo average by summing over $N_{\rm it}$ realizations of the above procedure.
\end{enumerate}
The above is computationally demanding, due to the large number of chained harmonic transforms. However, it is not infeasible (as demonstrated in \S\ref{sec: validation}), and not parametrically harder than the scalar case, with the principal added complication coming from the significantly larger number of fields in question.

\subsection{Ideal Limits}\label{subsec: tl-limits}
We now consider the limiting form of the trispectrum estimator in ideal settings. We caution that the resulting form for the normalization is not efficient computationally; a faster numerical approach is to use the Monte Carlo procedures of \S\ref{subsec: tl-norm}.

\subsubsection{Numerator}
Akin to before, we begin by writing the numerator explicitly in $T/E/B$ harmonic space in ideal conditions:
\beq
    \left[\F_4\widehat{t}\right]^u_\chi(\vec b,B)&\to& \frac{1}{4!}\sum_{\ell_im_iX_i}\frac{\partial \av{d^{X_1}_{\ell_1m_1}d^{X_2}_{\ell_2m_2}d^{X_3}_{\ell_3m_3}d^{X_4}_{\ell_4m_4}}}{\partial t^u_\chi(\vec b,B)}\bigg(h^{X_1}_{\ell_1m_1}h^{X_2}_{\ell_2m_2}h^{X_3}_{\ell_3m_3}h^{X_4}_{\ell_4m_4}\\\nonumber
    &&\qquad\qquad\qquad\qquad\,-\,\left[h^{X_1}_{\ell_1m_1}h^{X_2}_{\ell_2m_2}(-1)^{m_4}\delta^{\rm K}_{\ell_3\ell_4}\delta^{\rm K}_{m_3(-m_4)}\mathbb{C}^{-1,X_3X_4}_{\ell_3}+\text{5 perms.}\right]\\\nonumber
    &&\qquad\qquad\qquad\qquad\,+\,\left[(-1)^{m_2+m_4}\delta^{\rm K}_{\ell_1\ell_2}\delta^{\rm K}_{\ell_3\ell_4}\delta^{\rm K}_{m_1(-m_2)}\delta^{\rm K}_{m_3(-m_4)}\mathbb{C}^{-1,X_1X_2}_{\ell_1}\mathbb{C}^{-1,X_3X_4}_{\ell_3}+\text{2 perms.}\right]\bigg)^*,
\eeq
where we have evaluated the expectations analytically. After inserting the binned trispectrum definition, the first piece, involving four fields, takes the explicit form
\beq
    \left.\left[\F_4\widehat{t}\right]^u_\chi(\vec b,B)\right|_{\rm 4-field}&\to& \frac{\chi\,p_u}{\Delta_4^u(\vec b)}\sum_{LM}\Theta_L(B)(-1)^M\left(\sum_{\ell_1\ell_2m_1m_2}\Theta_{\ell_1}(b_1)\Theta_{\ell_2}(b_2)w^{L(-M)}_{\ell_1\ell_2m_1m_2}h^{u_1}_{\ell_1m_1}h^{u_2}_{\ell_2m_2}\right)\\\nonumber
    &&\qquad\qquad\,\times\,\left(\sum_{\ell_3\ell_4m_3m_4}\Theta_{\ell_3}(b_3)\Theta_{\ell_4}(b_4)w^{LM}_{\ell_3\ell_4m_3m_4}h^{u_3}_{\ell_3m_3}h^{u_4}_{\ell_4m_4}\right),
\eeq
with the additional parity restriction that $(-1)^{\ell_1+\ell_2+\ell_3+\ell_4}=\chi\,p_u$ Practically, this can be computed analogously to the non-ideal case (\S\ref{subsec: tl-num}) with
\beq
    \left.\left[\F_4\widehat{t}\right]^u_\chi(\vec b,B)\right|_{\rm 4-field}&\to& \frac{1}{2\Delta_4^u(\vec b)}\sum_{LM}(-1)^{M}\Theta_L(B)\\\nonumber
    &&\times\,\left\{A^{u_1u_2}_{b_1b_2}[d,d](L,-M)A^{u_3u_4}_{b_3b_4}[d,d](L,M)+(\chi \,p_u)\overline{A}^{u_1u_2}_{b_1b_2}[d,d](L,-M)\overline{A}^{u_3u_4}_{b_3b_4}[d,d](L,M)\right\},
\eeq
where $A$ and $\overline{A}$ are defined as before. This can be written as a real or imaginary part as in \eqref{eq: tau-re-im-form}.

For the term involving two-fields, we proceed similarly, but note that we can use the following relation to simplify the $m_i$ summation, given the Kronecker deltas and $3j$ symbols in the weights:
\beq
    \sum_{m_3m_4M}(-1)^{M}\tj{\ell_1}{\ell_2}{L}{m_1}{m_2}{-M}\tj{\ell_3}{\ell_4}{L}{m_3}{m_4}{M}(-1)^{m_4}\delta^{\rm K}_{m_3(-m_4)}\delta^{\rm K}_{\ell_3\ell_4} &=& (-1)^{\ell_1+\ell_3-m_1}\sqrt{\frac{2\ell_3+1}{2\ell_1+1}}\delta^{\rm K}_{L0}\delta^{\rm K}_{\ell_1\ell_2}\delta^{\rm K}_{\ell_3\ell_4}\delta^{\rm K}_{m_1(-m_2)}\nonumber\\
    \sum_{m_2m_4M}(-1)^{M}\tj{\ell_1}{\ell_2}{L}{m_1}{m_2}{-M}\tj{\ell_3}{\ell_4}{L}{m_3}{m_4}{M}(-1)^{m_4}\delta^{\rm K}_{m_2(-m_4)} &=& \delta^{\rm K}_{\ell_1\ell_3}\delta^{\rm K}_{\ell_2\ell_4}\delta^{\rm K}_{m_1(-m_3)}\frac{(-1)^{\ell_2+\ell_3+L-m_1}}{2\ell_1+1},
\eeq
assuming triangle conditions to be obeyed in both cases. Importantly, terms of the first form cannot contribute, since the weights vanish for $L=0$. The two-field term can thus be written
\beq\label{eq: m-i-2-field}
    \left.\left[\F_4\widehat{t}\right]^u_\chi(\vec b,B)\right|_{\rm 2-field}&\to& -\frac{\chi\,p_u}{\Delta_4^u(\vec b)}\sum_{LM}\Theta_L(B)(-1)^M\left(\sum_{\ell_1\ell_2m_1m_2}\Theta_{\ell_1}(b_1)\Theta_{\ell_2}(b_2)w^{L(-M)}_{\ell_1\ell_2m_1m_2}\right)\left(\sum_{\ell_3\ell_4m_3m_4}\Theta_{\ell_3}(b_3)\Theta_{\ell_4}(b_4)w^{LM}_{\ell_3\ell_4m_3m_4}\right)\nonumber\\
    &&\times\,\left[\av{h^{u_1}_{\ell_1m_1}h^{u_3}_{\ell_3m_3}}h^{u_2}_{\ell_2m_2}h^{u_4}_{\ell_4m_4}+\text{3 perms.}\right],
\eeq
where $(-1)^{\ell_1+\ell_2+\ell_3+\ell_4}=\chi\,p_u$ and the remaining permutations contract the pairs $2$--$4$, $1$--$4$ and $2$--$3$. Inserting the second relation of \eqref{eq: m-i-2-field} and simplifying, we find the simplified two-field term:
\beq
    \left.\left[\F_4\widehat{t}\right]^u_\chi(\vec b,B)\right|_{\rm 2-field}&\to&-\frac{\delta^{\rm K}_{\chi\,p_u,1}}{\Delta_4^u(\vec b)}\sum_{L}\Theta_L(B)\Theta_{\ell_1}(b_1)\Theta_{\ell_2}(b_2)\tj{\ell_1}{\ell_2}{L}{-1}{-1}{2}^2\frac{(2\ell_1+1)(2\ell_2+1)(2L+1)}{4\pi}\nonumber\\
    &&\,\times\,(-1)^{\ell_1+\ell_2+L}\bigg(\delta^{\rm K}_{b_1b_3}\delta^{\rm K}_{b_2b_4}\left[\mathbb{C}^{-1,u_1u_3}_{\ell_1}\left(\widehat{\mathbb{C}^{-1}_{\ell_2}}\right)^{u_2u_4}+\left(\widehat{\mathbb{C}^{-1}_{\ell_1}}\right)^{u_1u_3}\mathbb{C}^{-1,u_2u_4}_{\ell_2}\right]\nonumber\\
    &&\qquad\qquad\qquad\qquad\,+\,\delta^{\rm K}_{b_1b_4}\delta^{\rm K}_{b_2b_3}\left[\mathbb{C}^{-1,u_1u_4}_{\ell_1}\left(\widehat{\mathbb{C}^{-1}_{\ell_2}}\right)^{u_2u_3}+\left(\widehat{\mathbb{C}^{-1}_{\ell_1}}\right)^{u_1u_4}\mathbb{C}^{-1,u_2u_3}_{\ell_2}\right]\bigg),
\eeq
introducing the empirical power spectrum estimates
\beq
    \left(\widehat{\mathbb{C}^{-1}_{\ell}}\right)^{uu'} &=& \frac{\sum_{m}(-1)^{m}h^{u}_{\ell m}h^{u'}_{\ell(-m)}}{2\ell+1} \equiv \frac{\sum_{m}h^{u}_{\ell m}h^{u'*}_{\ell m}}{2\ell+1}.
\eeq
Finally, the zero-field term is simply obtained as $(-1/2)$ the expectation of the two-field term, yielding
\beq
    \left.\left[\F_4\widehat{t}\right]^u_\chi(\vec b,B)\right|_{\rm 0-field}&\to&\frac{\delta^{\rm K}_{\chi \,p_u,1}}{\Delta_4^u(\vec b)}\sum_{L}\Theta_L(B)\Theta_{\ell_1}(b_1)\Theta_{\ell_2}(b_2)\tj{\ell_1}{\ell_2}{L}{-1}{-1}{2}^2\frac{(2\ell_1+1)(2\ell_2+1)(2L+1)}{4\pi}\nonumber\\
    &&\,\times\,(-1)^{\ell_1+\ell_2+L}\bigg(\delta^{\rm K}_{b_1b_3}\delta^{\rm K}_{b_2b_4}\mathbb{C}^{-1,u_1u_3}_{\ell_1}\mathbb{C}^{-1,u_2u_4}_{\ell_2}+\delta^{\rm K}_{b_1b_4}\delta^{\rm K}_{b_2b_3}\mathbb{C}^{-1,u_1u_4}_{\ell_1}\mathbb{C}^{-1,u_2u_3}_{\ell_2}\bigg),
\eeq
As expected, the two- and zero-field parts of the estimator just involve the (Wiener filtered) power spectra, with weightings appropriate to the binning scheme. These terms generically vanish due to the requirement that $\chi\,p_u = (-1)^{\ell_1+\ell_2+\ell_3+\ell_4}\to 1$; as such, they contribute only to parity-even terms if there are an even number of $B$-modes in the spectrum, and parity-odd terms else. Under the additional assumption of parity-conserving power spectra, we find that the disconnected terms are important only for parity-even physics ($\chi=1$) in spectra with $p_u=1$ (e.g., $TTTE$, $EBEB$, etc.).

\subsubsection{Normalization}
We now turn to the Fisher matrix, which can be written explicitly in harmonic $T/E/B$ space as
\beq
    \mathcal{F}_{4,\chi\chi'}^{uu'}(\vec b,B,\vec b',B') &\to&  
    \frac{1}{4!}\sum_{\ell_im_iX_iX_i'}\left[\frac{\partial \av{d^{X_1}_{\ell_1m_1}\cdots d^{X_4}_{\ell_4m_4}}^*}{\partial t^u_\chi(\vec b,B)}\mathbb{C}^{-1,X_1X_1'}_{\ell_1}\cdots\mathbb{C}^{-1,X_4X_4'}_{\ell_4}\frac{\partial \av{d^{X_1'}_{\ell_1m_1}\cdots d^{X_4'}_{\ell_4m_4}}}{\partial t^{u'}_{\chi'}(\vec b',B ')}\right]^*,
\eeq
just as for the bispectrum. To proceed, we insert the reduced trispectrum definition and simplify via the following $3j$ relations:
\beq\label{eq: tl-fish-3j}
    \sum_{m_1m_2m_3m_4MM'}(-1)^{M+M'}\tj{\ell_1}{\ell_2}{L}{m_1}{m_2}{-M}\tj{\ell_3}{\ell_4}{L}{m_3}{m_4}{M}\tj{\ell_1}{\ell_2}{L'}{m_1}{m_2}{-M'}\tj{\ell_3}{\ell_4}{L'}{m_3}{m_4}{M'}&=&\frac{\delta^{\rm K}_{LL'}}{2L+1}\\\nonumber
    \sum_{m_1m_2m_3m_4MM'}(-1)^{M+M'}\tj{\ell_1}{\ell_2}{L}{m_1}{m_2}{-M}\tj{\ell_3}{\ell_4}{L}{m_3}{m_4}{M}\tj{\ell_1}{\ell_3}{L'}{m_1}{m_3}{-M'}\tj{\ell_2}{\ell_4}{L'}{m_2}{m_4}{M'} &=& (-1)^{\ell_2+\ell_3}\begin{Bmatrix} L &\ell_1 &\ell_2\\ L' & \ell_4 & \ell_3\end{Bmatrix}\\\nonumber
    \sum_{m_1m_2m_3m_4MM'}(-1)^{M+M'}\tj{\ell_1}{\ell_2}{L}{m_1}{m_2}{-M}\tj{\ell_3}{\ell_4}{L}{m_3}{m_4}{M}\tj{\ell_1}{\ell_4}{L'}{m_1}{m_4}{-M'}\tj{\ell_3}{\ell_2}{L'}{m_3}{m_2}{M'} &=& (-1)^{L+L'}\begin{Bmatrix} L &\ell_1 &\ell_2\\ L' & \ell_3 & \ell_4\end{Bmatrix},
\eeq
\citep[cf.,][]{Philcox:2023uwe}, which cover the three non-trivial permutations in the trispectrum definition \eqref{eq: tl-definition}. The Fisher matrix can thus be written as a sum of three parts, the first of which involves the first line of \eqref{eq: tl-fish-3j} and is thus diagonal in $B,B'$:
\beq
    \left.\mathcal{F}_{4,\chi\chi'}^{uu'}(\vec b,B,\vec b',B')\right|_{I} &\to&  
    \delta^{\rm K}_{BB'}\frac{(\chi\,p_u)\,\delta^{\rm K}_{\chi\, p_u,\chi'\,p_{u'}}}{\Delta_4^u(\vec b)\Delta_4^{u'}(\vec b')}\sum_{\ell_iL}\Theta_{\ell_1}(b_1)\cdots\Theta_{\ell_4}(b_4)\Theta_L(B)\frac{(2\ell_1+1)\cdots(2\ell_4+1)(2L+1)}{(4\pi)^2}\\\nonumber
    &&\,\times\,\tj{\ell_1}{\ell_2}{L}{-1}{-1}{2}^2\tj{\ell_3}{\ell_4}{L}{-1}{-1}{2}^2\left[\mathbb{C}^{-1,u_1u_1'}_{\ell_1}\cdots\mathbb{C}^{-1,u_4u_4'}_{\ell_4}\delta^{\rm K}_{b_1b_1'}\cdots \delta^{\rm K}_{b_4b_4'}+\text{7 perms.}\right],
\eeq
where we have additionally inserted the binned trispectrum definition and asserted that $(-1)^{\ell_1+\ell_2+\ell_3+\ell_4}=\chi\,p_u=\chi'\,p_{u'}$. Here, the permutations are those that preserve the $1$--$2$, $3$--$4$ pairings.

The second and third parts of the Fisher matrix are similar, but are not diagonal in $L,L'$, due to the $6j$ symbols present in \eqref{eq: tl-fish-3j}. This implies that they mix different binned components (as in the scalar scale), though we will find that $\vec b$ and $\vec b'$ must always be permutations of each other. The second part takes the explicit form:
\beq
    \left.\mathcal{F}_{4,\chi\chi'}^{uu'}(\vec b,B,\vec b',B')\right|_{II} &\to&  
    \frac{(\chi\,p_u)\,\delta^{\rm K}_{\chi\,p_u,\chi'\,p_{u'}}}{\Delta_4^u(\vec b)\Delta_4^{u'}(\vec b')}\sum_{\ell_iLL'}\Theta_{\ell_1}(b_1)\cdots \Theta_{\ell_4}(b_4)\Theta_L(B)\Theta_{L'}(B')\frac{(2\ell_1+1)\cdots(2\ell_4+1)(2L+1)(2L'+1)}{(4\pi)^2}\nonumber\\
    &&\,\times\,\tj{\ell_1}{\ell_2}{L}{-1}{-1}{2}\tj{\ell_3}{\ell_4}{L}{-1}{-1}{2}\tj{\ell_1}{\ell_3}{L'}{-1}{-1}{2}\tj{\ell_2}{\ell_4}{L'}{-1}{-1}{2}(-1)^{\ell_2+\ell_3}\begin{Bmatrix} L &\ell_1 &\ell_2\\ L' & \ell_4 & \ell_3\end{Bmatrix}\nonumber\\
    &&\,\times\,\left[\mathbb{C}^{-1,u_1u_1'}_{\ell_1}\mathbb{C}^{-1,u_2u_3'}_{\ell_1}\mathbb{C}^{-1,u_3u_2'}_{\ell_1}\mathbb{C}^{-1,u_4u_4'}_{\ell_4}\delta^{\rm K}_{b_1b_1'}\delta^{\rm K}_{b_2b_3'}\delta^{\rm K}_{b_3b_2'}\delta^{\rm K}_{b_4b_4'}+\text{7 perms.}\right],
\eeq
with permutations exchanging, e.g., $b_1'\leftrightarrow b_2'$, again asserting $(-1)^{\ell_1+\ell_2+\ell_3+\ell_4}=\chi\,p_u=\chi'\,p_{u'}$. Finally, the third contribution is given by
\beq
    \left.\mathcal{F}_{4,\chi\chi'}^{uu'}(\vec b,B,\vec b',B')\right|_{III} &\to&  
    \frac{(\chi\,p_u)\,\delta^{\rm K}_{\chi\,p_u,\chi'\,p_{u'}}}{\Delta_4^u(\vec b)\Delta_4^{u'}(\vec b')}\sum_{\ell_iLL'}\Theta_{\ell_1}(b_1)\cdots \Theta_{\ell_4}(b_4)\Theta_L(B)\Theta_{L'}(B')\frac{(2\ell_1+1)\cdots(2\ell_4+1)(2L+1)(2L'+1)}{(4\pi)^2}\nonumber\\
    &&\,\times\,\tj{\ell_1}{\ell_2}{L}{-1}{-1}{2}\tj{\ell_3}{\ell_4}{L}{-1}{-1}{2}\tj{\ell_1}{\ell_4}{L'}{-1}{-1}{2}\tj{\ell_3}{\ell_2}{L'}{-1}{-1}{2}(-1)^{L+L'}\begin{Bmatrix} L &\ell_1 &\ell_2\\ L' & \ell_3 & \ell_4\end{Bmatrix}\nonumber\\
    &&\,\times\,\left[\mathbb{C}^{-1,u_1u_1'}_{\ell_1}\mathbb{C}^{-1,u_2u_4'}_{\ell_1}\mathbb{C}^{-1,u_3u_3'}_{\ell_1}\mathbb{C}^{-1,u_4u_2'}_{\ell_4}\delta^{\rm K}_{b_1b_1'}\delta^{\rm K}_{b_2b_4'}\delta^{\rm K}_{b_3b_3'}\delta^{\rm K}_{b_4b_2'}+\text{7 perms.}\right].
\eeq

As for the bispectrum, $\mathcal{F}_4$ encodes correlations between different trispectra composed of different fields, though requires $\chi\,p_u = \chi'\,p_{u'}$, and, if the power spectra are parity-even, $p_u=p_{u'}$. The correlations between $B'\neq B$ imply that the Fisher matrix (and thus the covariance of the estimator) have a complex structure. This arises from a geometric degeneracy in the tetrahedron definition, since a tetrahedron on the two-sphere can be parametrized by either of two diagonals. This additionally makes the ideal Fisher matrix more work to implement, strictly requiring $\mathcal{O}(\ell_{\rm max}^6)$ operations. For this reason, the Monte Carlo procedure outlined in \S\ref{subsec: tl-norm} is usually preferred in practice.

\section{Validation}\label{sec: validation}
We now validate the power spectrum, bispectrum, and trispectrum algorithms derived above. All estimators are implemented in the publicly available \href{https://github.com/oliverphilcox/PolyBin}{\textsc{PolyBin}} package, \footnote{\href{https://github.com/oliverphilcox/PolyBin}{GitHub.com/OliverPhilcox/PolyBin}} the scalar version of which was described in \citep{Philcox:2023uwe}. The estimators make extensive use of the \textsc{HEALPix} \citep{Gorski:2004by} and \textsc{LIBSHARP} \citep{2013A&A...554A.112R} codes to manipulate data on the two-sphere and perform spin-weighted spherical harmonic transforms. Alongside the addition of tensor fields, the code has been extensively written since its genesis in \citep{Philcox:2023uwe}, optimizing for both CPU and memory efficiency, in part by careful avoidance of any unnecessary harmonic transforms (for example, if the mask is uniform). The scalar part of the new code has been explicitly tested against the old, to ensure calculations are consistent to machine precision, where appropriate.\footnote{Note that we use a different weighting convention for bispectra in this work to allow for parity-odd contributions, thus the bispectrum outputs differ by the appropriately averaged ratio of the two weights.} In the code (and in the plots below), we take the imaginary parts of any parity-breaking bispectrum and trispectrum such that the outputs are explicitly real.

\subsection{Set-Up}
We first present an overview of our testing methodology. To validate the estimators, we use simulated Gaussian realizations of the $T$, $Q$ and $U$ polarized CMB maps via \textsc{HEALPix}'s \texttt{synalm} routine \citep{Gorski:2004by}, with input lensed power spectra derived from the \textsc{class} code \citep{Blas:2011rf}, at the \textit{Planck} 2018 cosmology \citep{2020A&A...641A...6P}, and noise via the standard prescription of \citep{Hu:2001kj}. We additionally include a parity-breaking signal via the correlation coefficients $r_{\rm TB}=r_{\rm EB} = 0.5$; this facilitates testing of the parity-odd power spectrum sector and covariance matrices. All fields are generated at the desired testing resolution, here set to $N_{\rm side}=256$ for fast computation, thus we do not require pixel window functions, and we additionally ignore any harmonic space beam. For the window, we utilize the \textit{Planck} \texttt{GAL040} mask with $2\degree$ apodization \citep{Planck:2018yye}, which masks all but $\approx 34\%$ of the sky, practically removing the majority of power at $\ell\lesssim 5$.\footnote{As discussed in \citep{PhilcoxCMB}, an alternative approach is to include the mask only in the weighting function $\Si$, setting the window, $W$, to unity elsewhere in the algorithm. This may be preferable for complex masks containing an abundance of holes, and leads to somewhat expedited computation.} When Monte Carlo simulations are required, these are generated analogously to the mock data, with $N_{\rm it}=100$ iterations used by default. Whilst our testing suite is principally focused on CMB applications, we stress that it applies also to any other spin-$0$/spin-$2$ fields on the sphere, such as cosmic shear and its correlation with galaxy overdensity.

An important ingredient of the estimators is the weighting function $\Si$. As noted in \S\ref{sec: estimators}, the optimal choice is $\Si = \mathsf{C}^{-1}$, \textit{i.e.}\ the inverse pixel covariance. Here, we shall ignore the impact of the mask (noting that this is unity in the unmasked regions, ignoring apodization), and fix $\Si$ to be the diagonal-in-$\ell$ inverse harmonic-space covariance $\mathbb{C}^{-1}_\ell$, including the cross-covariances between all fields. As we see below, this is approximately optimal in practice for $\ell\gtrsim 5$. 

When testing the three-point estimators, it is useful to generate simulations with a known bispectrum, rather than restricting to GRFs. For this purpose, we generate random maps, $a_{\ell m}$, via a procedure similar to \citep{2011MNRAS.417....2S,Shiraishi:2014roa}, with the $T/E/B$-space redefinition
\beq
    a_{\ell m}^X \to a_{\ell m}^X + \sum_{\ell_2\ell_3m_2m_3YZ}\frac{1}{6}w^{\ell\ell_2\ell_3}_{mm_2m_3}b^{XYZ}_{\ell\ell_2\ell_3}h^{Y*}_{\ell_2m_2}h^{Z*}_{\ell_3m_3}.
\eeq
Here, $w^{\ell_1\ell_2\ell_3}_{m_1m_2m_3}$ is the bispectrum weighting function given in \S\ref{sec: defs}, $h^{X}_{\ell m} \equiv [\mathbb{C}_\ell^{-1}a]^X_{\ell m}$ is the inverse-covariance-weighted field and $b^{XYZ}_{\ell_1\ell_2\ell_3}$ is the desired bispectrum. At first order in $b^{XYZ}$, this satisfies
\beq
    \av{a_{\ell_1m_1}^Xa_{\ell_2m_2}^Ya_{\ell_3m_3}^Z} = w^{\ell_1\ell_2\ell_3}_{m_1m_2m_3}b^{XYZ}_{\ell_1\ell_2\ell_3},
\eeq
as desired, though we note that higher-order corrections may be important if $b^{XYZ}$ is large; furthermore, the power spectrum is distorted by an amount proportional to $\left(b^{XYZ}\right)^2$. If one assumes a separable bispectrum, $b^{XYZ}_{\ell_1\ell_2\ell_3} = \beta_{\ell_1}^X\beta_{\ell_2}^Y\beta_{\ell_3}^Z$, this can be practically implemented as a real-space integral:
\beq
    a_{\ell m}^X \to a_{\ell m}^X +\frac{1}{18}\beta^{X}_{\ell}\int d\hn\,\left[{}_{-1}\mathcal{O}(\hn){}_{-1}\mathcal{O}(\hn){}_{-2}Y^*_{\ell m}(\hn)-2{}_{-1}\mathcal{O}(\hn){}_{+2}\mathcal{O}(\hn){}_{+1}Y^*_{\ell m}(\hn)\right],
\eeq
defining
\beq
    {}_{s}\mathcal{O}(\hn) = \sum_{\ell m X}\beta^X_\ell h^X_{\ell m}\,{}_{s}Y_{\ell m}(\hn),
\eeq
making use of the aforementioned conjugate properties. As before, this can be evaluated via spin-weighted spherical harmonic transforms. We may additionally insert a factor of $[1\pm p_u(-1)^{\ell_1+\ell_2+\ell_3}]/2$ in order to inject input spectra of only a single parity.

In general, our estimators return measurements of the binned polyspectra, which, for comparison to theory, should be connected to the unbinned models.\footnote{\resub{A simple approximation is to evaluate the relevant theory polyspectrum in the bin center; this will often suffice in practice if the bins are relatively thin.}} For completeness, we list these relations below, derived by taking expectations of the ideal estimator and inserting the unbinned polyspectrum definitions given in \S\ref{sec: defs}. \resub{Firstly, the power spectrum becomes:}
\beq\label{eq: ideal-Cl-theory}
    \left[\mathcal{F}_2^{-1}C\right]^{{\rm th},u}(b) &=& \frac{1}{\Delta_2^u}\sum_{\ell}\Theta_\ell(b)(2\ell+1)\sum_{u'}S^{-1,u_1u_1'}_\ell S^{-1,u_2u_2'}_{\ell}C_\ell^{{\rm th},u'}\\\nonumber
    \mathcal{F}_2^{{\rm th},uu'}(b) &=& \frac{1}{\Delta_2^{u}\Delta_2^{u'}}\sum_\ell\Theta_\ell(b)(2\ell+1)\left[S^{-1,u_1u_1'}_\ell S^{-1,u_2u_2'}_{\ell}+S^{-1,u_1u_2'}_\ell S^{-1,u_2u_1'}_{\ell}\right],
\eeq
\resub{with $C^{\rm th,u}(b)\to C_\ell^{\rm th,u}$ in the narrow-bin limit. Notably, the sum over $u'$ in the first line is over \textit{all} pairs of fields, not just those in \eqref{eq: Cl-pairs}, since we have not inserted the binned definition; this additionally leads to the permutations in the second line of \eqref{eq: ideal-Cl-theory}. For the bispectrum, we find}
\beq
    \left[\mathcal{F}_3^{-1}b\right]^{{\rm th},u}_\chi(\vec b) &=& 
    \frac{1}{9}\frac{\chi\,p_u}{\Delta_3^u(\vec b)\Delta_3^{u'}(\vec b')}\sum_{\ell_1\ell_2\ell_3}\Theta_{\ell_1}(b_1)\Theta_{\ell_2}(b_2)\Theta_{\ell_3}(b_3)\frac{(2\ell_1+1)(2\ell_2+1)(2\ell_3+1)}{4\pi}\left[\tj{\ell_1}{\ell_2}{\ell_3}{-1}{-1}{2}+\text{2 cyc.}\right]^2\nonumber\\
    &&\qquad\qquad\,\times\,\sum_{u'}S^{-1,u_1u_1'}_{\ell_1}S^{-1,u_2u_2'}_{\ell_2}S^{-1,u_3u_3'}_{\ell_3}b_{\ell_1\ell_2\ell_3}^{{\rm th},u'}\nonumber\\
    \mathcal{F}_{3,\chi\chi'}^{{\rm th},uu'}(\vec b,\vec b') &=& \frac{1}{9}\frac{\chi\,p_u\,\delta_{\rm K}^{\chi\,p_u,\chi'\,p_{u'}}}{\Delta_3^u(\vec b)\Delta_3^{u'}(\vec b')}\sum_{\ell_1\ell_2\ell_3}\Theta_{\ell_1}(b_1')\Theta_{\ell_2}(b_2')\Theta_{\ell_3}(b_3')\frac{(2\ell_1+1)(2\ell_2+1)(2\ell_3+1)}{4\pi}\left[\tj{\ell_1}{\ell_2}{\ell_3}{-1}{-1}{2}+\text{2 cyc.}\right]^2\nonumber\\    &&\qquad\qquad\,\times\,\left[S^{-1,u_1u_1'}_{\ell_1}S^{-1,u_2u_2'}_{\ell_2}S^{-1,u_3u_3'}_{\ell_3}\delta^{\rm K}_{b_1b_1'}\delta^{\rm K}_{b_2b_2'}\delta^{\rm K}_{b_3b_3'}+\text{5 perms.}\right],
\eeq
\resub{where we restrict to triplets with $(-1)^{\ell_1+\ell_2+\ell_3}=\chi\,p_u$, and the $u'$ summation is taken over all triplets of fields. As before, this has the thin-bin limit $b^{{\rm th},u}\to b^{{\rm th},u}_{\ell_1\ell_2\ell_3}$. Finally, the binned trispectrum takes the form (cf.\,\citep{Philcox:2023uwe,PhilcoxCMB} for scalars)}
\beq
     \left[\F_4\widehat{t}\right]^{{\rm th},u}_\chi(\vec b,B) &=& 8\frac{\chi\,p_u}{\Delta_4^u(\vec b)}\sum_{\ell_iL}\Theta_{\ell_1}(b_1)\cdots\Theta_{\ell_4}(b_4)\Theta_L(B)\frac{(2\ell_1+1)\cdots(2\ell_4+1)(2L+1)}{(4\pi)^2}\\\nonumber
     &&\,\times\,\tj{\ell_1}{\ell_2}{L}{-1}{-1}{2}^2\tj{\ell_3}{\ell_4}{L}{-1}{-1}{2}^2\sum_{u'}S^{-1,u_1u_1'}_{\ell_1}\cdots S^{-1,u_4u_4'}_{\ell_4}t_{\ell_1\ell_2,\ell_3\ell_4}^{{\rm th},u'}(L)\\\nonumber
     \mathcal{F}_{4,\chi\chi'}^{{\rm th},uu'}(\vec b,B,\vec b',B') &=&  
     \delta^{\rm K}_{BB'}\frac{\chi\,p_u\,\delta^{\rm K}_{\chi\, p_u,\chi'\,p_{u'}}}{\Delta_4^u(\vec b)\Delta_4^{u'}(\vec b')}\sum_{\ell_iL}\Theta_{\ell_1}(b_1)\cdots\Theta_{\ell_4}(b_4)\Theta_L(B)\frac{(2\ell_1+1)\cdots(2\ell_4+1)(2L+1)}{(4\pi)^2}\\\nonumber
     &&\,\times\,\tj{\ell_1}{\ell_2}{L}{-1}{-1}{2}^2\tj{\ell_3}{\ell_4}{L}{-1}{-1}{2}^2\left[S^{-1,u_1u_1'}_{\ell_1}\cdots S^{-1,u_4u_4'}_{\ell_4}\delta^{\rm K}_{b_1b_1'}\cdots \delta^{\rm K}_{b_4b_4'}+\text{7 perms.}\right],
\eeq
\resub{summing over all quadruplets of fields $u'$, with the additional parity restriction that $(-1)^{\ell_1+\ell_2+\ell_3+\ell_4}=\chi\,p_u$.} Here, we have assumed that the off-diagonal terms in the Fisher matrix (which mix $L$ and $L'$) are subdominant \S\ref{subsec: tl-limits}). As noted in \citep{Hivon:2001jp}, when applied to real data, these relations are not quite exact, since the true expectation of the estimator includes the variation of the mask across the bins in the numerator and denominator. For higher-order statistics, such corrections are small and can generally be ignored.

\subsection{Power Spectrum}
To test the two-point estimators, we compute the binned power spectrum multipoles over an array of $40$ regularly-spaced bins between $\ell=2$ and $\ell=402$, for all six non-trivial combinations of $T$, $E$, and $B$, giving a total of $N_{\rm bin} = 240$ components. This is performed with and without the \textit{Planck} sky mask; in each case, we compute both the Fisher matrix, $\mathcal{F}_2$, and the estimator numerators, via the unwindowed prescription and the ideal limits.

\begin{table}[]
    \centering
    \begin{tabular}{l||c|c|c|c}
    \textbf{Mask} & \textbf{Ideal Fisher} & \textbf{Optimal Fisher} & \textbf{Ideal Num.} & \textbf{Optimal Num.}\\ 
    Uniform  & 19s & 18s $\times\,100$ & 1.6s & 1.6s\\
    Galactic & 19s & 45s $\times\,100$ & 1.6s & 1.6s
    \end{tabular}
    \caption{Computation times for ideal and optimal power spectrum estimators, for $N_{\rm side}=256$ and $40$ bins with $\ell\in[2,402)$. We show results for both the estimator numerators and Fisher matrices, with the latter computed from $N_{\rm it}=100$ Monte Carlo simulations. Timings are given for both a uniform mask and a \textit{Planck} sky mask; each example requires two harmonic transforms to define the numerator (one for spin $0$ and one for spin $2$), with the latter requiring an additional $\approx 2000$ transforms for each realization of the optimal Fisher matrix. This occurs since each of the $240$ elements of the map-space $Q_2$ derivative is estimated via a reverse harmonic transform (for both spins), then the $\Si$ weighting is applied, which requires a further transformation.}
    \label{tab: Cl}
\end{table}

In Tab.\,\ref{tab: Cl}, we list the runtime of the optimal and ideal estimators. Both with and without the mask function, this is dominated by computation of the Fisher matrix, which must be estimated for each pair of bins and fields from a suite of $N_{\rm it}$ GRF realizations. In the presence of a non-trivial window, this requires harmonic transforms for each element, giving a total of $2\times 3\times N_{\rm bin}\times N_{\rm it}$ (counting number of spins, number of transforms, number of bins, and number of simulations). Whilst this may be somewhat expensive for high-resolution maps, we stress that it need be computed only once for a given survey geometry and fiducial cosmology, \textit{i.e.}\ it is independent of the data. For a uniform mask, no harmonic transforms are needed since all computations can be performed in $T/E/B$ harmonic-space. Computation of the estimator numerators is straightforward, requiring only pair of transforms to define the filtered data, and subsequent harmonic-space summations in each bin. For high-resolution maps at a given $N_{\rm side}$, computation of this part is essentially independent of $N_{\rm bin}$ (unlike for the higher-point functions).

\begin{figure}
    \centering
    \includegraphics[width=\textwidth]{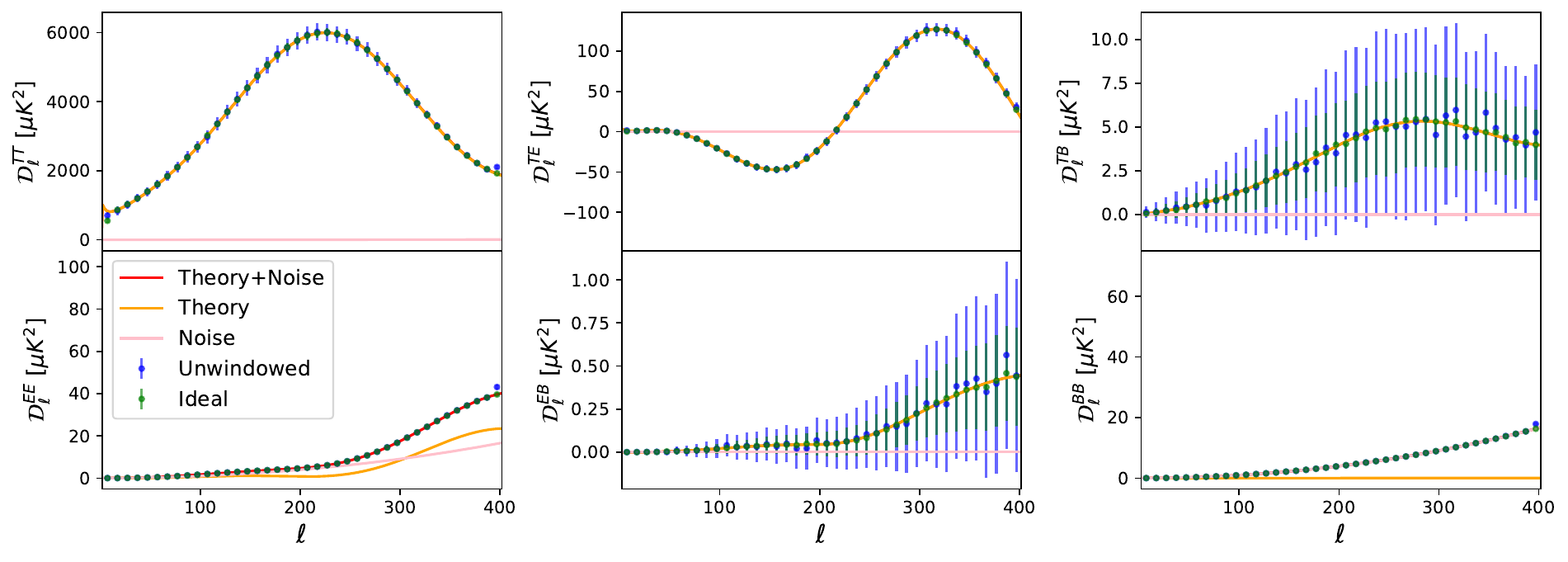}
    \includegraphics[width=\textwidth]{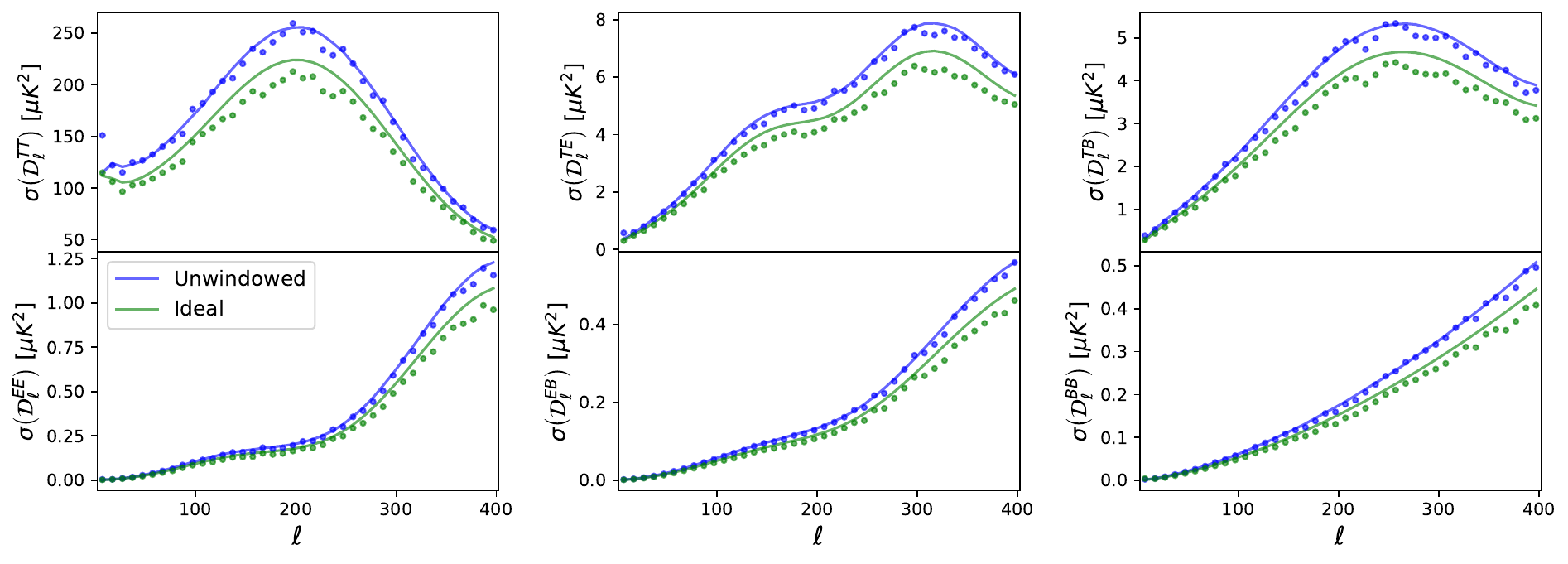}
    \caption{Measured and theoretical power spectra from a suite of simulations, obtained using the unwindowed and idealized estimators of this work implemented in \textsc{PolyBin}. Simulations are constructed using a \textit{Planck} cosmology and noise model and a galactic sky mask, but include additional parity-odd correlations for testing. We show measurements from $1000$ GRF simulations, with Fisher matrices (needed for the unwindowed/optimal estimators) constructed from $100$ Monte Carlo realizations. The top panels show the measured spectra using the two estimators alongside the theoretical expectations, rescaled by $\ell(\ell+1)/2\pi$. In the bottom panel, we plot the estimator errors (points) and the diagonal of the (unwindowed or ideal) Fisher matrix. For the ideal estimators, we rescale the means and variances by factors of $\av{W^2}/\av{W}^2$ and $\av{W^4}/\av{W^2}^2$ respectively, to account for the mask effects; this holds only approximately, thus the ideal variances (green) may not precisely match their predictions.  If the estimator is unbiased and optimal, the mean of the unwindowed estimator mean will match theory, and the variance will match the inverse Fisher matrix; this is demonstrated in the above.}
    \label{fig: Cl-data}
\end{figure}

Fig.\,\ref{fig: Cl-data} displays the output binned power spectrum measurements from the \textsc{PolyBin} code against theoretical expectations, all computed using a galactic mask. For each of the six spectra considered, we find excellent agreement between empirical and predicted spectra, both for the unwindowed estimators and their ideal equivalents. The last bin of the unwindowed estimators is somewhat biased; this is as expected, since we do not correctly account for mask-induced correlations between it and the next bin. In practice, one would always compute the spectrum across a slightly wider range of $\ell$-bins than used in the final analysis, allowing such effects to be nulled.

Turning to the variances, the unwindowed results show excellent agreement between the empirical variances and the inverse Fisher matrix predictions. This indicates that our estimators are (approximately) minimum-variance, and thus optimal in the Gaussian regime. The idealized estimators show fair agreement with their idealized Fisher matrix predictions, though we note that this prediction does not fully account for the spatial dependence of the window, and is thus only approximate. As in \citep{Philcox:2023uwe}, the unwindowed estimators appear to have higher variances than their idealized equivalents: this is due to an anticorrelation between neighboring bins, and does not lead to reduced signal-to-noise (\textit{i.e.}\ it does not imply suboptimality).

\begin{figure}
    \centering
    \includegraphics[width=0.5\textwidth]{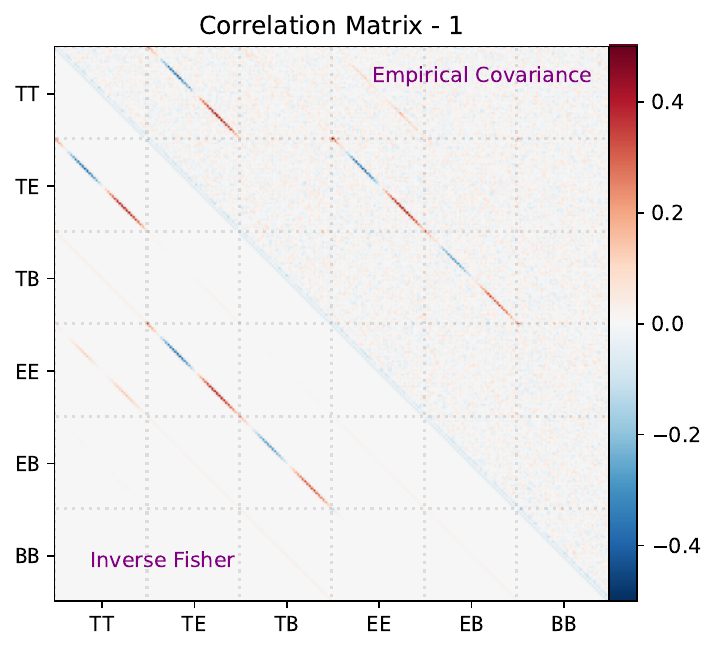}
    \caption{Correlation matrices corresponding to the unwindowed power spectrum measurements shown in Fig.\,\ref{fig: Cl-data}. The lower half matrix shows the theoretical correlation obtained from the inverse Fisher matrix, whilst the upper half shows that extracted from a suite of $1000$ mocks. Subboxes show correlations between individual spectra, with $\ell$-bins stacked from lowest (top left) to highest (bottom-right). We find excellent agreement between theory and data, and observe off-diagonal correlations from both couplings between different polarized components and the survey mask.}
    \label{fig: Cl-cov}
\end{figure}

In Fig.\,\ref{fig: Cl-cov}, we show the correlation structure of the optimal estimator, both theoretically (from the inverse Fisher matrix) and empirically (from the simulation suite). As before, we find excellent agreement, with no discernible difference seen between the two matrices (except for noise). We observe strong off-diagonal signals induced by physical $TE$, $TB$, and $EB$ correlations (with only the first expected from conventional physics), whilst the effects of masking lead to a leakage of power between adjacent bins, as evidenced by the negative elements alongside the diagonal. The similarity of both variances and correlations is an important verfication of our estimator and its optimality. Finally, we note that our results are stable to the choice of hyperparameters; reducing to just $N_{\rm it}=10$ Monte Carlo iterations used to estimate the Fisher matrix yields an error of $(0.0\pm 0.3)\sigma$ across all bins, implying that our choice of $N_{\rm it}=100$ is conservative.


\subsection{Bispectrum}
As for the power spectrum, we validate the bispectrum estimators by ensuring that (a) the mean of the estimators matches the theoretical expectation, (b) the variance is close to the inverse Fisher matrix, \textit{i.e.}\ the optimal limit. For this purpose, we analyze simulations both with and without an injected three-point function, noting that the estimators are strictly minimum-variance only in the Gaussian limit. For this test, we compute the bispectrum in $N_\ell=9$ $\ell$-bins equally spaced in $\ell^{3/2}$ (chosen to evenly distribute any primordial signal-to-noise \citep{Kalaja:2020mkq}), with $\ell\in[2,440)$.\footnote{The exact bin edges are fixed to $\{2,   5,  30,  68, 114, 167, 227, 293, 364, 440\}$.} We utilize the same binning for both squeezed and non-squeezed triangles, though note that this is not required by \textsc{PolyBin}; alternative choices may be preferred if one is testing a model that peaks in squeezed configurations (such as local-type primordial non-Gaussianity). In practice, we drop the first bin after the full bispectrum is computed (thus accounting for bin leakage), since this is heavily contaminated by the window function and generally non-Gaussian; a similar procedure was used for the trispectrum in \citep{PhilcoxCMB}. Gaussian simulations are generated as before, with non-Gaussianity incorporated for even $\ell_1+\ell_2+\ell_3$ via the dimensionless factorized bispectrum $b^{XYZ}_{\ell_1\ell_2\ell_3}=\beta_{\ell_1}^X\beta_{\ell_2}^Y\beta_{\ell_3}^Z$, with $\beta_\ell^T = (2\times 10^{-6})e^{-(\ell-2)/40}$, $\beta_{\ell}^E = \sqrt{C_{\ell=80}^{\rm EE}/C_{\ell=80}^{\rm TT}}\,\beta_\ell^T$, and $\beta_\ell^B=0$, set by initial testing to ensure a signal of appropriate strength. We compute bispectra for all 20 combinations of fields and parities (cf.\,\S\ref{sec: bl}), leading to a total of $N_{\rm bin}=2140$ bins (after dropping modes with $\ell<5$).

Before presenting the bispectrum results, it is useful to understand their scalings by estimating the number of spherical harmonic transforms involved in each part of the estimator. The numerator requires $2\times 3\times N_\ell$ transforms to define the ${}_{\pm 1}H$ and ${}_{\pm 2}H$ fields (one per spin, one per field, and one per $\ell$-bin), then the three-field term is evaluated using a real-space summation. When the one-field term is included, this is enhanced by a factor of $N_{\rm it}$, yielding a total of $\approx 6N_{\rm it}N_\ell$ transforms for the numerator. For the Fisher matrix, each realization requires $N_{\rm bin}$ harmonic transforms to form the $Q_3$ arrays, which are then combined in harmonic- or real-space, with the latter requiring an additional set of transforms to apply the $\Si$ weighting and the mask. As such, the total Fisher matrix scaling is $\mathcal{O}(N_{\rm it}N_{\rm bin})$, whilst the scaling of the numerator is $\mathcal{O}(N_{\rm it}N_\ell)$.\footnote{Note that the total number of bispectrum bins, $N_{\rm bin}$, scales as $N_\ell^3$ if one considers all possible triangle configurations.} Finally, we note that the computation of the Fisher matrix may be practically limited by computational memory limits rather than runtime. Each Monte Carlo iteration requires $Q_3$ to be computed in each bin, for two random fields, and for two weightings ($\Si$ and $\mathsf{A}^{-1}$). After this, the arrays are combined as an inner product, requiring $4N_{\rm bin}$ maps to be held in memory. For $N_{\rm side}=256$ and the above binning configuration, this requires $\approx 120\,\mathrm{GB}$ of memory, which is well within the reach of most clusters.\footnote{Note that the algorithm is constructed so as to reduce memory requirements for all steps apart from the (irreducible) $Q_3$ maps, \textit{i.e.}\ it minimizes storage and recomputation of intermediate products.} For higher-resolution studies, one may wish to use wider bins to reduce the memory consumption, or, in the most restrictive limit, compute the inner product element-wise, leading to an algorithm with quadratic complexity rather than linear.

\begin{table}[]
    \centering
    \begin{tabular}{l||c|c|c|c|c}
    \textbf{Mask} & \textbf{Ideal Fisher} & \textbf{Optimal Fisher} & \textbf{Ideal Num.} & \textbf{Optimal Num. [3-field]} & \textbf{Optimal Num. [full]} \\ 
    Uniform & 10h & 770s $\times$ 100 & 37s & 37s & 230m\\
    Galactic & 10h & 1280s $\times$ 100 & 36s & 36s & 158m
    \end{tabular}
    \caption{Runtimes of the ideal and unwindowed bispectrum estimators, applied to every combination of fields and parities across $9$ linear bins with $\ell\in[2,440)$, giving $2448$ bins (some of which are later dropped). The ideal estimator requires six harmonic transforms for each linear bin (for the two spins, and three $T/E/B$ maps), with the remaining computation involving only real-space summation; however, the normalization requires an $\mathcal{O}(\ell_{\rm max}^3)$ sum, which takes an additional $\approx 10$ hours to compute on a 40-core machine. The same scaling holds for the three-field term of the optimal estimator; however, the one-field term requires the above transformations for each of $N_{\rm it}=100$ Monte Carlo realizations (including permutations of data and fields), leading to the significantly larger computation time. For the optimal Fisher matrix, we require around $4\,600$ harmonic transforms per realization for a uniform mask (to define the $Q_3$ derivatives); this increases to around $54\,000$ for a non-trivial mask, since one must perform forward and reverse transforms to apply $W$ to $Q_3$, for both its spin-$0$ and spin-$2$ components, and for two possible weightings.}
    \label{tab: Bl}
\end{table}

Computation times for the bispectrum estimator are shown in Tab.\,\ref{tab: Bl}, when applied to both a \textit{Planck} galactic mask and a trivial uniform mask. With our choice of binning, each Monte Carlo realization of the Fisher matrices require $\approx 10-20$ minutes to compute, with the windowed computation taking longer due to the additional harmonic transforms required. Due to our low $N_{\rm side}$, the transforms are not strongly rate-limiting; this is evidenced by the non-trivial mask computations being only a factor of two slower than the uniform case, despite involving $\approx 10\times$ more harmonic transforms. In this example, computing the Fisher matrix via Monte Carlo summation is actually more efficient than direct computation of the idealized form. The latter involves an $\mathcal{O}(\ell_{\rm max}^3)$ summation, which took $\approx 10$ hours on a $40$-core machine. The estimator numerators are fast to compute: the three-field term takes under a minute per simulation (for all $\sim 2500$ bins), whilst the one-field correction multiplies this by a factor around $N_{\rm it}=100$. The latter computation can be somewhat expedited by pre-processing the Monte Carlo simulations and holding them in memory (which itself required $\approx 4$ minutes); if memory is a limitation, this can be dropped, such that each simulation is analyzed only as necessary. As discussed above, the action of the one-field term is to reduce the variance only on ultra-large scales and does not affect the mean; for studies focussing on smaller scales, this contribution can be dropped for significant computational gain. As such, the fastest bispectrum estimator is the optimal form (with a Monte Carlo Fisher matrix), but without the one-field term. 

\begin{figure}
    \centering
    \includegraphics[width=\textwidth]{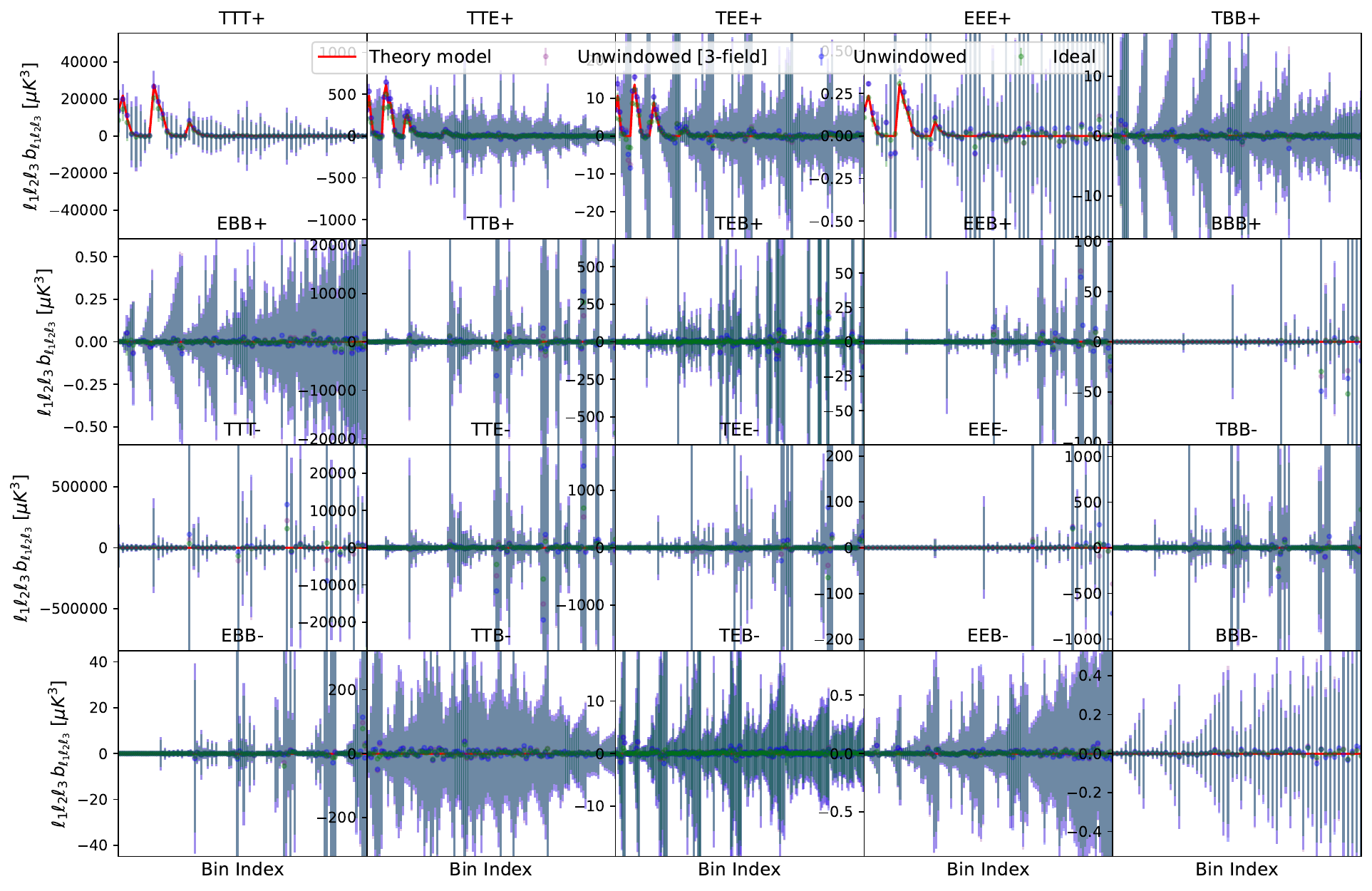}
    \caption{Measured and theoretical bispectra from a suite of masked simulations, measured using the unwindowed and idealized estimators of this work. This figure is analogous to Fig.\,\ref{fig: Cl-data}, but includes all ten non-trivial bispectrum field configurations and both parities (indicated by the titles, with $\pm$ indicating $\chi=\pm 1$). In each subpanel, the bins are stacked in one dimension and increase in magnitude from the left to the right of the figure. We show the mean bispectra measured from $1000$ simulations from both types of estimators, and compare this to the injected parity-even signal (red curve). We find that the theory is well recovered in all relevant cases, with null results seen in correlators for which is does not contribute. The errors strongly depend on the bin configuration, with particularly large values seen if $p_u\chi=-1$ and all three bins are equal (since this term vanishes in the thin-bin limit).}
    \label{fig: Bl-data}
\end{figure}

\begin{figure}
    \centering
    \includegraphics[width=\textwidth]{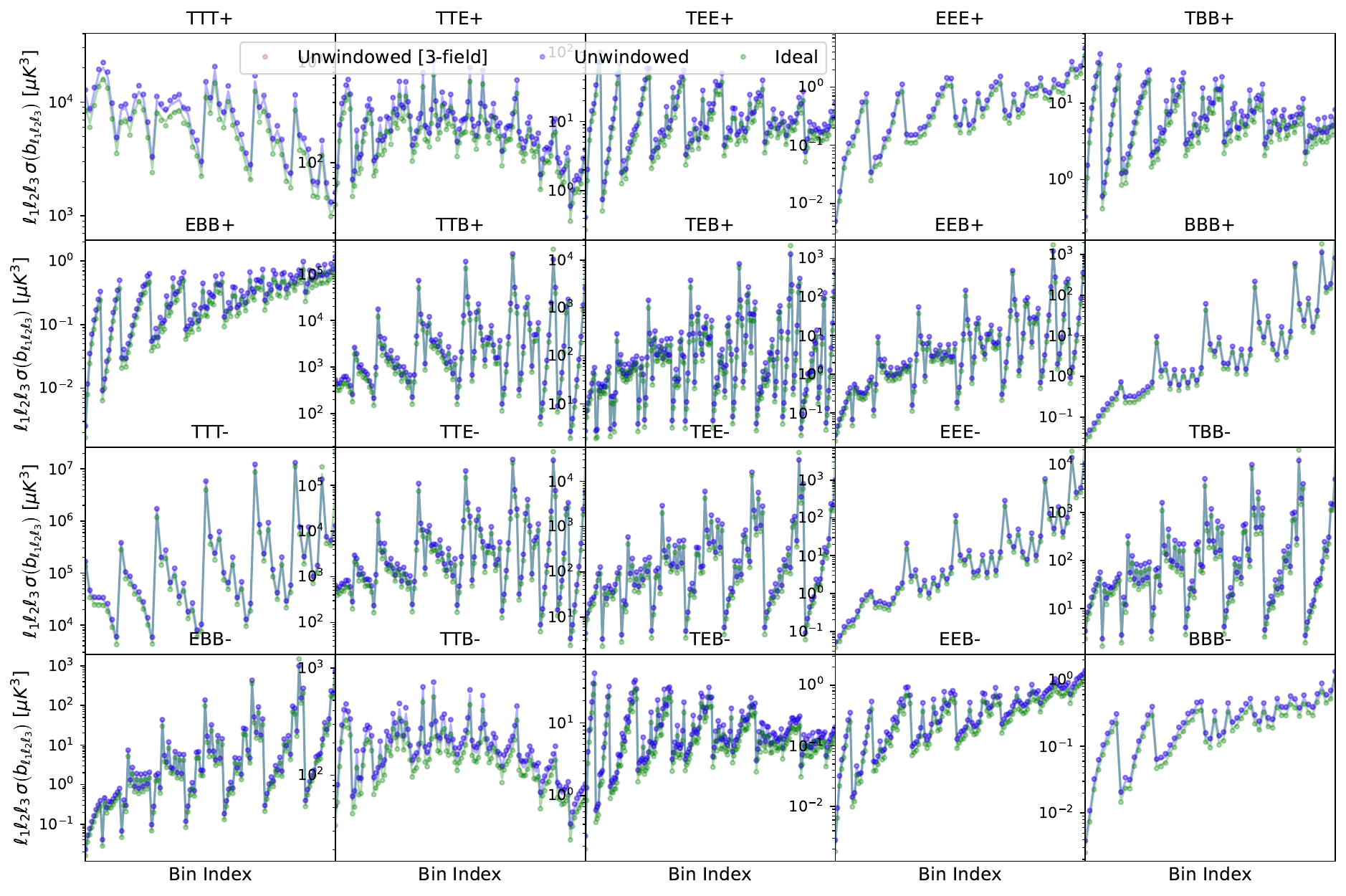}
    \caption{As Fig.\,\ref{fig: Bl-data}, but displaying the empirical bispectrum variances calculated from GRF simulations; these are in good agreement with the expected variances across all correlators, implying that our estimators are close to optimal. We find that removal of the one-field term in the estimators (purple) significantly enhances the variances on large scales.}
    \label{fig: Bl-var}
\end{figure}

Fig.\,\ref{fig: Bl-data} shows the bispectra measured with our pipeline, averaged over $1000$ realizations. We find excellent agreement between theory and data for both types of estimator across all fields and parities. This demonstrates that our algorithm can correctly recover an injected bispectrum, \textit{i.e.}\ it is unbiased. The various data-points show a large dynamic range; this is due to the large range of $\ell$ ($5\leq\ell<440$), and the differences in amplitude between $T$, $E$, and $B$. We note that some $\chi\,p_u=-1$ data-points have very large scatter; this corresponds to bins with $b_1=b_2=b_3$, which have no contribution from odd $\ell_1+\ell_2+\ell_3$ in the thin-bin limit. 

In Fig.\,\ref{fig: Bl-var}, we display the empirical bispectrum variances from GRF simulations (always including a Galactic mask). We find excellent agreement between the empirical unwindowed variances and the inverse Fisher predictions for all correlators across a huge range of amplitudes; this implies that our estimator is close to minimum-variance, as expected.\footnote{Repeating the exercise with a uniform mask finds exquisite agreement in all bins, as expected.} As for the power spectrum, the ideal estimator shows slightly reduced variances compared to the optimal form; this is again sourced by an anticorrelation between neighboring bins. These are relatively well modeled by the ideal Fisher matrix (multiplied by $\av{W^6}/\av{W^3}^2$ for mask $W$), though exact agreement is not guaranteed in this case. If one pushes to larger scales than plotted ($\ell<5$), both estimators become suboptimal; this occurs since the bispectrum becomes mask dominated, thus our diagonal-in-$\ell$ $\Si$ weighting is far from the true inverse pixel covariance. In this regime, a more nuanced weighting, for instance invoking conjugate-gradient-descent inversion, may be of use. Finally, we note that on these range of scales, the one-field term has little effect; for more complex masks, it may serve to reduce the large-scale variance, particularly when coupled with an accurate estimate of $\Si$.\footnote{The utility of the one-field term depends also on the contributions from various scales within each bin, and thus the Gaunt factor weights $w^{\ell_1\ell_2\ell_3}_{m_1m_2m_3}$. Whilst this factors out for thin bins with $\Delta \ell=1$, it may lead to some weighting schemes being preferred in practice (such as the $\tjo{\ell_1}{\ell_2}{\ell_3}$ weights used for parity-even-only analyses).}

\begin{figure}
    \centering
    \includegraphics[width=0.7\textwidth]{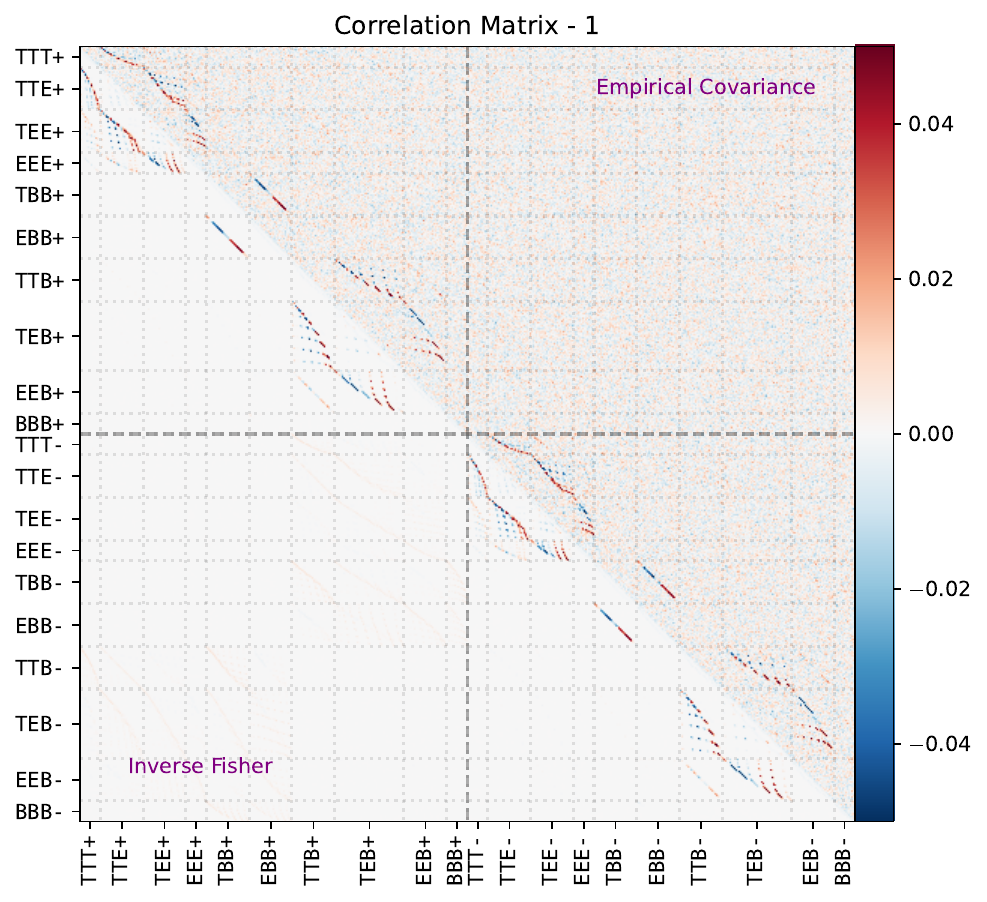}
    \caption{Correlation matrices corresponding to the unwindowed bispectrum measurements shown in Fig.\,\ref{fig: Bl-data}. As in Fig.\,\ref{fig: Cl-cov}, the lower half matrix shows the theoretical correlation obtained from the inverse Fisher matrix, whilst the upper half shows that extracted from a suite of $1000$ mocks. We demarcate components with $\chi=1$ and $\chi=-1$ by dashed lines, with individual subboxes showing the correlations between pairs of bispectra. The covariance matrix has a complex structure sourced by different polarized components and the survey mask, which is excellently reproduced by the theoretical prediction.}
    \label{fig: Bl-cov}
\end{figure}

Fig.\,\ref{fig: Bl-cov} shows the structure of the (inverted) optimal Fisher matrix, alongside the empirical correlations estimated from the $1000$ GRF simulations. We observe a complex geometric structure coupled across bins and fields; this arises from the power spectrum correlations between $T$, $E$, and $B$ (which break parity in our testing suite). That this is far from diagonal implies that computing the polarized binned bispectra is non-trivial; the results would be biased if one assumes a diagonal normalization term (which is often the case in unpolarized analyses \citep[e.g.,][]{Coulton:2019bnz}). If one zooms in sufficiently, a complex structure within each submatrix becomes visible; \resub{this is induced both by intrinsic correlations between fields and the non-uniform mask}. The latter effect contributes at the $\sim 5\%$ level (coupling adjacent bins), but is significantly amplified if the bin width is reduced. In all cases, the correlation structure is well-recovered from the suite of simulations again validating our estimator. Finally, we note that similar results can be obtained with significantly fewer realizations used to create the Fisher matrix; reducing from $N_{\rm it}=100$ to $N_{\rm it}=10$ changes the measured bispectra by at most $0.007\sigma$.

\subsection{Trispectrum}
Finally, we validate the trispectrum estimators. Our procedure is essentially identical to that for the lower-order statistics; we check both the mean and the variance of the estimator using suites of simulations with and without a Galactic mask. To check for bias, we consider the individual four-, two- and zero-field terms in the estimator explicitly rather than attempting to inject a trispectrum directly; since the full estimator is the sum of these components and the odd-parity estimator differs from the even only by a factor of $(-1)^{\ell_1+\ell_2+\ell_3+\ell_4}$, this is a sufficient test (and indeed, that used in \citep{Philcox:2023uwe}). To keep computation times reasonable, we restrict to $4$ $\ell$-bins evenly spaced in $\ell^{5/2}$, with $\ell_{\rm min}=2$ and $\ell_{\rm max}=483$,\footnote{These are defined by the bin edges $\{2, 5,  62, 212, 483\}$.} and use the same binning for all legs, though note that \textsc{PolyBin} can use different binning to highlight squeezed (large $\ell_2,\ell_4$) and flattened (small $L$) configurations. As before, we drop the first $\ell$-bin in the final analysis (after accounting for bin leakage), since it is mask-dominated, and our choice of $\Si$ weighting is far from optimal in this case. Here, we consider both parities, but now restrict to field configurations containing at most one $B$ mode. This emulates physical scenarios when one is interested in computing scalar physics from $T$ and $E$ modes, but wishes to limit $E$-to-$B$ mode leakage at leading order. In total, we use $12$ fields and $2$ parities, giving a total of $820$ bins in the output data-vector. Finally, we compute the ideal trispectra only from $100$ GRF simulations due to its large computational requirements (but use $1000$ for the unwindowed estimators). In all cases, we use $100$ Monte Carlo realizations to compute the disconnected contributions.

The computational scalings of the trispectrum are akin to those of the bispectrum. First, one computes the ${}_{\pm 1}H$ maps, requiring $\mathcal{O}(N_\ell)$ harmonic transforms, then obtains the $A(L,M)$ fields, which scales as $N_\ell^2$. The four-field term of the trispectrum numerator is computed as a harmonic-space summation, yielding an overall complexity of $\mathcal{O}(N_\ell^2)$. The two- and zero-field terms require computation of $H$ and $A$ for $N_{\rm it}/2$ pairs of simulations, thus the final scaling is $\mathcal{O}(N_{\rm it}N_\ell^2)$. This is slower than the bispectrum, but notably much reduced compared to the total number of trispectrum bins (which scales as $N_\ell^5$). For the Fisher matrix, $Q_4$ requires a harmonic transform for each $A$ field and $L$-bin (\textit{i.e.}\ a total of $\sim N_\ell^3$), then a harmonic transform to form $Q_{4,\ell m}^X$, giving a total of $\mathcal{O}(N_\ell^4)$ operations. In the presence of a non-trivial mask, these must be additionally transformed into real-space, yielding a total complexity of $\mathcal{O}(N_{\rm it}N_{\rm bin})$ from $N_{\rm it}$ realizations. As before, this requires memory proportional to $N_{\rm bin}$ (precisely $8N_{\rm bin}$ times the dimensionality of map, due to the various permutations and weightings required), which may be limiting in practice. Whilst the above demonstrates that trispectrum computation is expensive, it is not parametrically harder than the bispectrum, and, furthermore, computation of the full Fisher matrix is usually faster than the ideal equivalent, since the latter requires a summation over $\mathcal{O}(\ell_{\rm max}^6)$ individual multipoles (including correlations between $L$ bins).

\begin{table}[]
    \centering
    \begin{tabular}{l||c|c|c|c|c}
    \textbf{Mask} & \textbf{Ideal Fisher} & \textbf{Optimal Fisher} & \textbf{Ideal Num.} & \textbf{Optimal Num. [4-field]} & \textbf{Optimal Num.} \\ 
    Uniform & $>7$ days & 22m $\times$ 100 & 705m & 15s & 78m \\
    Galactic & $>7$ days & 36m $\times$ 100 & 708m & 14s & 92m\\
    \end{tabular}
    \caption{Computation times for ideal and unwindowed trispectrum estimators, as in Tabs.\,\ref{tab: Cl}\,\&\,\ref{tab: Bl} . Here, we use $4$ linear bins with $\ell\in[2,483)$ across $24$ combinations of fields and parities, giving $1996$ bins (of which we later remove the low-$\ell$ contributions). Here, the four-field (disconnected plus connected) term of the trispectrum estimator requires $\approx 100$ harmonic transforms to define the numerator (from each computation of $H$ and $A$ fields), with the estimation of the disconnected (\textit{i.e.}\. Gaussian) terms increasing this by a factor proportional to $N_{\rm it}=100$ (with a coefficient larger than one, due to the extra permutations present in these terms). The ideal estimator requires \textit{much} more time to compute than the optimal form; this is due to the internal $\mathcal{O}(\ell_{\rm max}^3)$ summations in the two- and zero-field terms, which make this estimator impractical in realistic settings. The Fisher matrix requires $65\,000$ harmonic transforms per realization (reducing to $13\,000$ transforms if the mask is uniform), due to computation of the $Q_4$ derivatives. In this instance, computation of the ideal Fisher matrix is not feasible (even at $\ell_{\rm max}=483$), due to the internal $\mathcal{O}(\ell_{\rm max}^6)$ summation over Wigner $6j$ symbols. This remains true even if one assumes an (inexact) diagonal approximation, which reduces the sum to $\mathcal{O}(\ell_{\rm max}^5)$. }
    \label{tab: Tl}
\end{table}

The trispectrum runtimes are enumerated in Tab.\,\ref{tab: Tl}. Our conclusions are similar to those of the power spectrum and bispectrum; the four-field term of the estimator is fast to compute, with a factor of $N_{\rm it}$ slow-down when we add the two- and zero-field terms. In contrast to the bispectrum, these terms \textit{must} be present in the estimator if $\chi\,p_u=1$; else, the trispectrum is dominated by the Gaussian power spectrum contributions. This additionally implies that $N_{\rm it}$ must be relatively large and the fiducial spectrum must be close to the truth, else residual Gaussian pieces will dominate the error (though this is second order in $C_\ell^{\rm sim}-C_\ell^{\rm true}$). To compute such terms with the idealized trispectrum estimator, an $\mathcal{O}(\ell_{\rm max}^3)$ summation over Wigner $3j$ symbols is required; in practice, this is computationally expensive (taking $10$ hours in this example), thus the ideal form is unlikely to be of practical use. For our choice of binning, the optimal trispectrum Fisher matrix requires only $\approx 2\times$ the computation time of the bispectrum result; this is well within the capabilities of most clusters. As before, this is usually limited by memory; at $N_{\rm side}=256$ with $1996$ bins (before removal of low-$\ell$ bins), a total of $210\,\mathrm{GB}$ of data products is stored in memory for efficient computation. On a high-memory node, computations requiring tenfold more memory would also be practical; beyond this, one must compute the outer products element-wise, leading to an algorithm with quadratic complexity, as for the bispectrum. Finally, we note that the ideal trispectrum Fisher matrix is impractical to compute, even for relatively low $\ell_{\rm max}$. With our choice of binning, neither the ideal Fisher matrix nor its diagonal limit could be computed within a few node-days; this further supports the windowed estimators, which avoid having to perform such a sum. 

\begin{figure}
    \centering
    \includegraphics[width=\textwidth]{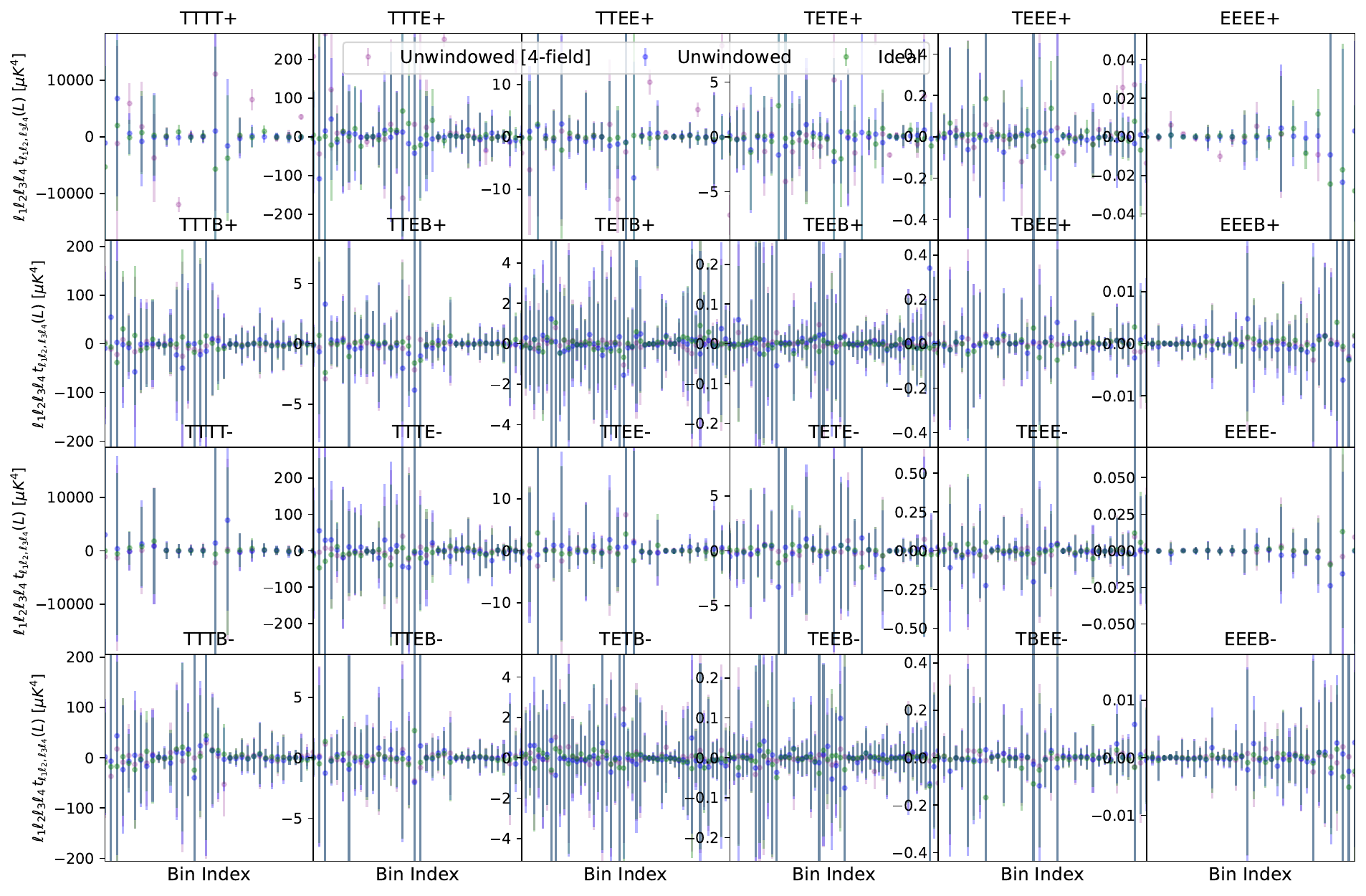}
    \caption{Comparison of measured and theoretical trispectra from a suite of masked simulations, across a wide range of bins, fields, and parities. Since no trispectrum is injected, the mean of the estimators should be zero; the null results in the figure show this holds true in practice. Strong detections of the disconnected trispectrum can be seen in the four-field term; further analysis of these contributions is shown in Fig.\,\ref{fig:  Tl-data-gauss}. Some parity-odd bins appear to have very large variances; these correspond to configurations that vanish in the exact thin-bin limit due to the restriction to odd $\ell_1+\ell_2+\ell_3+\ell_4$. These were excluded from previous work \citep{Philcox:2023uwe}, but included here for full generality.}
    \label{fig: Tl-data}
\end{figure}

\begin{figure}
    \centering
    \includegraphics[width=\textwidth]{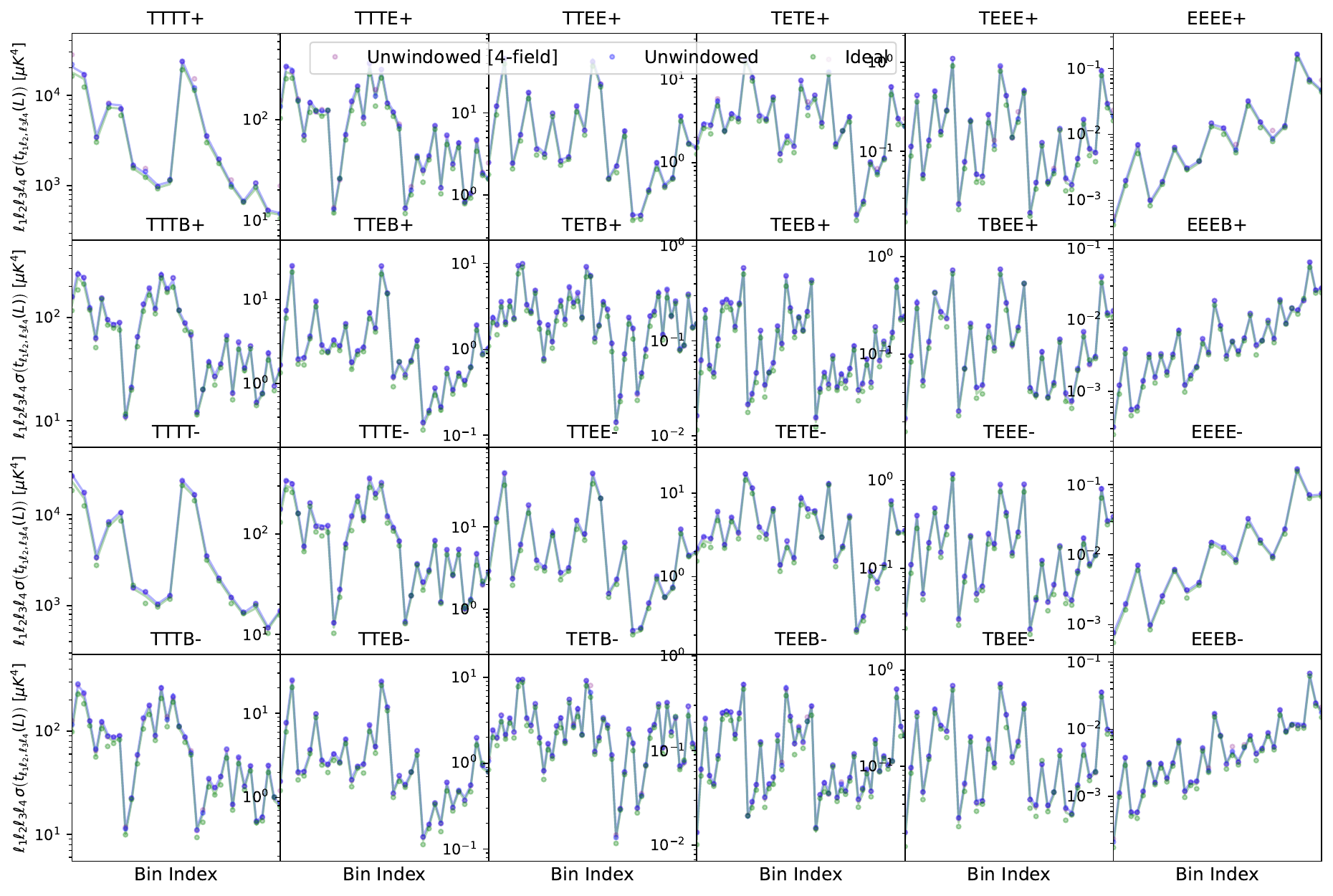}
    \caption{As Fig.\,\ref{fig: Tl-data}, but displaying the theoretical and empirical trispectrum variances. We find excellent agreement across the range of scales considered ($5\leq \ell<483$), implying that the unwindowed forms are approximately minimum variance, and thus close-to optimal.}
    \label{fig: Tl-var}
\end{figure}

Figs.\,\ref{fig: Tl-data}\,\&\,\ref{fig: Tl-var} display the trispectra estimated from our GRF simulations and their variances. In all cases, we find trispectra consistent with zero; this is as expected, and implies that the subtraction of the disconnected trispectrum components is proceeding correctly. 
Here, we find that $N_{\rm it}=100$ is sufficient to recover unbiased trispectra, noting that the errorbar on the two- and zero-field contributions scales as $N_{\rm it}^{-1/2}$. Turning to the variances, we again find that the unwindowed estimates are in excellent agreement with predictions from the inverted Fisher matrix, across all scales considered. As discussed above, the ideal Fisher matrices cannot be computed explicitly; here, we obtain them by applying the optimal pipeline without a window function (which is exact, in the limit of infinite Monte Carlo simulations). These are in fair agreement with the empirical variances from the ideal estimators (when weighting by $\av{W^8}/\av{W^4}^2$), though we again stress that this relation is only expected to be approximate. For a trivial mask, the ideal and optimal estimator are in excellent agreement, as expected.

\begin{figure}
    \centering
    \includegraphics[width=0.7\textwidth]{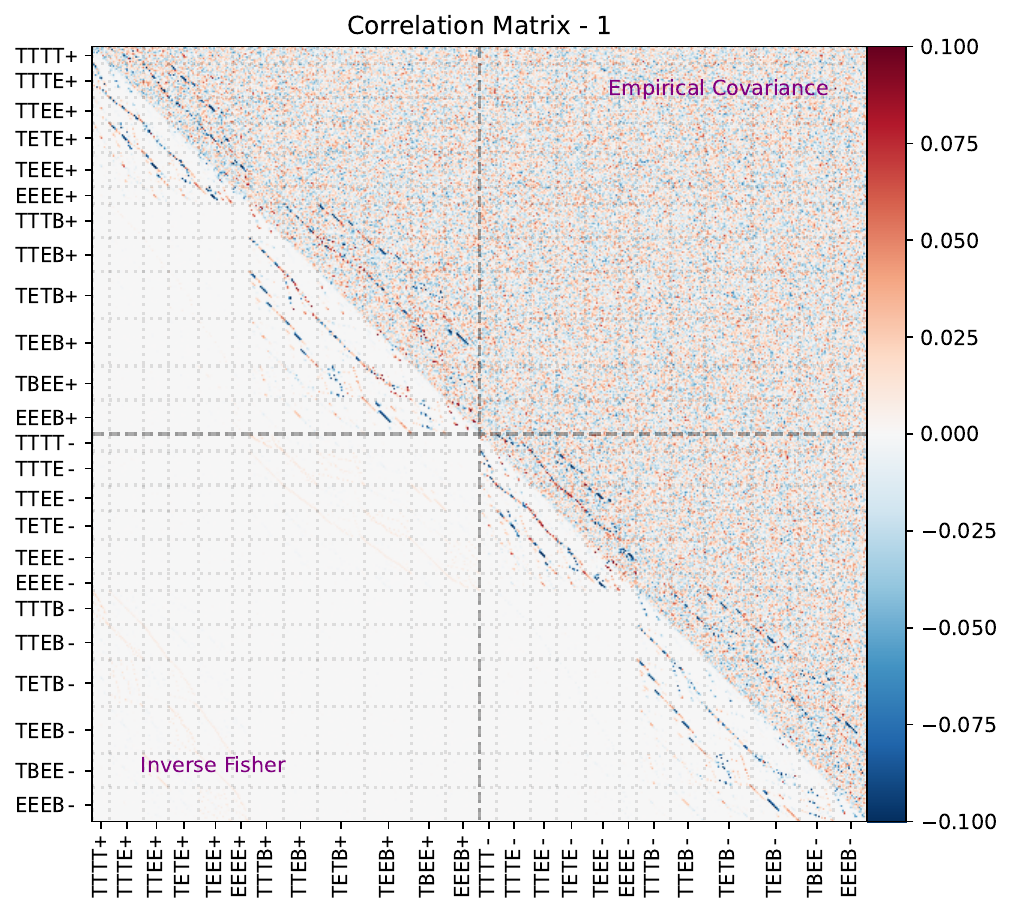}
    \caption{As Fig.\,\ref{fig: Bl-cov}, but comparing the correlation matrix of the unwindowed trispectrum measurements to that predicted from the inverse Fisher matrix shown in Figs.\,\ref{fig: Tl-data}\,\&\,\ref{fig: Tl-var}. Although the empirical data (obtained from $1000$ mocks) is somewhat noisy, we find excellent agreement for all scales and parities considered.}
    \label{fig: Tl-cov}
\end{figure}

The correlation structure is considered in detail in Fig.\,\ref{fig: Tl-cov}. We find significant covariance at $\mathcal{O}(10\%)$ between different combinations of fields, principally sourced by the intrinsic $TE$, $TB$ and $EB$ power spectra in our testing suite. Spectra with $\chi=1$ and $\chi=-1$ are largely uncorrelated, but we observe highly non-trivial patterns within each trispectrum configuration. As for the scalar case \citep{Philcox:2023uwe}, this is partly due to an internal degeneracy in the trispectrum definition due to the two possible tetrahedron diagonals; further correlations also arise from the masking, which correlates adjacent-in-$\ell$ bins (which are not always adjacent in one-dimensional projections). We find no evidence for deviation of the inverse Fisher matrix from the empirical correlations; furthermore, the complex structure of this matrix is important to take into account any studies requiring the statistical properties of trispectra. Finally, we note that our Fisher matrices are highly converged; using only $N_{\rm it} = 10$ simulations to define the Fisher matrix changes the trispectra by at most $0.018\sigma$, thus our choice is overly conservative.

\begin{figure}
    \centering
    \includegraphics[width=\textwidth]{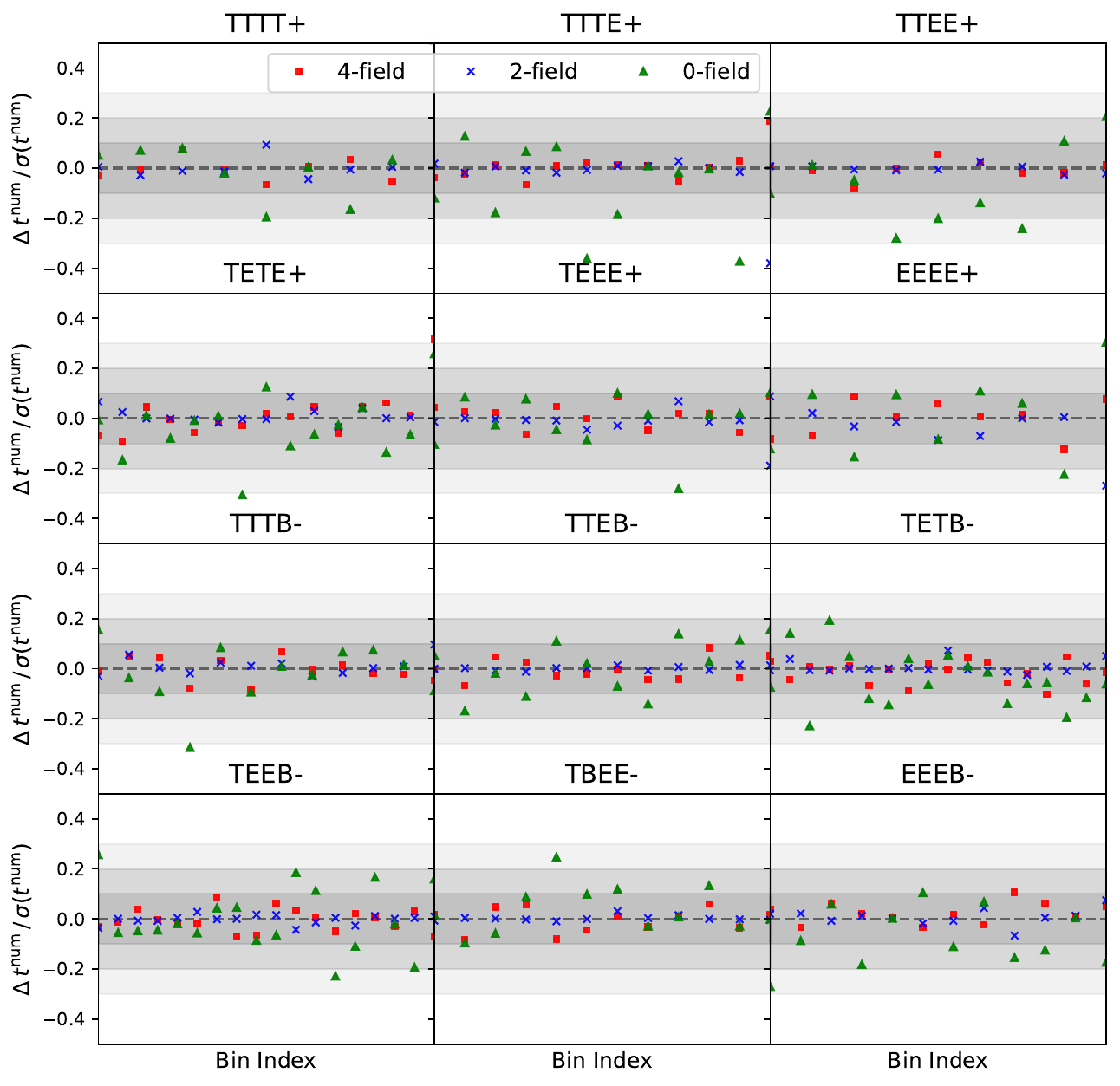}
    \caption{
    Comparison of the four-, two- and zero-field terms entering the unwindowed trispectrum estimator, averaged over 1000 GRFs with a uniform mask. In each case, we compare the measured numerator to the theoretical prediction given in \eqref{eq: GRF-disconnected-tl}, normalizing by the standard deviation of the full statistic. Results are displayed for all combinations of fields and bins where the disconnected term is non-zero. In all cases, we find good agreement though note that the zero-field term exhibits largest variation since this is an average over only $N_{\rm it}=100$ simulations, as opposed to the $1000$ used in the four-field term. This serves to validate that the trispectrum estimators are unbiased (noting that parity-odd estimators differ only by a sign, and the full estimator is the sum of the above components).}
    \label{fig: Tl-data-gauss}
\end{figure}

Whilst the above results demonstrate that the trispectrum estimators do not induce spurious signals, they do not sufficiently demonstrate that our pipeline is unbiased (though agreement of empirical and theoretical variances is itself a strong constraint). To further test the approach, we consider the individual components of the trispectrum applied to unwindowed GRF data. In this regime, we have a definite prediction of the four-, two- and zero-field components, with $\mathbb{E}[\mathcal{F}_4\widehat{t}_0] = \mathbb{E}[\mathcal{F}_4\widehat{t}_4] = -(1/2)\mathbb{E}[\mathcal{F}_4\widehat{t}_2]$, and 
\beq\label{eq: GRF-disconnected-tl}
    \mathbb{E}\left[\F_4\widehat{t}_0\right]^u_\chi(\vec b,B)&=&\frac{\delta^{\rm K}_{\chi \,p_u,1}}{\Delta_4^u(\vec b)}\sum_{L}\Theta_L(B)\Theta_{\ell_1}(b_1)\Theta_{\ell_2}(b_2)\tj{\ell_1}{\ell_2}{L}{-1}{-1}{2}^2\frac{(2\ell_1+1)(2\ell_2+1)(2L+1)}{4\pi}\nonumber\\
    &&\,\times\,(-1)^{\ell_1+\ell_2+L}\bigg(\delta^{\rm K}_{b_1b_3}\delta^{\rm K}_{b_2b_4}\mathbb{C}^{-1,u_1u_3}_{\ell_1}\mathbb{C}^{-1,u_2u_4}_{\ell_2}+\delta^{\rm K}_{b_1b_4}\delta^{\rm K}_{b_2b_3}\mathbb{C}^{-1,u_1u_4}_{\ell_1}\mathbb{C}^{-1,u_2u_3}_{\ell_2}\bigg),
\eeq
upon realization averaging. This is non-zero only for particular configurations, in particular pairs of equal bins with non-vanishing two-point correlators. Fig.\,\ref{fig: Tl-data-gauss} displays all such contributions relative to the theoretical model, and, in every case, we find good agreement, which implies that the sum of the three contributions is consistent with zero (as previously noted). The dominant source of noise here is the zero-field (data-independent term); for a single simulation, this is relatively small compared to the scatter in the four-field term, but contributes an additional source of variance that should be included in any likelihood analysis. Since the parity-breaking spectra not shown above differ only by a factor of $(-1)^{\ell_1+\ell_2+\ell_3+\ell_4}$, this test should be sufficient to convince us that our pipeline is unbiased.

\section{Summary}\label{sec: summary}
In this work, we have presented minimum-variance estimators for the polyspectra of scalar and tensor fields defined on the two-sphere, derived by extremizing the theoretical likelihood of the dataset. We have considered both the general forms (with an arbitrary weighting scheme $\Si$) and the idealized limits, and, in all cases, given practical algorithms and efficient code for their computation. Such approaches can be used to measure binned correlators for a range of spin-$0$ and spin-$2$ fields, and take into account the effects of leakage between different bins, polarizations, and parities, due to effects such as intrinsic correlations, survey masks, and inpainting. All estimators have been verified using suites of Gaussian and non-Gaussian simulations, and are found to yield unbiased results, with variances approaching the optimal limits. Furthermore, they can be efficiently computed using spin-weighted spherical harmonic transforms and Monte Carlo summation (with normalization matrices requiring only $\mathcal{O}(10)$ realizations), with the higher-point optimal estimators becoming significantly more practical to implement than their idealized equivalents.

It is interesting to consider potential applications of the above techniques. A particularly promising avenue is the investigation of inflationary higher-point functions, as traced by large-scale CMB temperature and polarization anisotropies. Whilst some of this work has already been performed in the context of \textit{Planck} \citep[e.g.,][]{2006JCAP...05..004C,Planck:2015zfm,Komatsu:2003iq,Planck:2019kim,Duivenvoorden:2019ses,2014A&A...571A..24P,PhilcoxCMB}, many further studies are possible, and allow for novel connections to theoretical disciplines such as the `Cosmological Collider' program, through the empirical search for signatures of primordial particle interactions \citep[e.g.][]{Arkani-Hamed:2015bza}. Much of this work will benefit from the addition of the trispectrum, in particular its polarized components, which to our knowledge, have not been previously been considered. For the \textit{Planck} bispectrum, addition of polarization modes to the temperature three-point function \resub{was found to the roughly double the sensitivity to primordial physics, as evidence through bounds on the non-Gaussianity parameter $f_{\rm NL}$} \citep{Planck:2019kim}; a similar result can likely be obtained \resub{for the associated parameters $g_{\rm NL}$ (and possibly $\tau_{\rm NL}$) arising in the four-point function}. Measurements of the parity-odd $E$ mode trispectrum (and its combination with $T$, e.g., $TTTE$), would bolster the sensitivity to sources of primordial parity-violation, improving upon the results of \citep{PhilcoxCMB,Cabass:2022rhr}. The $B$ modes could also be of use, and would allow connection between inflationary gravitational wave theory and observations \citep[e.g.,][]{Bonifacio:2022vwa,Cabass:2021fnw,Pajer:2020wnj}.

Though this paper has been framed in the language of CMB analyses, this is not a restriction, and the technology contained within can be similarly applied to any other spin-zero and/or spin-two field defined on the two-sphere. Another interesting cosmological candidate is the measurement of higher-order cosmic shear correlators, and their connection to the spin-zero galaxy overdensity. Historically, most analysis of higher-order shear correlators has been performed in position space, rather than harmonic space; the techniques developed in this work allow for efficient computation of the shear bispectra and beyond, in a manner that can be simply connected to observations. This obviates common problems such as the difficulty of convolving theory models with masks for statistics beyond the power spectrum, and should facilitate tighter constraints to be wrought on cosmological and astrophysical parameters, in combination with conventional two-point analysis. Finally, we hope that applications may be found beyond cosmology; many other fields consider stochastic spherical data, such as climate science and geology, and could potentially benefit from such techniques.

\acknowledgments
{\small
\noindent We thank Adri Duivenvoorden, Maresuke Shiraishi, and Masahiro Takada for insightful discussions. OHEP is a Junior Fellow of the Simons Society of Fellows and thanks Kavli IPMU for hosting a visit during which this work was started. The author is supported calorficially by 7/11 (Japanese edition) and thanks the hospitality team at the Simons Foundation for demonstrating that there is such a thing as a free lunch.
}

\appendix

\bibliographystyle{apsrev4-1}
\bibliography{refs}

\end{document}